%% file: PandaX-III_CDR.tex
\documentclass{article}
\pdfoutput=1
\usepackage{authblk}
\usepackage[verbose,bindingoffset=0in,top=1in,headheight=48pt,headsep=20pt,bottom=1in,footskip=20pt,left=1.25in,right=1.25in]{geometry}
\usepackage{lineno, url}
\usepackage{multirow}
\usepackage{array}
\usepackage{bm}
\usepackage[load-configurations=abbreviations,detect-all=true]{siunitx}
\usepackage{amsmath, amssymb}
\usepackage{mhchem}
\usepackage{indentfirst}
\usepackage{fancyhdr}

\usepackage{graphicx}  
\DeclareGraphicsExtensions{.eps,.pdf,.jpg,.png}
\graphicspath{{./fig/}}

\input{defs.tex}

\title{PandaX-III: Searching for Neutrinoless Double Beta Decay with High Pressure $^{136}$Xe  Gas Time Projection Chambers}

\input{PANDAX-III_collaboration.tex}

\date{\today}
\begin{document}

\maketitle

\begin{abstract}
Searching for the Neutrinoless Double Beta Decay (NLDBD) is now regarded as the topmost promising technique to explore the nature of neutrinos after the discovery of neutrino masses in oscillation experiments.
PandaX-III (Particle And Astrophysical Xenon Experiment III) will search for the NLDBD of \xeots at the China Jin Ping underground Laboratory (CJPL). 
In the first phase of the experiment, a high pressure gas Time Projection Chamber (TPC) will contain 200 kg, 90\% \xeots enriched gas operated at 10 bar. 
Fine pitch micro-pattern gas detector (Microbulk Micromegas) will be used at both ends of the TPC for the charge readout with a cathode in the middle.
Charge signals can be used to reconstruct tracks of NLDBD events and provide good energy and spatial resolution. 
The detector will be immersed in a large water tank to ensure $\sim$5~m of water shielding in all directions. 
The second phase, a ton-scale experiment, will consist of five TPCs in the same water tank, with improved energy resolution and better control over backgrounds.
\end{abstract}
\input{Introduction.tex}

\input{overview.tex}

\input{TPC.tex}

\input{HighPressureVessel.tex}

\input{GasSystem.tex}

\input{WaterTank.tex}

\input{Electronics.tex}

\input{RadScreening}

\input{Simulation.tex}

\input{Sensitivity.tex}

\input{FutureRnD.tex}

\input{Conclusions.tex}

\input{Acknowledgment.tex}

\bibliographystyle{JHEP}
\bibliography{PandaX-III_CDR}

\end{document}

%% file: defs.tex
\usepackage{xspace}

\newcommand{\TM}{\emph{Topmetal}\xspace}

\newcommand{\TMIIm}{\mbox{\emph{Topmetal-II\raise0.5ex\hbox{-}}}\xspace}
\newcommand{\TMS}{\mbox{\emph{Topmetal-S}}\xspace}

\newcommand{\Piii}{PandaX-III\xspace}
\newcommand{\Qbb}{\ensuremath{Q_{\beta\beta}}\xspace}
\newcommand{\mbb}{\ensuremath{m_{\beta\beta}}\xspace}

\newcommand{\AmAlpha}{$^{241}$Am\xspace}
\newcommand{\xeots}{\ce{^{136}Xe}\xspace}
\newcommand{\utte}{$^{238}$U\xspace}
\newcommand{\thttt}{$^{232}$Th\xspace}
\newcommand{\cose}{$^{60}$Co\xspace}

\newcommand{\tltze}{$^{208}$Tl\xspace}

\newcommand{\rnttz}{$^{220}$Rn\xspace}
\newcommand{\natt}{\ce{^{22}Na}\xspace}
\newcommand{\csots}{\ce{^{137}Cs}\xspace}

\newcommand{\ckky}{c/(keV$\cdot$kg$\cdot$y)\xspace}

\DeclareSIUnit\keV{keV}
\DeclareSIQualifier\ee{ee}
\DeclareSIQualifier\nr{nr}
\DeclareSIUnit\pe{p.e.}

\usepackage{array}
\newcolumntype{L}[1]{>{\raggedright\let\newline\\\arraybackslash\hspace{0pt}}m{#1}}
\newcolumntype{C}[1]{>{\centering\let\newline\\\arraybackslash\hspace{0pt}}m{#1}}
\newcolumntype{R}[1]{>{\raggedleft\let\newline\\\arraybackslash\hspace{0pt}}m{#1}}

%% file: PANDAX-III_collaboration.tex
\author[1]{Xun Chen}
\author[1]{Changbo Fu}
\author[1]{Javier Galan}
\author[1]{Karl Giboni}
\author[1]{Franco Giuliani}
\author[1]{Linghui Gu}
\author[1]{Ke Han\footnote{Corresponding Author: Ke Han, ke.han@sjtu.edu.cn}}
\author[1, 10]{Xiangdong Ji} 
\author[1]{Heng Lin}
\author[1]{Jianglai Liu}
\author[1]{Kaixiang Ni}
\author[1]{Hiroki Kusano}
\author[1]{Xiangxiang Ren}
\author[1]{Shaobo Wang}
\author[1]{Yong Yang}
\author[1]{Dan Zhang}
\author[1]{Tao Zhang}
\author[1]{Li Zhao} 

\author[2]{Xiangming Sun}

\author[3]{Shouyang Hu}
\author[3]{Siyu Jian}
\author[3]{Xinglong Li}
\author[3]{Xiaomei Li}
\author[3]{Hao Liang}
\author[3]{Huanqiao Zhang}
\author[3]{Mingrui Zhao}
\author[3]{Jing Zhou}

\author[4]{Yajun Mao}
\author[4]{Hao Qiao}
\author[4]{Siguang Wang}
\author[4]{Ying Yuan}

\author[5]{Meng Wang}

\author[6]{Amir N. Khan}
\author[6]{Neill Raper}
\author[6]{Jian Tang}
\author[6]{Wei Wang}
 
\author[7]{Jianing Dong}
\author[7]{Changqing Feng}
\author[7]{Chen Li}
\author[7]{Jianbei Liu}
\author[7]{Shubin Liu}
\author[7]{Xiaolian Wang}
\author[7]{Danyang Zhu}
  
\author[8]{Juan F. Castel}
\author[8]{Susana Cebri\'an}
\author[8]{Theopisti Dafni}
\author[8]{Javier G. Garza}
\author[8]{Igor G. Irastorza}
\author[8]{Francisco J. Iguaz}
\author[8]{Gloria Luz\'on}
\author[8,1]{Hector Mirallas}

\author[9]{Stephan Aune}
\author[9]{Eric Berthoumieux}
\author[9]{Yann Bedfer}
\author[9]{Denis Calvet}
\author[9]{Nicole d'Hose} 
\author[9]{Alain Delbart}
\author[9]{Maria Diakaki}
\author[9]{Esther Ferrer-Ribas}
\author[9]{Andrea Ferrero}
\author[9]{Fabienne Kunne}
\author[9]{Damien Neyret}
\author[9]{Thomas Papaevangelou}
\author[9]{Franck Sabati\'e}
\author[9]{Maxence Vanderbroucke}
 
\author[10]{Andi Tan}
 
\author[11]{Wick Haxton}
\author[11]{Yuan Mei}
 
\author[12]{Chinorat Kobdaj}
\author[12]{Yu-Peng Yan}
 
\affil[1]{INPAC and Department of Physics and Astronomy, Shanghai Jiao Tong University, Shanghai Laboratory for Particle Physics and Cosmology, Shanghai 200240, China }

\affil[2]{College of Physical Science and Technology, Central China Normal University,  Wuhan 430079, China}

\affil[3]{China Institute of Atomic Energy, Beijing 102413, China }

\affil[4]{School of Physics, Peking University, Beijing 100871, China}
\affil[5]{School of Physics and Key Laboratory of Particle Physics and Particle Irradiation (MOE), Shandong University, Jinan 250100, China }
\affil[6]{School of Physics and Engineering, Sun Yat-Sen University, Guangzhou 510275, China}
  
\affil[7]{Department of Modern Physics, University of Science and Technology of China, Hefei 230026, China;
State Key Laboratory of Particle Detection and Electronics, University of Science and Technology of China, Hefei 230026, China}
\affil[8]{University of Zaragoza, C/P Cerbuna 12 50009, Zaragoza, Spain }
\affil[9]{IRFU, CEA, Universit\'e Paris-Saclay, F-91191 Gif-sur-Yvette, France}
\affil[10]{Department of Physics,  University of Maryland, College Park, MD 20742, USA}
\affil[11]{ Nuclear Science Division, Lawrence Berkeley National Laboratory, Berkeley, CA 94720, USA }
\affil[12]{Center for Excellence in High Energy Physics and Astrophysics, Suranaree University of Technology, Nakhon Ratchasima 30000, Thailand}

%% file: Introduction.tex

\section{Physics Motivation}
\label{sec:intro}

The Standard Model (SM) of elementary particle physics has been extremely successful in explaining the matter contents of the Universe and their fundamental interactions~\cite{PDG2014}.
However, one glaring exception is about the nature of neutrinos.
The formulation of the SM has been accomplished with massless neutrinos, while data from a wealth of past and ongoing neutrino experiments like Super-Kamiokande~\cite{SK:1998}, SNO~\cite{SNO:2002} and many others
(see, e.g.,~\cite{Xing:2011zza, Barger:2012pxa} and references within) have firmly proven, since 1998, the phenomenon of neutrino oscillations.
The most natural explanation of neutrino oscillations is a non-differential coupling with the form of a mass matrix.
Massive neutrinos give rise to a host of new exciting possibilities beyond the physics of the SM, which are yet to be explored.

One of the questions about neutrinos is how their masses are generated.
One possibility is that the neutrinos interact with Higgs fields like their charged fermion partners.
The left-handed neutrinos turn into right-handed ones through such interactions.
In this case, they are called Dirac neutrinos.
However, the right-handed neutrinos and the left-handed anti-neutrinos are SM singlets, and if they exist, their creation cross sections in the SM processes must be negligible up to the energy scales and luminosities which have been probed to date.
Moreover, the tiny values of neutrino masses imply extreme fine-tuning of the couplings between Higgs fields and neutrinos, which are suppressed by $\gtrsim$12 orders of magnitude compared to the natural size!

An interesting possibility arises from the fact that the neutrinos are electrically neutral fermions: a new interaction is allowed which couples the left-handed neutrino to the right-handed anti-neutrino field. This self-interaction generates a new mass term, different from that of Dirac fermions.
The existence of such a mass term implies that neutrinos can have one half as many degrees of freedom as Dirac neutrinos, or equivalently, that neutrinos and anti-neutrinos are identical.
Since the possibility that, more generally, neutral fermions may have this self-interaction was first suggested by E. Majorana~\cite{MajoranaPaper} in 1937, such neutrinos are called Majorana fermions.
So far, no fundamental Majorana fermion has been identified in nature, although there are some evidences for composite Majorana fermions in condensed matter systems~\cite{Elliott:2014iha, Ryu:2010zza}.

If the neutrinos are Majorana fermions, the lepton number conservation --- an accidental symmetry in the SM --- is manifestly violated~\cite{Avignone:2007fu}.
Discovery of lepton number non-conservation will be an important breakthrough in fundamental physics, with far-reaching implications.
For example, the matter dominance over anti-matter in the observable universe is a profound puzzle in science.
Understanding this puzzle requires, among others, lepton/baryon number-violating processes which have yet to be established~\cite{Luty:1992un}.

A leading experimental method to uncover the possible Majorana nature of the neutrinos is through the discovery of a rare nuclear weak-decay process called neutrinoless double beta decay (NLDBD)~\cite{Avignone:2007fu}.
Certain even-N and even-Z nuclei cannot undergo the normal beta decay by emitting a single electron (positron) and an anti-neutrino (neutrino) because the required kinematic phase space is not available. However, they can decay through second-order weak interactions in which two electrons (positrons) and two anti-neutrinos (neutrinos) are emitted simultaneously.
This SM double beta decay (DBD) process has been observed experimentally.
If neutrinos are Majorana particles, a second DBD mode becomes possible: a nucleus can decay to its daughter by emitting just two electrons (or positrons) without anti-neutrinos (neutrinos), hence the name NLDBD.
In the process, the Majorana neutrino is exchanged between the two neutrons (protons) emitting the two e$^-$ (e$^+$).
As NLDBD manifestly violates lepton number conservation, its observation will be a clear signal that the neutrinos are their own antiparticles.

One of the important features of the NLDBD is its sensitivity to the absolute scale of the neutrino masses. If the neutrinos are Majorana particles, their mass is the constant of the neutrino self-coupling mentioned above, so the NLDBD rate is proportional to the squared effective Majorana neutrino mass $\langle m_{\beta\beta}\rangle$,
\begin{equation}
    \left(T^{0\nu}_{1/2}\right)^{-1}  = G^{0\nu} |M^{0\nu}|^2 \frac{\langle m_{\beta\beta}\rangle^2}{m_e^2}
\end{equation}
where $G^{0\nu}$ is the phase-space factor and $M^{0\nu}$ is the nuclear matrix element.
$\langle m_{\beta\beta}\rangle$ is defined as $|\sum_i U_{ei}^2 m_i|$, where $U_{ei} $ is the electron neutrino mixing matrix elements into $i$-th eigenstate with eigenmass $m_i$. Thus by knowing the decay rate, the nuclear matrix element, and matrix elements $U_{ei}$, and the mass differences, one can determine the absolute neutrino mass scale. It turns out that $\langle m_{\beta\beta} \rangle$ is very sensitive to the so-called neutrino mass hierarchy, i.e., whether the smallest neutrino mass is $m_1$ (normal hierarchy) or $m_3$ (inverted hierarchy).
There are several on-going experiments such as JUNO ~\cite{An:2015jdp} RENO-50~\cite{RENO-50} and DUNE~\cite{LBNE} which aim to determine the mass hierarchy, and we probably will know the answer in the next five to ten years.

Given the scientific importance of NLDBD, many groups around the world have spent decades on perfecting experimental technologies for making a discovery.
For a review of the current status of various experiments, see the joint US DOE/NSF panel report to the Nuclear Science Advisory Committee of November 2015~\cite{NLDBD_NSAC}.
The leading experiments in the world now include CUORE~\cite{Artusa:2014lgv}, EXO-200~\cite{Albert:2014awa}, GERDA~\cite{Agostini:2016iid}, KamLAND-Zen~\cite{KamLAND-Zen:2016pfg}, Majorana~\cite{Abgrall:2013rze}, and SNO+~\cite{Andringa:2015tza}, etc.
The best limit published so far is by the KamLAND-Zen group on the NLDBD half-life of $^{136}$Xe, $T_{1/2}^{0\nu}>1.07\times10^{26}$y~\cite{KamLAND-Zen:2016pfg}.
It is generally agreed that the immediate goal of the worldwide NLDBD experiments shall be to determine the nature of neutrinos when the neutrino mass spectrum has the inverted hierarchy.
To reach this goal, one needs to have at least one ton of a candidate isotope.
Therefore, in the 2015 long-range plan of the US nuclear science community, the second recommendation to the US funding agencies is "the timely development and deployment of a U.S.-led ton-scale neutrinoless double beta decay experiment"~\cite{Geesaman:2015fha}.

To make a successful ton-scale experiment, one has to make major efforts in at least the following areas: 1) successful procurement of a ton of isotope, 2) excellent energy resolution in the detector, 3) superb background reduction, to the level 0.1 event/ton/year in the region of interest, 4) novel ways to distinguish true NLDBD events from the background.

\xeots is one of the most common isotopes for NLDBD experiments, due to its relatively rich abundance (8.9\%) in natural xenon, and to the relatively low cost of enrichment. In fact, there is already one ton of enriched \xeots in the world, which is an important advantage. It also has relatively high two-electron endpoint ($Q_{\beta\beta}$= 2458\,keV), and a large lifetime for the two-neutrino mode (T$_{1/2}$$\sim$2.3$\cdot$10$^{21}$ years, as recently measured by EXO-200~\cite{Albert:2013gpz} and KamLAND-Zen~\cite{KamLANDZen:2012aa}), which reduces the overlap of the populations of the two neutrinos and the neutrinoless modes. The major backgrounds are gammas coming from $^{214}$Bi and from $^{208}$Tl, related to the $^{238}$U and $^{232}$Th decay chains. In fact, the gamma line from $^{214}$Bi at 2.448 MeV is only 10 keV lower than the NLDBD threshold, and separating this background from the potential physical events represents a major challenge.
So far the NLDBD experiments using the \xeots isotope include EXO-200 and its upgraded version nEXO~\cite{Pocar:2015ota}, KamLAND-Zen~\cite{KamLAND-Zen:2016pfg}, and NEXT~\cite{Martin-Albo:2015rhw}.
The EXO-200 experiment employed a liquid time projection chamber (TPC) as detector, and so does the nEXO design.
KamLAND-Zen dissolves \xeots in liquid scintillator and detects the final-state electrons through its scintillation.
The NEXT collaboration in Spain uses high-pressure gas TPCs, in which proportional scintillations from the ionized electrons are observed with PMTs.
Two important advantages of the gas TPC are that the intrinsic energy resolution can be much better than in the liquid TPC, and that the electron tracks in the gas TPC can be used to distinguish uniquely NLDBD events from backgrounds.

In this conceptual design report, we describe the PandaX-III project, a plan to realize a ton-scale \xeots NLDBD experiment using high-pressure gas TPCs in the China Jinping Underground Laboratory (CJPL).
Unlike the NEXT experiment which also employs the gas TPC,
the PandaX-III TPC will read out the ionized electrons directly, using at first Microbulk Micromegas, followed by improvements based both on continued R\&D on Micromegas and on new technologies, such as the \TM.
The first 200 kg module is under construction and is expected to be deployed in the next 2-3 years. In the following section, we give a brief overview of PandaX-III project.

%% file: overview.tex

\section{The Project Overview}
To search for the NLDBD mode of \xeots, \Piii experiment will use a technology of high-pressure low-background gas TPCs which was developed for the experiments in the early 1990's~\cite{Wong:1991fg, Zanotti:1991vh, Vuilleumier:1993zm}
and recently for the NEXT experiment~\cite{Martin-Albo:2015rhw}, while there has been a little follow up during in-between 20 years.
This type of experiment is challenging for a number of reasons.
First of all, there is a conflict between high-pressure vessel and low-background.
To reach the lowest background possible, one has to be creative in finding materials and methods to produce a high-pressure container while keeping the environment clean.
The second problem is the implementation of large readout planes.
One could use wire planes to read the ionization electrons, but the large electronic noise deteriorates the energy resolution, which is critical for the NLDBD search.
There are other problems such as applying high-voltages on gas TPC, etc.

However, the gas TPC technology for NLDBD does provide two huge advantages that are hard to ignore.
First of all, the event energy resolution in the gas is intrinsically better than in the liquid, due to the very limited electron-ion recombination in gas phase~\cite{BOLOTNIKOV1997360}.
In fact, just reading the ionization electrons, the gas energy resolution is nearly ten times better than that in the liquid.
Of course, combining scintillation light and ionized electrons together can ameliorate the problem to a certain extent in the latter case.
The second advantage is the capability of imaging the electron tracks, which is impossible for all liquid and solid detectors.
This advantage could be proven critical for identifying the NLDBD event by event and rejecting the gamma background effectively.
Monte Carlo simulations predict that the tracking information will help to suppress the background event rate by a factor of 20 to 100 times in a relatively simple realization of the track recording and reconstruction scheme~\cite{Cebrian:2013mza}.

\begin{figure}[tbp]
\centering
\includegraphics[width=0.45\textwidth]{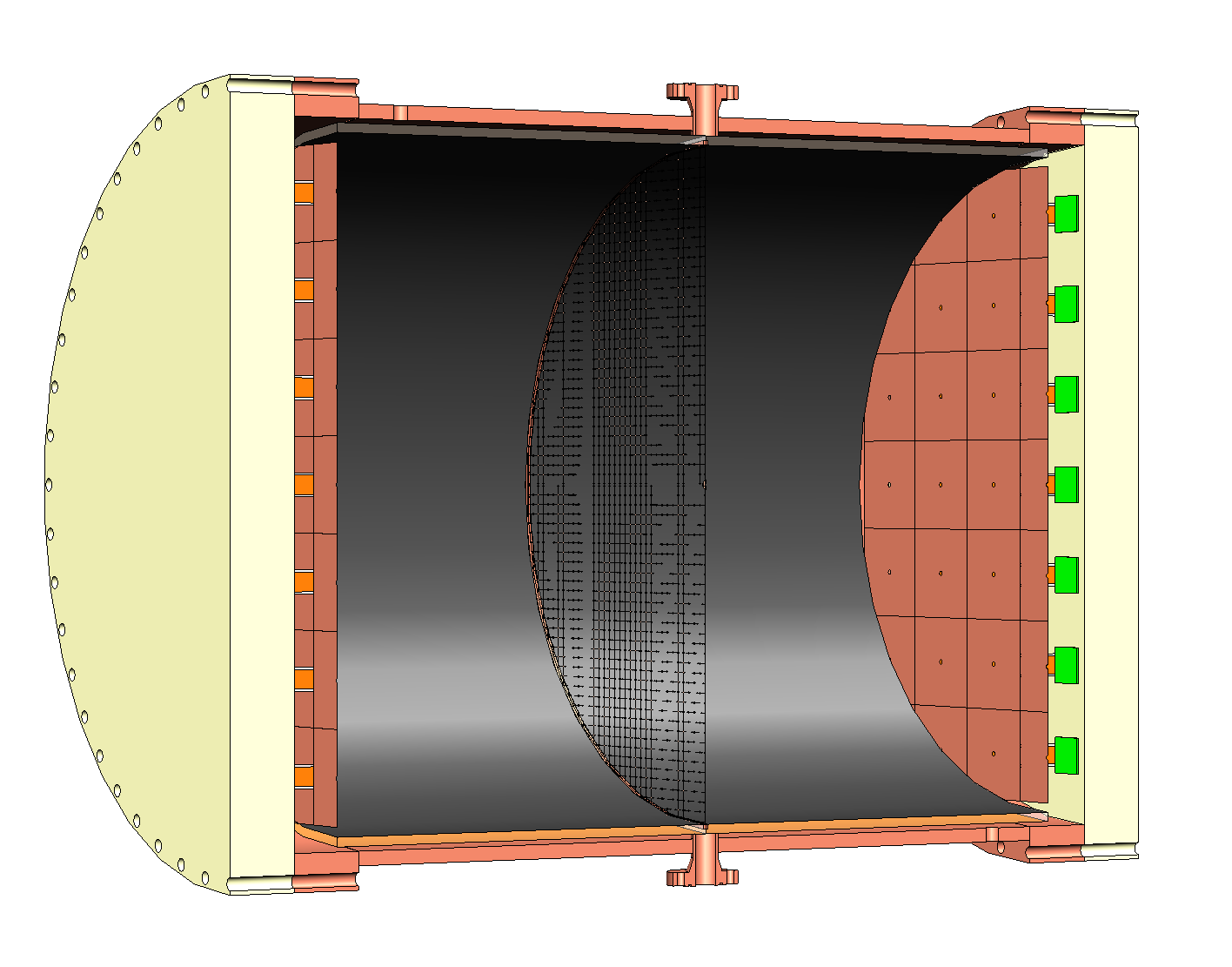}
\includegraphics[width=0.48\textwidth]{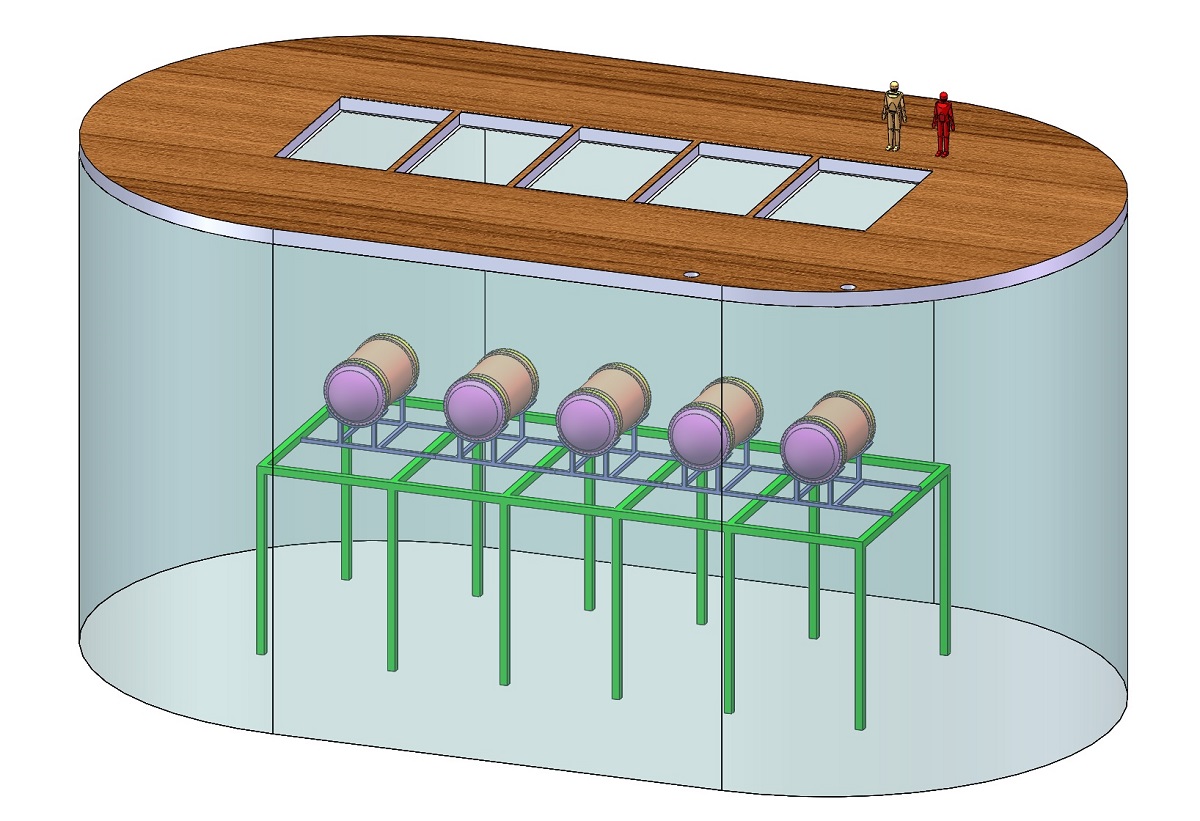}
\caption{(Left) Schematic of the \Piii TPC. (Right) Illustration of full \Piii program, with 5 TPCs immersed in a large ultra-clean water tank.}
\label{fig:detector}
\end{figure}

In a nutshell, the \Piii concept to build a ton scale \xeots NLDBD experiment is
\begin{itemize}
\item
    {to read out the TPC events using the ionization electrons, in order to avoid the background from PMT's (Fig.~\ref{fig:detector} (Left);}
\item
    {to use 90\% enriched \xeots with 1\% TMA (trimethylamine) mixture, the latter of which improves the signal quality and suppresses diffusion of drift electrons;}
\item
    {to use existing application-specific integrated circuit (ASIC) to read out the tracking and energy information;}
\item
    {to use Oxygen-Free High Conductivity (OFHC) copper to make the high-pressure vessel of the first module;}
\item{to construct a ton-scale experiment consisting of additional 200-kg modules with increasing sophistication (Fig.~\ref{fig:detector} (Right).}
\end{itemize}
These considerations are quite different from other realizations of gas TPCs, such as NEXT.

About 10~bar of xenon gas with 90\% enriched \xeots will be enclosed in an OFHC copper pressure vessel of cylindrical shape.
Its length is about 2~m and diameter about 1.5~m, resulting a total volume about 3.5~m$^3$.
A symmetrical TPC will be placed in the vessel with the cathode in the middle and two anode planes on the two ends, setting up two drift regions for ionized electron tracks.
Each drift region is about 1~m long, and the potential difference between anode and cathode will be at 100kV level.
The field cages will be made of radio-pure materials with as little dead space as possible to fully utilize \xeots.

To reach the energy resolution required in NLDBD experiments, we will use a new type of readout, Micro-MEsh Gaseous Structure or Micromegas, which has been developed in the last two decades~\cite{Giomataris:1995fq}.
A particular realization of Micromegas, called Microbulk Micromegas (MM)~\cite{Andriamonje:2010zz}, is of specific interest for rare event searches.
It has been extensively studied within the T-REX project~\cite{Irastorza:2015dcb, Irastorza:2015geo} at the University of Zaragoza.
Tests at 10~bar of Xe + TMA mixture have shown energy resolution at the NLDBD Q-value down to 3\% FWHM~\cite{Gonzalez-Diaz:2015oba}.
Thus in the first phase of the PandaX-III experiment, we will use specially designed $20\times20$~cm$^2$ MM modules, which will be manufactured at CERN.
At the same time, we will carry out active R\&D in MM to improve the quality of the readout planes.
Moreover, a new type of electron readout, called \TM~\cite{TopmetalII-2016}, has been under development by our collaborators at Lawrence Berkeley National Laboratory and Central China Normal University.
It has the potential to greatly improve the energy resolution to below 1\% FWHM.

The readout electronics is a challenge due to the radiopurity requirement and the large number of channels.
Moreover, timing and energy resolutions are also important concerns.
For the first module, we used AGET chips from the Saclay group, developed for other MM applications~\cite{ref:AGET_ASIC}.
We plan to improve them according to our refined application requirements in the future.
The front-end electronics will be shielded by 15~cm thick copper end-caps, and resulting a much reduced background contribution.

The radiopurity of the high pressure vessel is an important concern, as it is bulky and very close to the TPC.
OFHC copper has been chosen as the baseline material to make the vessel
In the past, most of the measurements on the radioactive contents in OFHC cooper yielded upper limits only.
More recently, the Majorana collaboration has improved the measurement technology and finds that some commercial OFHC cooper has uranium and thorium content at the 0.1 part per trillion~(ppt) level in mass, which is about a thousand time cleaner than the cleanest stainless steel vessel~\cite{Abgrall:2016cct}.

The copper vessel will be immersed in a ultra-clean water shield at CJPL Hall B4.
CJPL is the deepest underground lab in the world, where the cosmic ray rate is about 1$\mu$/week/m$^2$. 
Hall B4 is 65~m in length, 14~m in width, and the highest point is about 14~m. 
Civil engineering construction of the hall itself was finished in March, 2016. 
Since then, the PandaX group at Shanghai Jiao Tong University (SJTU) has commissioned  the civil engineering to excavate a pit for the construction of a water pool on the floor of the hall. 
The civil engineering of the pool was finished in June 2016. 
The ultra-clean water system will be implemented in the next year or so.
Moreover, a radon-free clean room will be set up to assemble the TPC.
The inner part of the experimental hall is a clean room of class 10000.
Dedicated class-1000 clean room will be setup for detector assembly and commissioning.

The remaining part of this report is organized as follows.
In Section 3, we lay out the details of the TPC, which is the central part of the detector, and report on the status of the ongoing prototype TPC work.
Section 4 discusses details of the high pressure (HP) vessel, of its design and its radio-purity level, while section 5 is about the high pressure gas system which is used for maintaining the required pressure level inside the vessel.
In section 6, we present the ultra-clean water system as an effective shield to background events from the experimental hall and instruments in the laboratory.
The electronics and data acquisition system is described in detail in section 7.
Section 8 is about material assay techniques to control the radioactive backgrounds coming from different parts of the detector.
In section 9, we examine our background budget with the help of two Geant4-based Monte Carlo simulation packages. 
Signal efficiency and topological reconstruction are also discussed in this section.
Section 10 describes the projected sensitivity based on detector performance and background budget. 
The last section focuses on future R\&D efforts, aiming to improve the detector performance of future modules. 

%% file: TPC.tex

\section{The TPC working at \Piii}
\label{sec:TPC}

\begin{figure}[tb]
  \centering
    \includegraphics[width=0.8\linewidth]{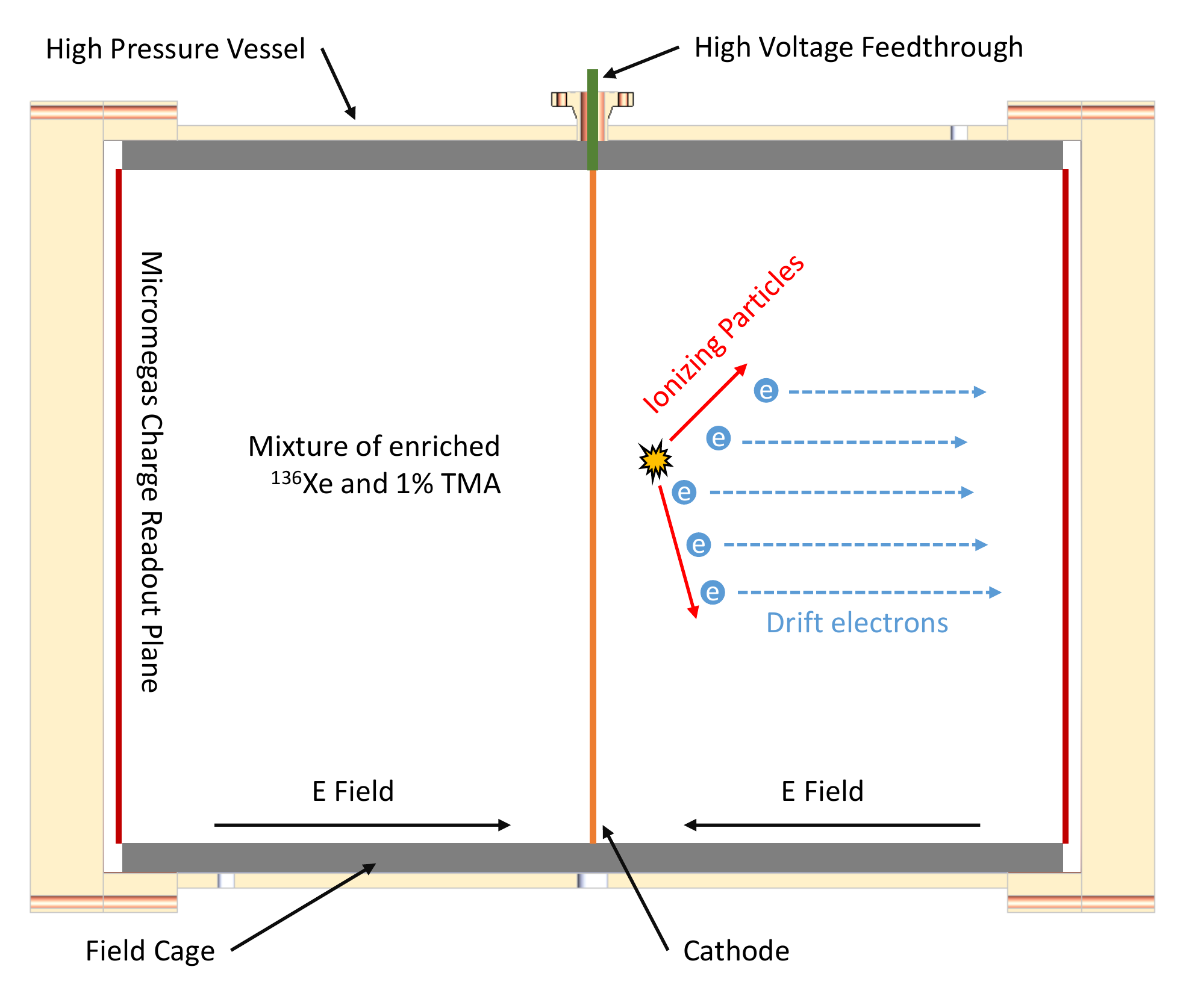}
    \caption{Schematic of the \Piii TPC, with major components listed.
    }
    \label{fig:tpc}
\end{figure}

In a gas TPC, an incident particle ionizes the gas along its trajectory and produces secondary electrons, which drift along an $E$ field and are measured by a X-Y readout plane as shown in Fig. 2.
The drift time of electrons indicates the relative Z coordinate.
So a complete 3D image (X, Y, Z) of the trajectory is recorded.
The charge signal collected can also be used to reconstruct the energy loss of the particle along its track.

The central component of the \Piii detector is a High pressure gas TPC (HpgTPC).
The cylindrical TPC features a symmetric design with cathode in the middle and charge readout planes at the two ends, as shown in Fig.~\ref{fig:tpc}.
For each half of the TPC, its active volume is about 1~m long and 1.5~m in diameter.
With a negative high voltage up to 100~kV on the cathode, the electric field pulls drift electrons to the two ends.
A field cage along the cylindrical barrel sets the boundary conditions for a homogeneous drift field in the active volume.
Drift electrons are then collected by charge readout planes at the two ends, which consist of tiles of Scalable Radio-pure Readout Modules (SR2M).

Each SR2M covers $20\times20$ cm$^2$ of readout area with Micromegas~\cite{Giomataris:1995fq} gas detectors.
Micromegas measures both the X, Y coordinates as well as the timing profile of arriving drift electrons.
The spatial and temporal resolutions of Micromegas are on the order of 1~mm and 0.1~$\mu$s, as determined by practical pixel pitch size and electronics sampling rate respectively.
The HpgTPC is enclosed in a radio-pure high pressure vessel, which holds 200 kg of xenon gas (90\% enriched \xeots) at 10 bar.
Xenon gas is mixed with 1\% TMA (trimethylamine) to improve signal quality in the Micromegas in terms of energy resolution and electron diffusion~\cite{Cebrian:2012sp, Alvarez:2013oha}.

\subsection{The field cage}
In our symmetrical design (see Fig.~\ref{fig:detector} (Left)), the cylindrical field cage is separated in the middle by the cathode on negative HV and delineated by charge readout planes at the two ends near ground potential.
Each field cage consists of an outer dielectric barrel and voltage-dividing resistors or resistive films inside the barrel.

\begin{figure}[tbp]
\centering
\includegraphics[width=0.45\textwidth]{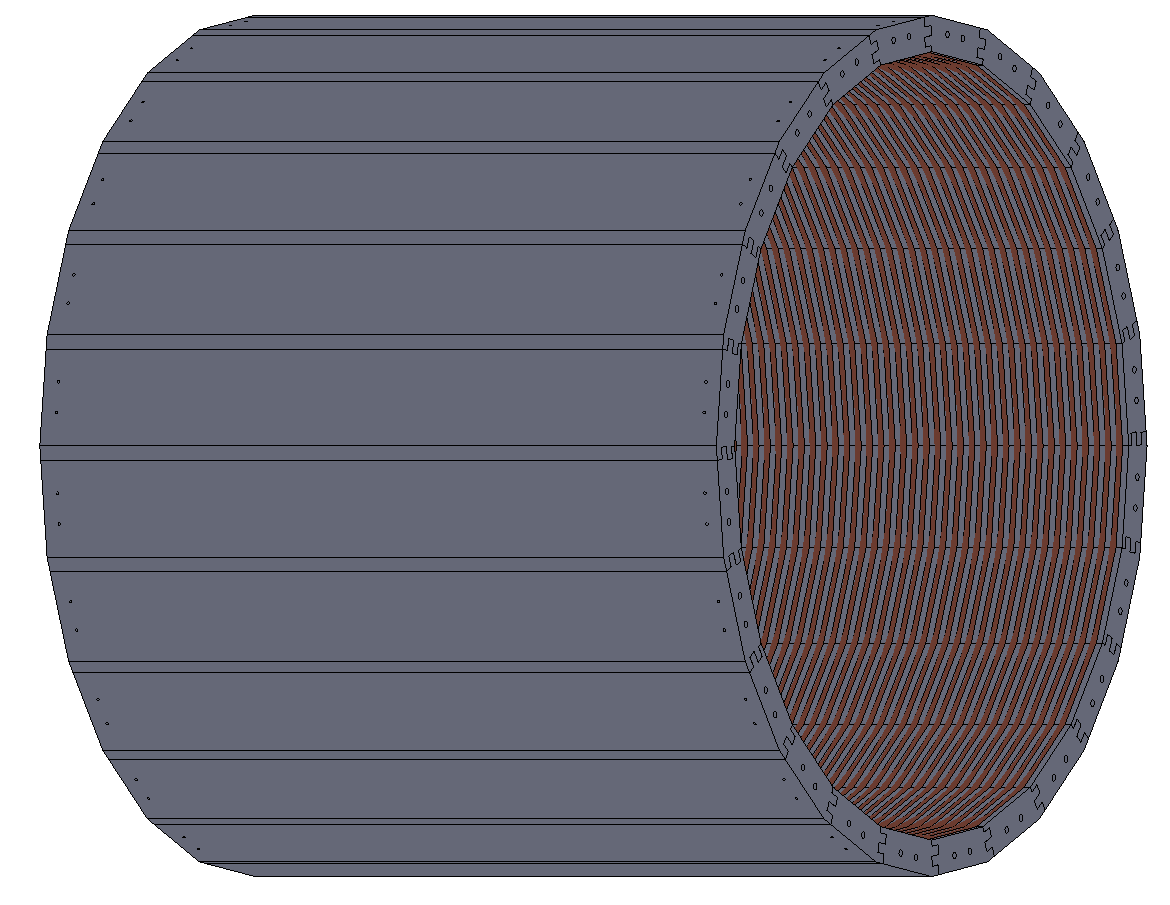}
\hspace{0.5cm}
\includegraphics[width=0.5\textwidth]{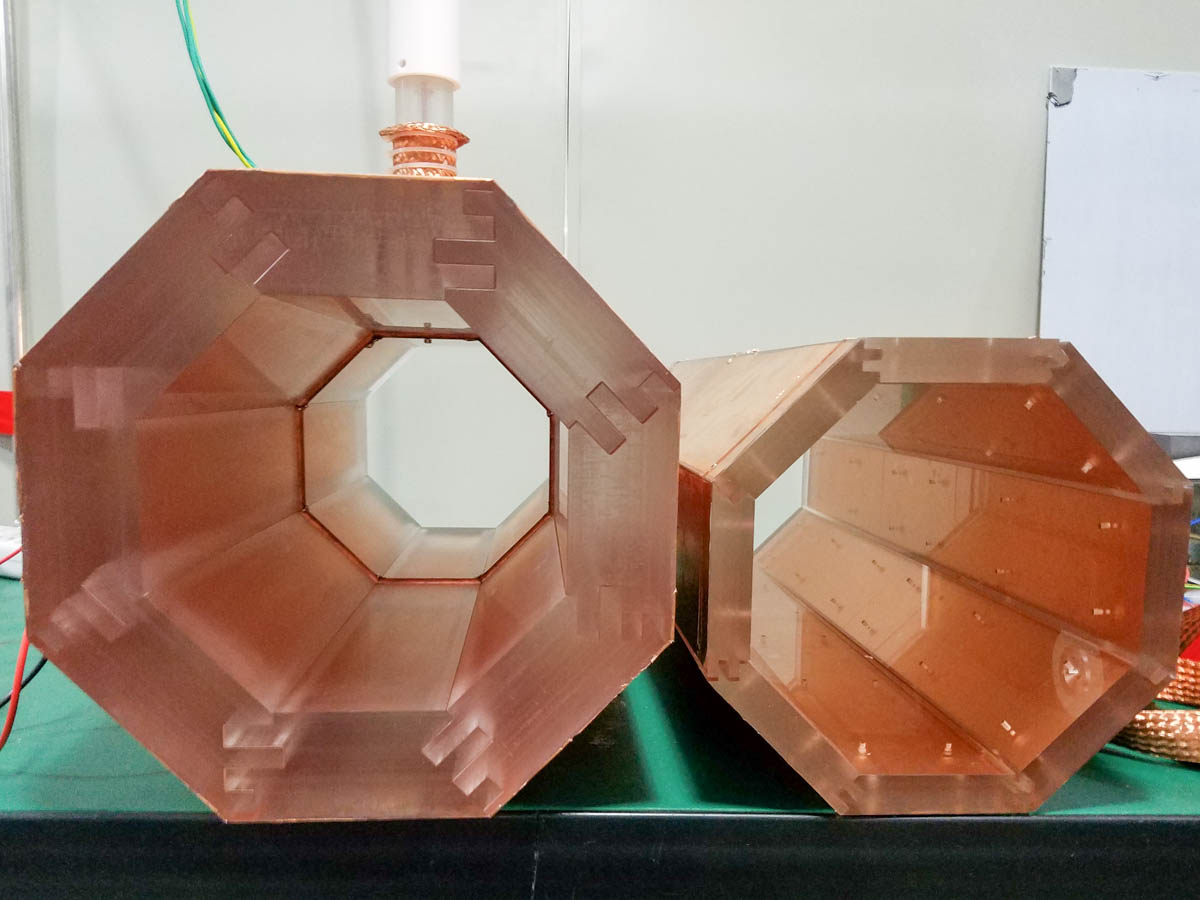}
\caption{(Left) An example design of TPC field cage with 24 acrylic side wedges. (Right) Pictures of acrylic mock-up field cage made of 8 interlocking side panels. The two setups are of different panel thicknesses.}

\label{fig:acrylicFC}
\end{figure}

Considering the relatively small dielectric strength of xenon gas, the field cage would be surrounded by more than 10~cm of xenon if no dielectric shield is used, which is extremely wasteful, especially for the enriched \xeots being used in \Piii .
In our design we use an outer barrel of 50~mm Teflon (or other dielectric material such as acrylic or nylon) to shield the cathode potential.
For each half of the field cage, the barrel consists of 24 interlocking wedges, and each piece is 1~m long and about 20~cm wide.
The cathode mesh is mounted on D-shaped OFHC copper rods, which are embedded at the joint of two acrylic barrel.
On one of the wedges a 5~mm copper rod penetrates the wedge and connects the cathode ring with a HV feedthrough.
An example setup of the TPC field cage with 24 side wedges is illustrated in Fig.~\ref{fig:acrylicFC} (Left).
Two small (25cm inner diameter) acrylic mock-up setups (Fig.~\ref{fig:acrylicFC} (Right)) were constructed to validate the design.
In each of the mock-up, 8 acrylic wedges were used.
The two set-ups differ in the thickness of the side wedges (2~cm and 5~cm).
Inside the thicker barrel, the cathode ring made of 8 copper rods can be seen in the picture.
The mock-up setups are being tested for high voltage tolerance and preliminary results are promising.

To provide the intermediary potentials for the field shaping, we envisage a novel design with the resistors and electrode structures directly deposited on the side wedges.
Since this method is not yet tested we have an interchangeable classical design with discrete copper shaping rings and voltage-dividing surface mount (SMD) resistors, which we call the classical design.
More details about the two options are as follows.

\begin{description}
\item[Classical design]
The classical copper shaping ring plus resistors design have been widely used in low background TPCs, for example, PandaX-II dark matter experiment~\cite{Tan:2016zwf}.
In the classical design the electrodes are formed out of 4~mm diameter copper rods pressed to 3~mm in one direction, giving it an oval shape.
The rods are then forced into grooves in the panels cut by a 3~mm ball end mill.
A groove along the panel contains a chain of giga-ohm level SMD high-voltage resistors connecting the copper rods.
Additional chains of resistors may be added for redundancy.
We have built for 20-kg scale prototype TPC (see Section~\ref{sec:prototype}) with field shaping structure very similar to the classical design.

\item[Novel Design]
This preferred implementation of the field shaper is much simpler conceptually.
On each dielectric panel there is just one giant resistor deposited over the full length and width of the panel.
The resistive coating connects to the cathode at one end and to the grounded mesh of the anode structure on the other.
It is equivalent to a classical structure with an infinite number of shaping electrodes.
The challenge is to identify an appropriate material for coating and the best method of deposition.
The ideal coating should be radioactively clean and homogeneous over the dimensions of the dielectric panel.
It should also adhere well to the substrate under the high pressure xenon + TMA environment over a long period time.
Diamond-Like-Carbon (DLC) coating has been used successfully for micro-patterned gas detectors such as Micromegas~\cite{Ochi:2014rda}.
We will construct a detailed field simulation to validate design parameters, especially the tolerance for non-uniformity of film resistivity.
The targeted high resistivity, which is a key requirement for TPC field shaping, will be explored with film deposition thickness and/or patterning in the film.
With special attention to material radiopurity, we will investigate different commercial films, including germanium- or DLC-coated polyimide and carbon loaded Kapton.
\end{description}

\subsection{High Voltage feedthrough system}

High Voltage (HV) feedthrough system is designed to provide the cathode plane with up to -100~kV high voltage.
As illustrated in Fig.~\ref{fig:tpc}, the HV feedthrough utilizes a small port on the side of high pressure barrel and connects the cathode plane inside to the HV power supply outside.
The major technical challenge is to minimize sparks in the TPC.
It is also non-trivial to have a feedthrough to withstand 10~bar pressure, especially considering the radiopurity requirement.

\begin{figure}[tbp]
\centering
\includegraphics[width=0.6\textwidth]{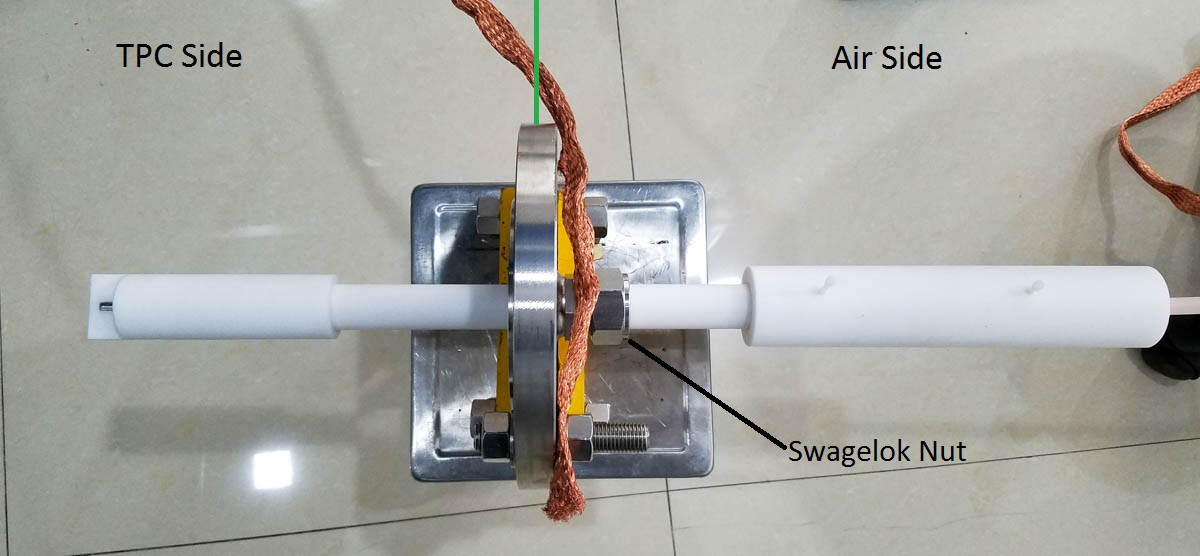}
\caption{High-voltage feedthrough for the prototype TPC. The TPC side is on the left and air side on the right. The Swagelok nut for the compression fit is visible near the flange.}
\label{fig:prototypeHV}
\end{figure}

A HV feedthrough for a single-ended prototype TPC (discussed later in Section~\ref{sec:prototype}) was constructed using a compression seal approach.
One inch Swagelok nut is used as illustrated in Fig.~\ref{fig:prototypeHV}.
A stainless steel (SS) rod with a diameter of 6~mm is pressed into a Teflon tube, which has an outside diameter of 1~inch
The Teflon tube was then clamped using a Swagelok nut and a gasket ring from air side.
The feedthrough passed the leakage test with 15~bar of nitrogen gas pressure.
Inside the prototype high pressure vessel, the SS rod pressed on to the cathode plane from below.
We were able to apply -70~kV of high voltage while the TPC is exposed in lab air and up to -95~kV in nitrogen gas environment at 10~bar.

Another prototype feedthrough was designed to work with acrylic prototype field cage.
In Fig~\ref{fig:acrylicFC} (Right), one can see the HV connection on the thicker acrylic barrel.
The feedthrough was pressed to the cathode copper bar on the side.
Preliminary test demonstrated positive results from this feedthrough design.
The HV feedthrough for \Piii will be based on the two prototypes.

\subsection{Charge readout plane}

The \Piii anode planes at both ends of the TPC are equipped with charge sensing, finely-patterned Micromegas readout modules, which provide measurement of both energy and topology of the events in the TPC.
The Micromegas concept was invented by I. Giomataris at CEA-Saclay about 20 year ago~\cite{Giomataris:1995fq} and since then it has become one of the most successful developments among Micro-Pattern Gas Detectors (MPGD), and it is increasingly being applied in many particle and nuclear physics experiments.
Like other MPGDs, Micromegas can be used as the readout planes of gas TPCs, outperforming conventional multi-wire planes in a number of aspects.
Recently, Micromegas have gained a wide attention for rare event detection applications~\cite{Irastorza:2011hh,Dafni:2012fi} because of its unique features such as energy resolution, stability and homogeneity of response, radiopurity, and etc.

\begin{figure}[tbp]
\centering
    \includegraphics[width=0.52\textwidth]{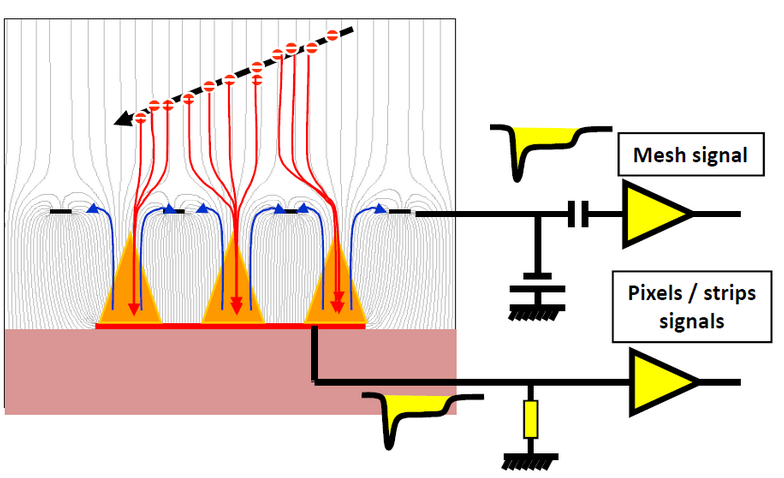}\hfill
    \includegraphics[width=0.47\textwidth]{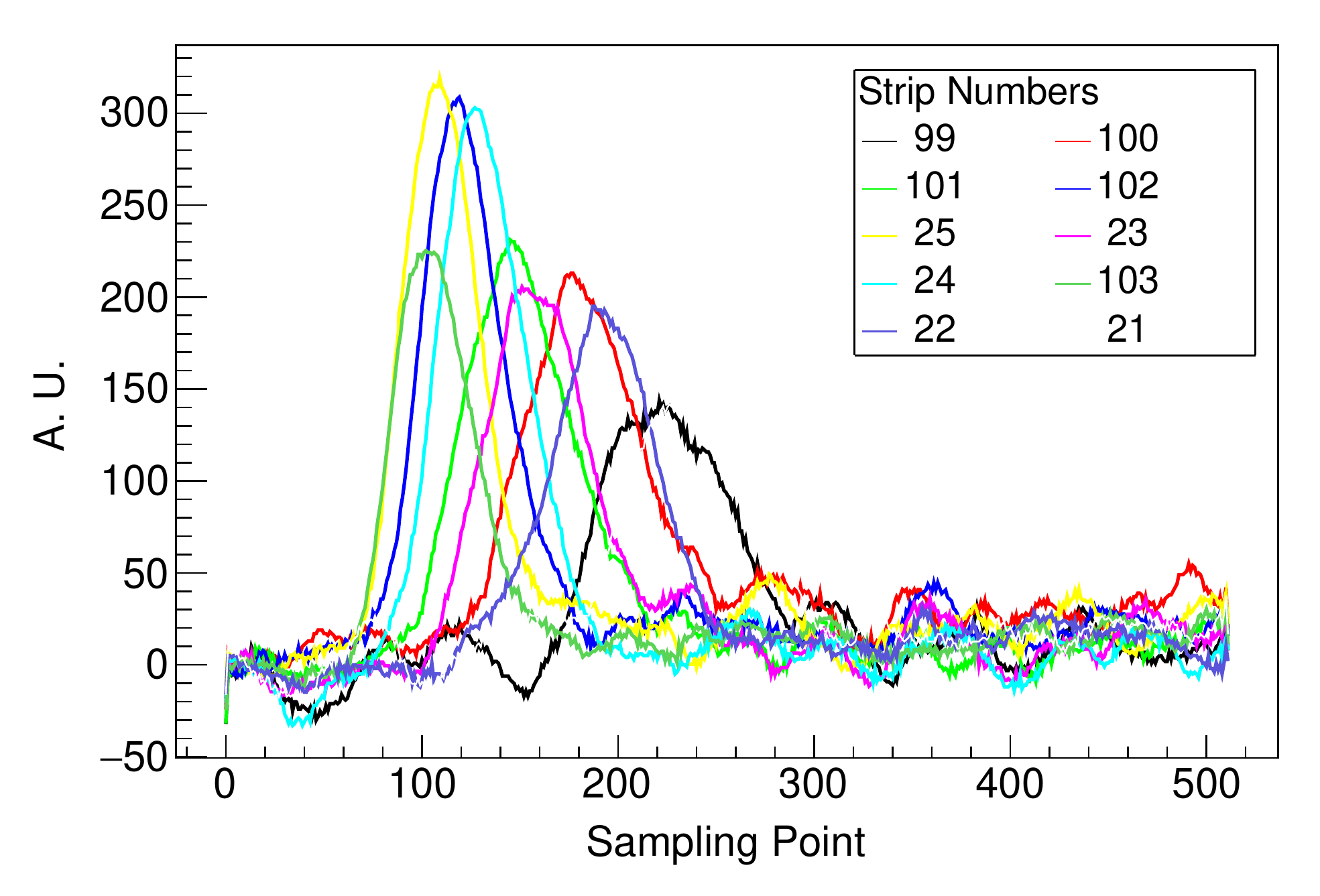}
    \caption{
    (Left) Working principle of Micromegas gas detector.
	 When ionization electrons reach the amplification gap between the micro-mesh and the readout electrode, a large number of electron/ion pairs are created by the avalanche effect.
	Electron signals and ion signals are induced on anode and mesh respectively by the movement of charges.
	Energy can be reconstructed from anode as well as mesh signals.
	Electron signals from pixelated anodes also include topological information of the ionization electron tracks.
	The electron signal is much shorter than the ionic one because of their large mobility in gas.
	So the electron signal is used to reconstruct precisely the arrival time of the particles (timing).
	Figure is taken from Ref.~\cite{Aune:2013pna}.
	(Right) Sample electron signal pulses from anodes.
	Pulses from different readout anode channels (X-Y strips in this example) carry the charge and timing info.
	The sampling rate is 10 MHz.
    }
    \label{fig:Micromegas}
\end{figure}

Micromegas devices are characterized by a micro-mesh suspended over the pixelated anode plane by some insulator pillars (or supporting structure), forming a thin gap (of about 25 to 150 $\mu$m) where the charge amplification takes place (See Fig.~\ref{fig:Micromegas}).
When used as readout of TPCs, the ionization electrons trigger the avalanche in the amplification gap.
Detectable signals in the anode and the mesh are induced by the movement of charges in between.
Since mobility of electrons in gas is over 100 times larger than that of ions, the electron signal is much shorter and can be used to measure the timing profile of primary charge.
While the anode is usually patterned (e.g. pixelated) and therefore provides topological information on the primary charge cloud, the mesh is common to all or several pixels, and therefore provides a redundant reading of the same avalanches with the possibility of integrating the charge over a wider area.

\subsubsection{Microbulk Micromegas}

\begin{figure}[tbp]
\centering
    \includegraphics[height=0.28\textwidth]{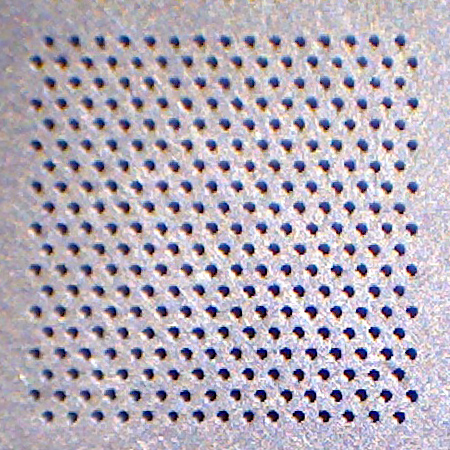}\hfill
    \includegraphics[height=0.28\textwidth]{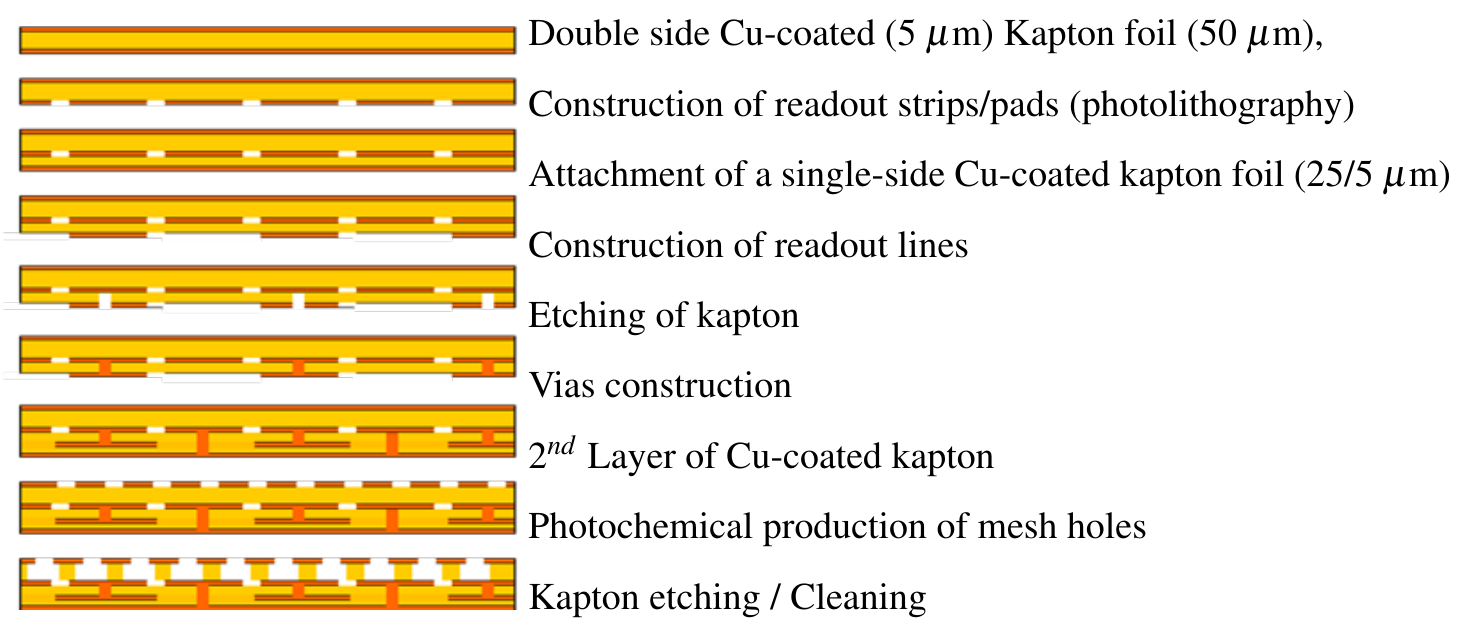}
    \caption{
    (Left) A picture of mesh of a Microbulk Micromegas detector, produced by CERN.
    The area corresponds to a pixel on the anode plane.
    The pixel pitch, defined as the diagonal line of the square, is 3~mm.
    (Right) Manufacturing process of a Microbulk with a 2D readout scheme.
    Figure is taken from Ref.~\cite{Andriamonje:2010zz}.
    }
    \label{fig:Microbulk}
\end{figure}

The \Piii readouts will be Micromegas of the Microbulk type~\cite{Andriamonje:2010zz}.
This technique, originally developed by CEA and CERN, provides all-in-one readouts out of doubly copper-clad Kapton foils.
The mesh is etched out of one of the copper layers of the foil, and the Micromegas gap is created by removing part of the Kapton by means of appropriate chemical baths and photo-lithographic techniques (see Fig.~\ref{fig:Microbulk}).
The fabrication technique has been developed substantially during the previous years and the resulting readouts have very appealing features, outperforming previous generations of Micromegas in several aspects.
The mechanical homogeneity of the gap and mesh geometry is superior, and thus the excellent energy resolutions.
Microbulk Micromegas have obtained the best energy resolution among all MPGDs and enjoy extremely good radiopurity levels~\cite{Andriamonje:2010zz, Cebrian:2010ta}.
Values around 11\% FWHM for the 5.9 keV peak of $^{55}$Fe, are the benchmark result routinely achieved by small Microbulk readouts in Ar + Isobutane mixture at atmospheric pressure~\cite{Iguaz:2012ur}, while 13\% FWHM are typically reached in larger finely pixelated readouts~\cite{Aune:2013pna}.

To cover the needed readout surface, a scheme of tessellation of $20\times20$\,cm$^2$ Microbulk modules is foreseen, in order to remain within conservative fabrication limits.
One such single Micromegas unit, the Scalable Radiopure Readout Module (SR2M) is described in subsection~\ref{sec:SR2M} below.
This readout configuration represents an optimal choice regarding the different experimental parameters of interest in the double beta decay experiment: energy resolution, topological quality, radiopurity and scalability.
The choice minimizes risk as it relies to a large extent on tested solutions and existing operational experience with Micromegas in high pressure Xenon, so minimal preparatory work is needed and a relatively fast construction plan towards a 200 kg detector can be envisaged.
Micromegas readouts are in addition a cost-effective and robust solution for instrumenting large areas.
Key technological elements of the readout have been demonstrated in the R\&D performed during the previous years within the T-REX project at the University of Zaragoza~\cite{Irastorza:2015dcb, Irastorza:2015geo, Cebrian:2012sp, Cebrian:2010nw}.

\subsubsection{The Scalable Radiopure Readout Module (SR2M)}
\label{sec:SR2M}
\begin{figure}[tbp]
\centering
\includegraphics[height=7cm]{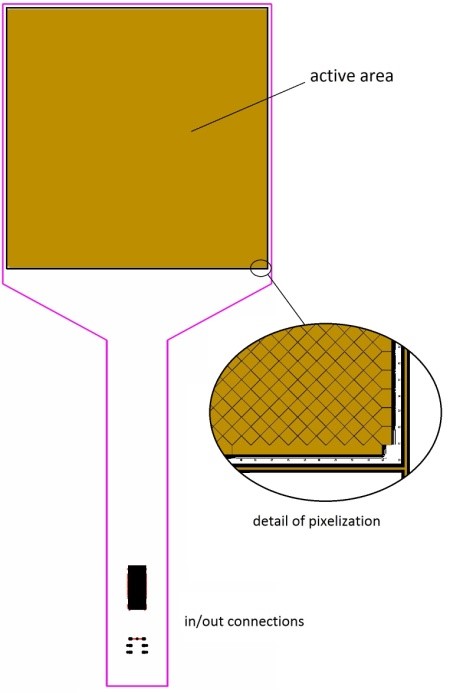} \hspace{1cm}\includegraphics[height=7cm]{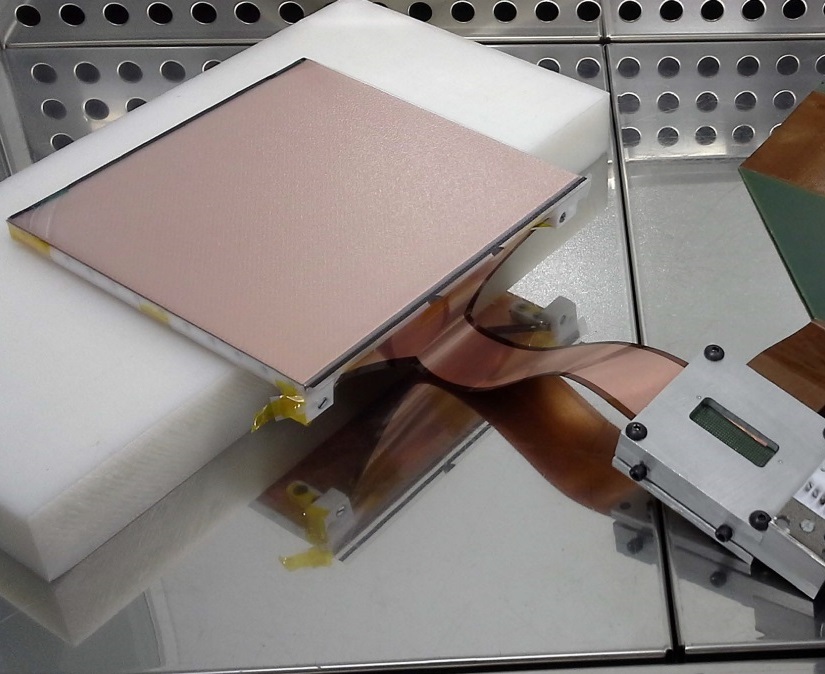}
\caption{Design  (Left) and a picture (Right) of the SR2M Micromegas with  20~cm$\times$20~cm active area,  with 3 mm pitch anode patterning, as well as the flat cable extension to route the signals to a connector. }
\label{fig:SR2M_pic}
\end{figure}

The two circular readout planes of both end-caps consist of a tessellation of smaller identical Microbulk Micromegas modules, dubbed SR2Ms.
Each of the SR2M and the associated mechanics, wiring out of the chamber and electronics will be replicated, facilitating the engineering of the full size detector.
The size of each SR2M Microbulk Micromegas will be a relatively small 20~cm$\times$20~cm area, well within technical fabrication limits.
Each of the SR2M units will be supported by a structure in high purity copper.
The readout pattern will follow an X-Y design with strips of 3~mm pitch, summing up to 128 channels per module (64 each direction), as shown in Fig.~\ref{fig:SR2M_pic} Left.
The two readout planes will consist of about 80 modules and will thus add up to about 10000 channels.
Each module has a dead margin of $\sim$1~mm by the edges, due to fabrication limitations.
Besides this, another dead area appears due to assembling tolerance of 1~mm between modules, summing a total of 3~mm distance between active areas of consecutive modules.
To mitigate the border effects, a thin electrode surrounding the active area of each module will be engraved (the ``rim'' electrode) and independently powered at a voltage a few volts above the mesh voltage.
This rim gently pushes the drift lines away from the dead areas and drives them toward the active pixels.

The 128 signal channels of each SR2M unit are extracted via a flexible cable that is the continuation of the very Microbulk foil (see Fig.~\ref{fig:SR2M_pic}), avoiding the need of connectors, soldering or any other material close to the sensitive volume.
The cable is bent backwards as soon as it goes out of the Micromegas, without interfering with the tessellation, and bringing the signals far enough from the readout so that any additional interfaces or feedthroughs could be shielded from the  active volume.

\subsection{Operation of Microbulk in xenon and TMA}

\begin{figure}[tbp]
\centering
\includegraphics[width=0.4\textwidth]{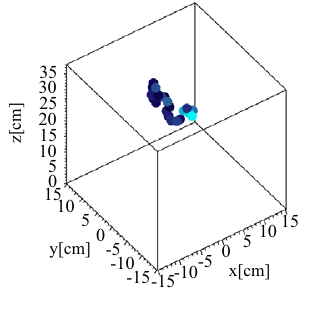}
\includegraphics[width=0.58\textwidth]{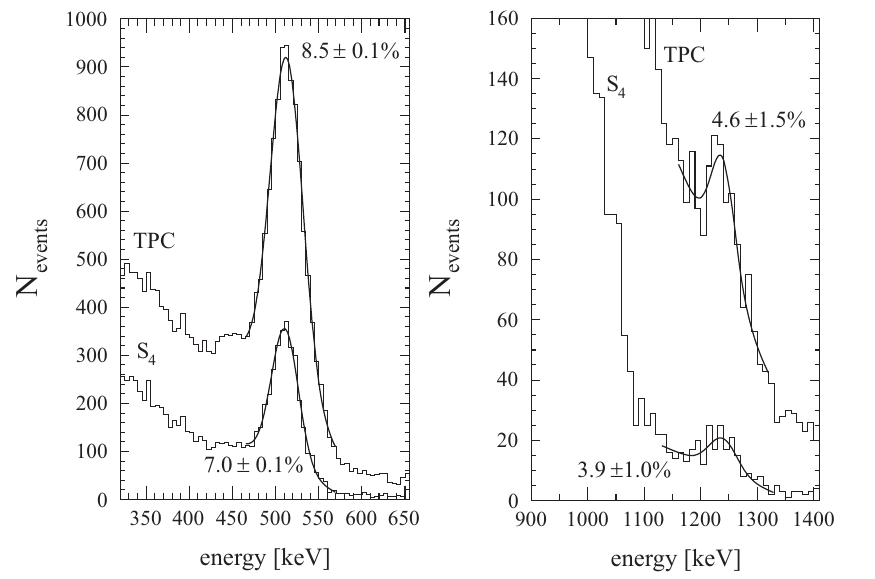}
\caption{(Left) a sample of 3D tracks from 1274 keV gamma of \natt. (Right) energy resolution obtained for the 511 keV and 1274 keV peaks of \natt. Figures are from~\cite{Gonzalez-Diaz:2015oba}.}
\label{fig:trackReso1274}
\end{figure}

The xenon+TMA mixture enhances substantially the performance of Micromegas in high pressure Xe and offers more stable operating conditions of the chamber.
TMA forms a Penning mixture with Xe, which translates to higher gain at the same voltage, higher maximum gains, and better energy resolution.
Data with small MM readouts in a 2-liter T-REX chamber has demonstrated that satisfactory operation is obtained in high pressure Xe+1\%TMA mixture, achieving a gas gain of 3000 (400) at 1 (10) bar~\cite{Cebrian:2012sp, Herrera:2013qda}.
The energy resolution measured is 7.3\% (9.6\%) FWHM at 22 keV at 1 (10) bar~\cite{Herrera:2014fsb}.
If extrapolated to the \xeots \Qbb energy, these values correspond to close to 1\% FWHM, showing high promise for application to double beta decay searches.

\begin{figure}[tbp]
\centering
\includegraphics[width=0.98\textwidth]{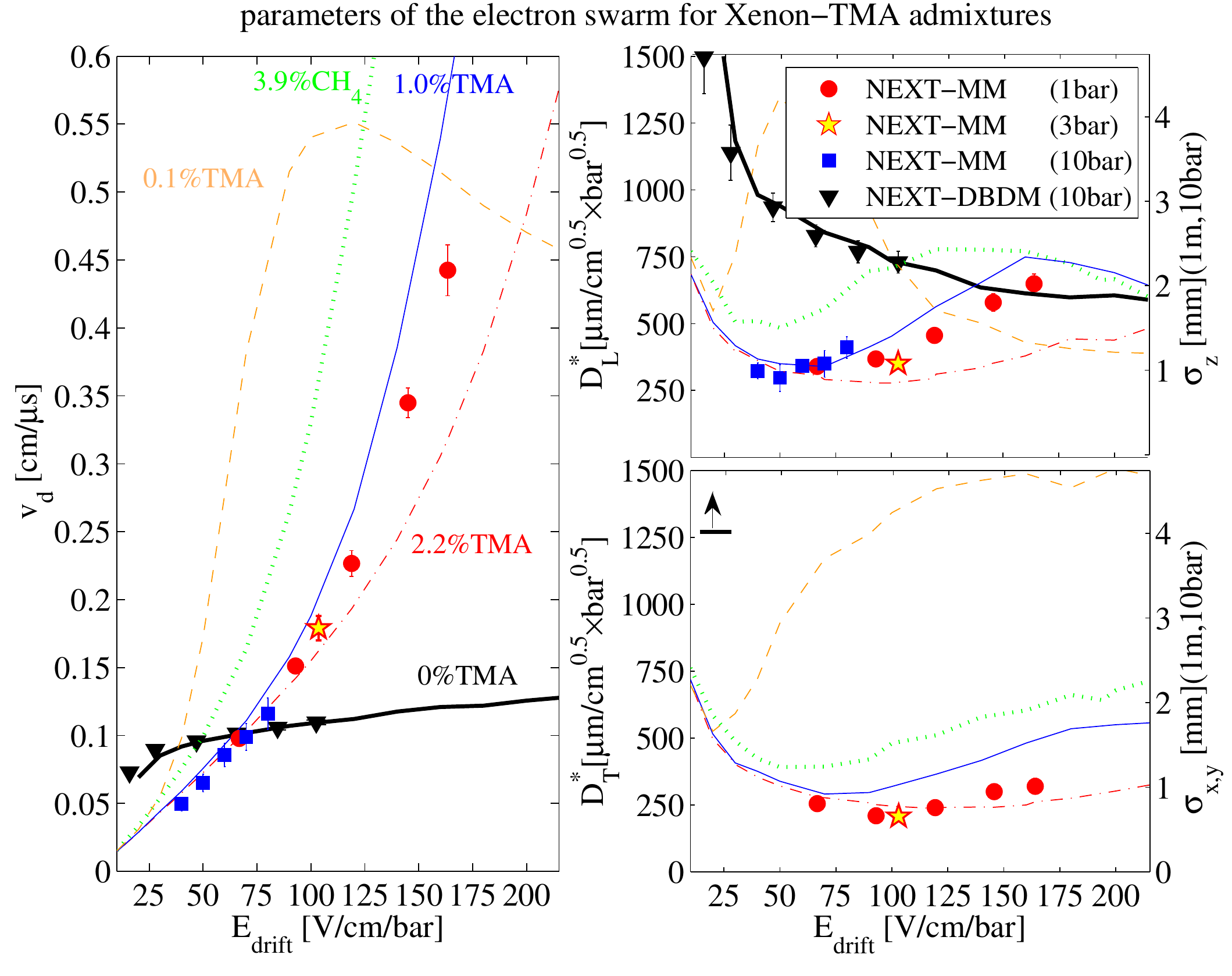}
\caption{Experimental measurements (dots) of drift velocity (Left) and longitudinal and transversal diffusion (Right) in Xe-TMA mixtures, compared with other gas mixtures and with Magboltz simulated results (lines). The plot is taken from Ref.~\cite{Gonzalez-Diaz:2015oba} where we refer for details on the data.}
\label{fig:TMA_diffusion}
\end{figure}

A very important advantage of using Xe-TMA is the very low levels of electron diffusion. Both longitudinal and transversal diffusion have been measured experimentally~\cite{Gonzalez-Diaz:2015oba} for a number of gas parameters (pressure and TMA concentration). Fig.~\ref{fig:TMA_diffusion} summarizes the measurements and compares them with microscopic simulations done with Magboltz. A typical value is 300 $\mu$m\,cm$^{-1/2}$\,bar$^{1/2}$ for the longitudinal diffusion and 250 $\mu$m\,cm$^{-1/2}$\,bar$^{1/2}$ for the transversal one for a drift field of 750 V/cm and 1\% of TMA at 10 bar. These values correspond to a factor 10 and 3 better (respectively for the transversal and longitudinal diffusions) that the ones in pure Xenon. Just to illustrate these numbers, this means that a point-like charge, after drifting one meter, will end up as a 1\,mm sigma diffused cloud. This allows to preserve the topological information of more complex extended tracks, and is a very important feature for $\beta\beta$ searches. To take advantage to this fact, correspondingly high readout granularities are needed. While the baseline for the proposed readout is a pitch of 3 mm, studies to optimize this choice are ongoing~\cite{Galan:2015tgl}.

More realistic operation was tested in the larger T-REX prototype TPC, which is able to host an active volume of 24 liter of 10 bar gas (more than of ~1 kg of xenon)~\cite{Alvarez:2013oha, Alvarez:2013kqa, Gonzalez-Diaz:2015oba}.
In this volume high energy ($\sim$MeV) electron tracks are fully contained and therefore calibration at energies closer to \Qbb and with extended tracks is possible.
It reproduces, in what the readout is concerned, the same operating conditions of a full scale TPC for NLDBD.
The tests performed in this chamber, as part of the NEXT-MM program, involved a Microbulk Micromegas of 30 cm diameter composed of 4 circular sectors.
The readout was pixelated with 8 mm side pads, making a total of 1152 channels.
Each channel was independently read with a DAQ chain based on the AFTER chip.
The Microbulk meshes were read as well and used as trigger.

A $^{22}$Na source was used to produce electron ionization tracks of 511 keV and 1274 keV in the body of the detector.
An example of reproduced $\sim$10 cm 3D tracks from recorded 1274 keV events are shown in Fig.~\ref{fig:trackReso1274} Left.
The topological quality is very good, only limited by the relatively large pixel size.
Despite this the Bragg peak is easily identified visually in most of the tracks.
The energy of the event is obtained by summing up the charge received in all active pixels of the track, after appropriate time integration, calibration and equalization of channels.
As shown in Fig.~\ref{fig:trackReso1274} Right, the energy resolution obtained for extended tracks are of 7\%  and 4\% FWHM for 511 keV and 1274 keV respectively, at 10 bar Xe+1\%TMA.
This corresponds to a remarkable resolution of 3\% FWHM at \Qbb.
The reason why this number is still a factor $\sim$3 worse than the ``ideal'' value derived from the low-energy small-scale tests presented above is attributed to a number of instrumental limitations identified in the set-up and affecting extended tracks.
They include: electronics under-sampling, crosstalk and a number of connectivity issues.
Despite the fact that future iterations in the readout design will possibly override, at least partially, such limitations, and potentially approach the final figure down to the sub-\% level, we remain in the present document with the conservative figure of 3\% FWHM at \Qbb for the nominal energy resolution of the first \Piii TPC module.

\subsection{Calibration system}
\begin{figure}[tbp]
\centering
\includegraphics[width=0.4\textwidth]{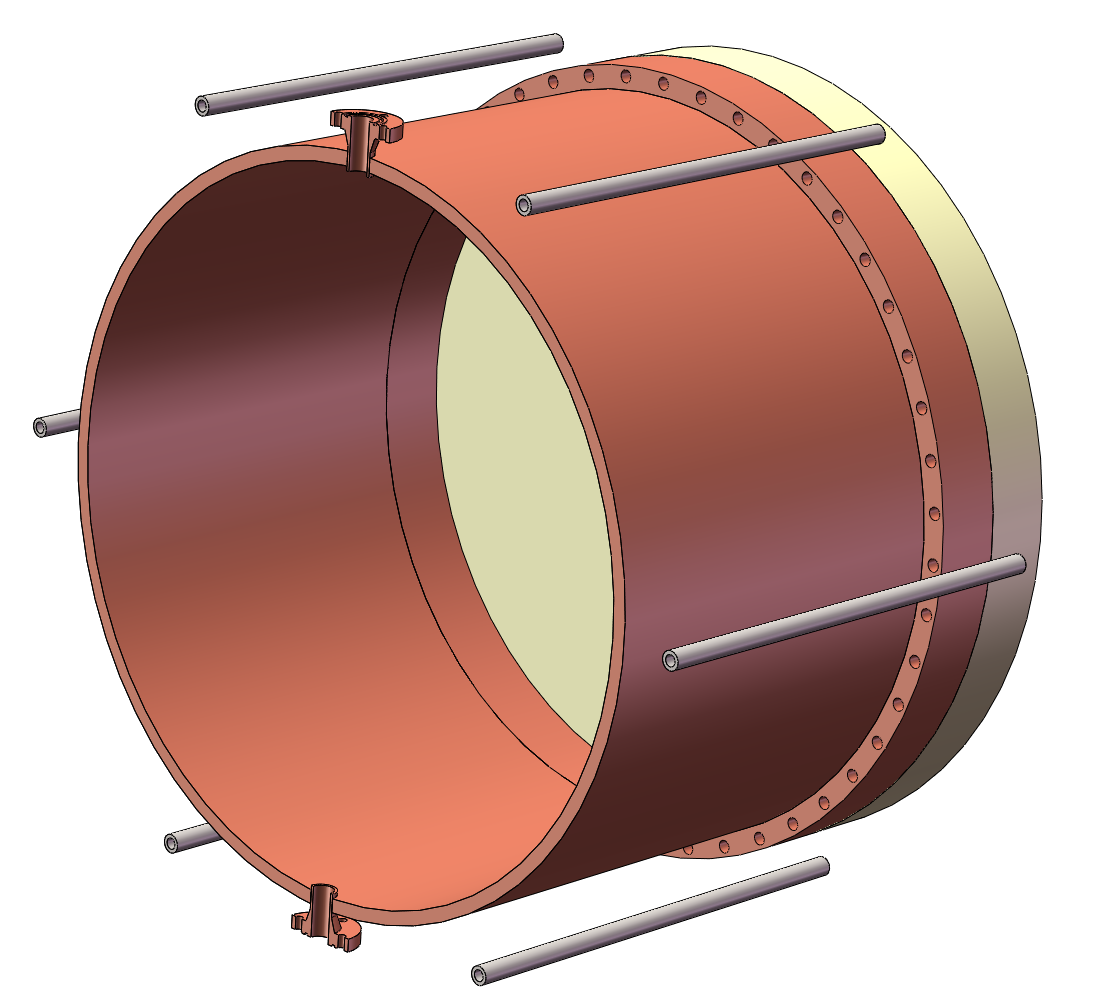}
\includegraphics[width=0.45\textwidth]{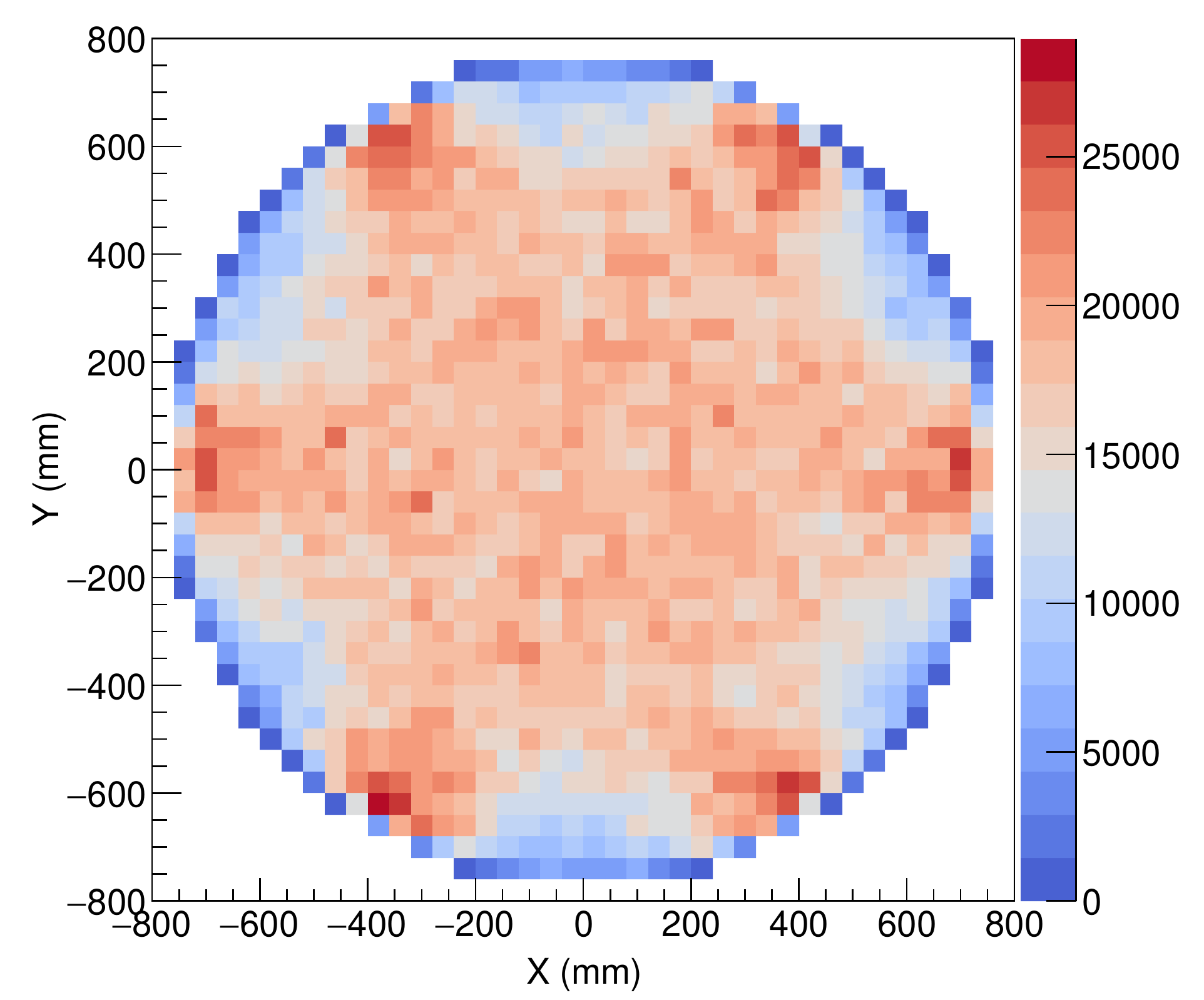}
\caption{(Left) Cross-sectional illustration of 6 calibration tubes positioned around the \Piii copper vessel. (Right) Illumination of the active volume from \thttt calibration source tubes. In the plot, an energy threshold of 500~keV is applied.
}
\label{fig:cal_illustration}
\end{figure}
\begin{figure}[tbp]
\centering
\includegraphics[width=0.6\textwidth]{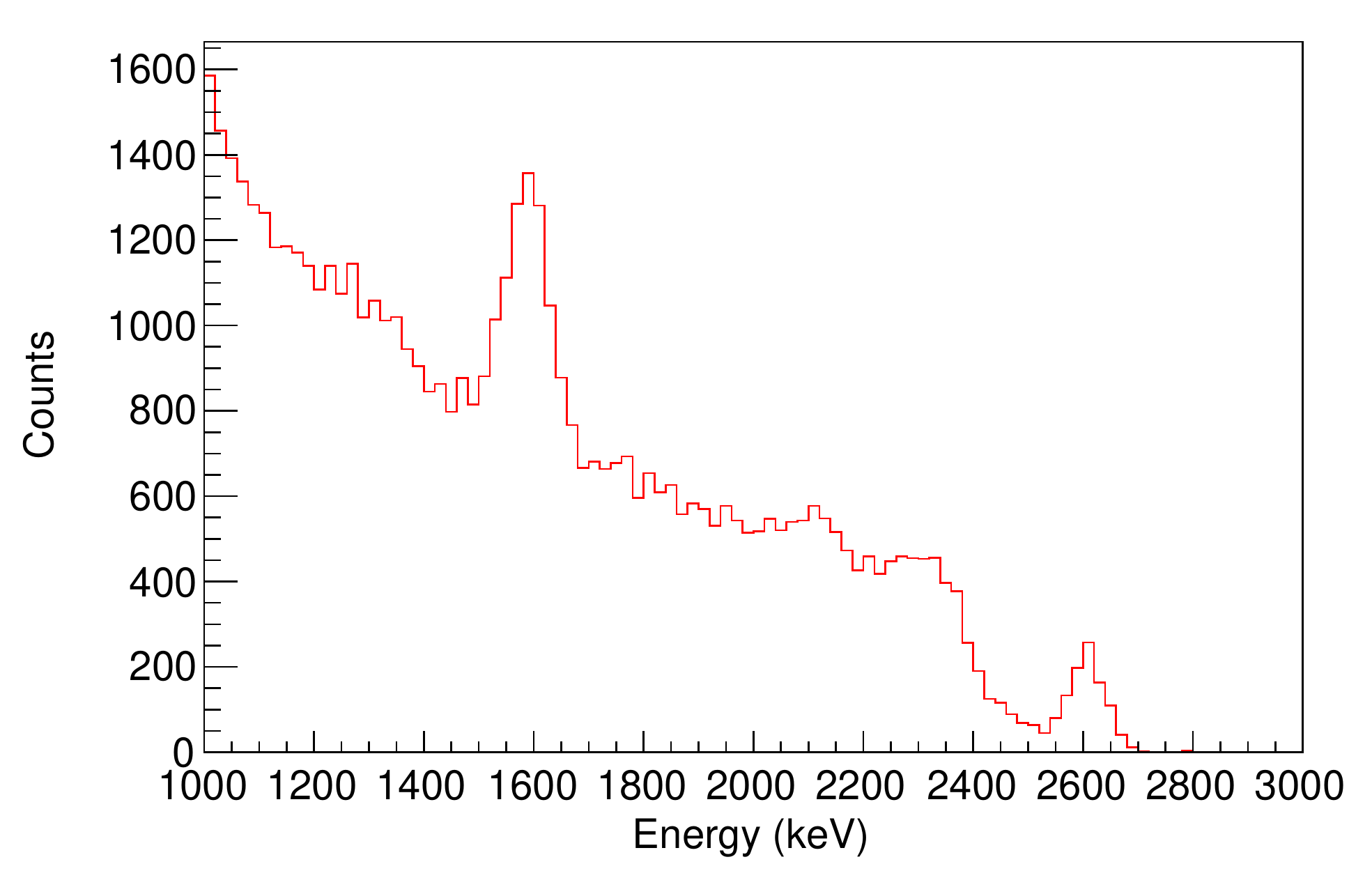}

\caption{Simulated energy spectrum with 3\% energy resolution from external \thttt calibration sources. The 2615~keV peak and the double-escape peak are the most prominent features.
}
\label{fig:cal_spec}
\end{figure}

Energy calibration of the \Piii TPC at the Q-value (2548~keV)  is required for an unambiguous detection (or exclusion at certain half-life limit) of NLDBD signal.
At lower energy range, energy calibration helps measure precisely the half-life of two neutrino double beta decay and other more exotic process (such as decay with Majorons) with spectrum fit.
Rigorous and regular calibration routine with properly designed hardware is an important task.
External gamma source \thttt is the primary calibration source with its many gamma lines in MeV range.
The prominent 2615~keV line from its daughter nuclei \tltze is of particular importance.
Other external gamma sources such as \cose and \csots as well as internal gamma, x-ray, and alpha sources will also be used.

We will use  gamma calibration source outside the high pressure vessel. In Fig.~\ref{fig:cal_illustration} (Left), 6 external stainless steel calibration tubes are evenly spaced and positioned 90~cm from the axle of the TPC barrel.
Inside the SS tubes, thoriated calibration source string can be inserted and sealed.
During calibration, those strings are lowered from the top of the water pool (see Section~\ref{sec:water} for more details) and secured in position with magnetic latches (not shown in the figure).
During normal physics data taking, the strings will be moved away and shielded from our detectors.
Fig.~\ref{fig:cal_illustration} (Right) shows the illumination of 6 thoriated strings from a Geant4-base simulation.
An energy threshold of 500~keV is used to include gamma peaks and Compton continuum.
Illuminations of specific gamma peaks, such as the 2615~keV, are similar qualitatively to the figure shown.
Acceptable uniformity of illumination is achieved in the current design.
Bins along the circumference of the TPC active area show lower counting rate mainly because of the artifact of plotting square pixels around an arc.
Should a better uniformity be needed, more calibration tubes can be added easily around the TPC vessel.
Alternatively, we can have the tubes rotating back and forth by small angles to improve the uniformity.
Simulated energy spectrum in the range of 1000 to 3000~keV from the external \thttt sources are shown in Fig.~\ref{fig:cal_spec}.
A FWHM of 3\% is assumed in the simulation.
The 2615~keV peak of \tltze and double-escape peak are the most prominent.
The single-escape peak is also visible but less prominent.

External Calibration tubes can accommodate other gamma source strings with \cose, \csots, and a combination of multiple sources.
Fast decaying \rnttz gas can also be delivered inside the TPC~\cite{Kobayashi:2016vwj}.
\rnttz and its daughters nuclei provide multiple alpha sources above the NLDBD Q-value.
The other advantages for \rnttz as internal source are that there are no long-lived radioactive daughter nuclei and it's relatively easy to  extra \rnttz from the TPC gas mixture.

\subsection{Prototype TPC}
\label{sec:prototype}
\begin{figure}[tbp]
\centering
\includegraphics[height=0.5\textwidth]{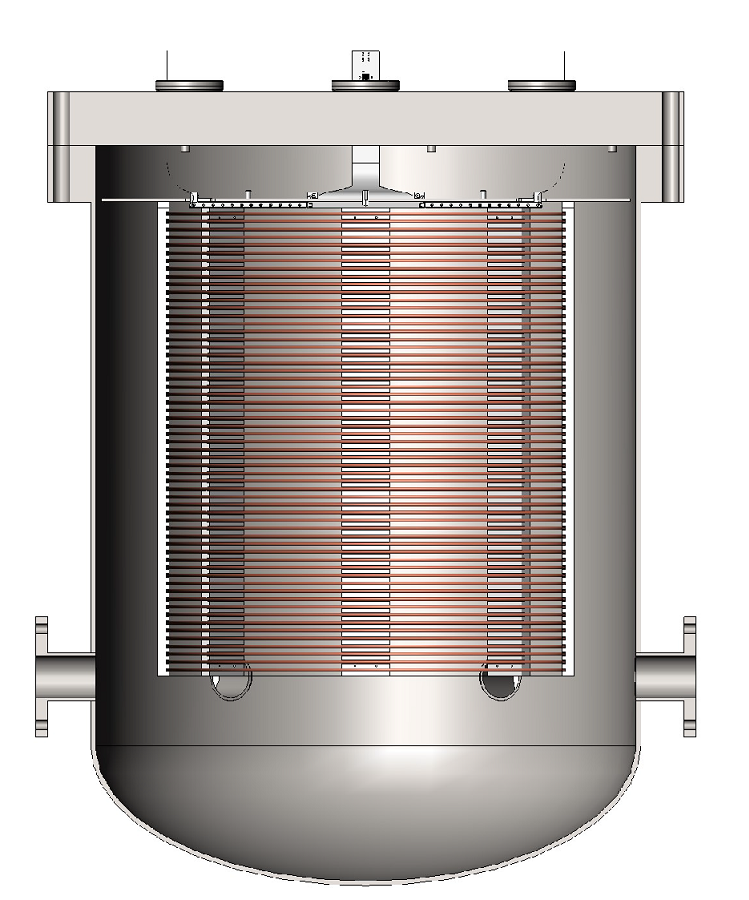}
\includegraphics[height=0.4\textwidth]{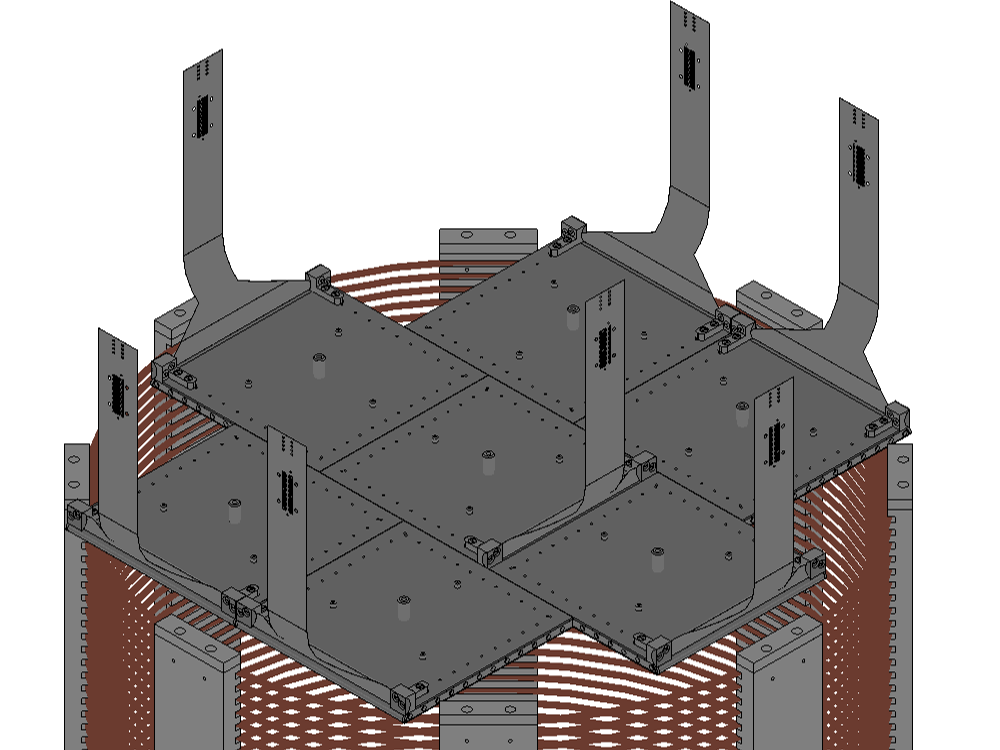}
\caption{(Left) Sectional view of the prototype TPC inside a stainless steel high pressure vessel.
	(Right) Illustration of 7 SR2Ms which cover the most of the active area of the TPC.
	At the four corners with no SR2M coverage, we will add PMTs to read out the scintillation light signal (while without TMA as a quencer).
	Currently only the center SR2M is installed in the prototype setup.}
\label{fig:prototype}
\end{figure}

\begin{figure}[tbp]
\centering
\includegraphics[width=0.45\textwidth]{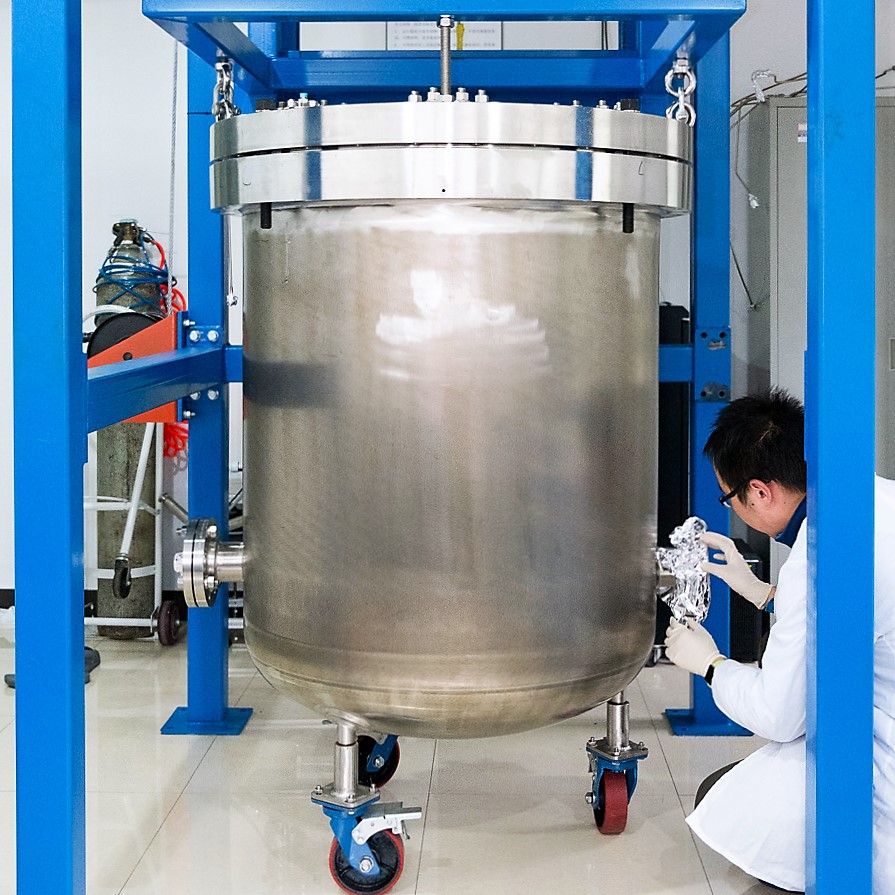}
\hspace{1em}
\includegraphics[width=0.45\textwidth]{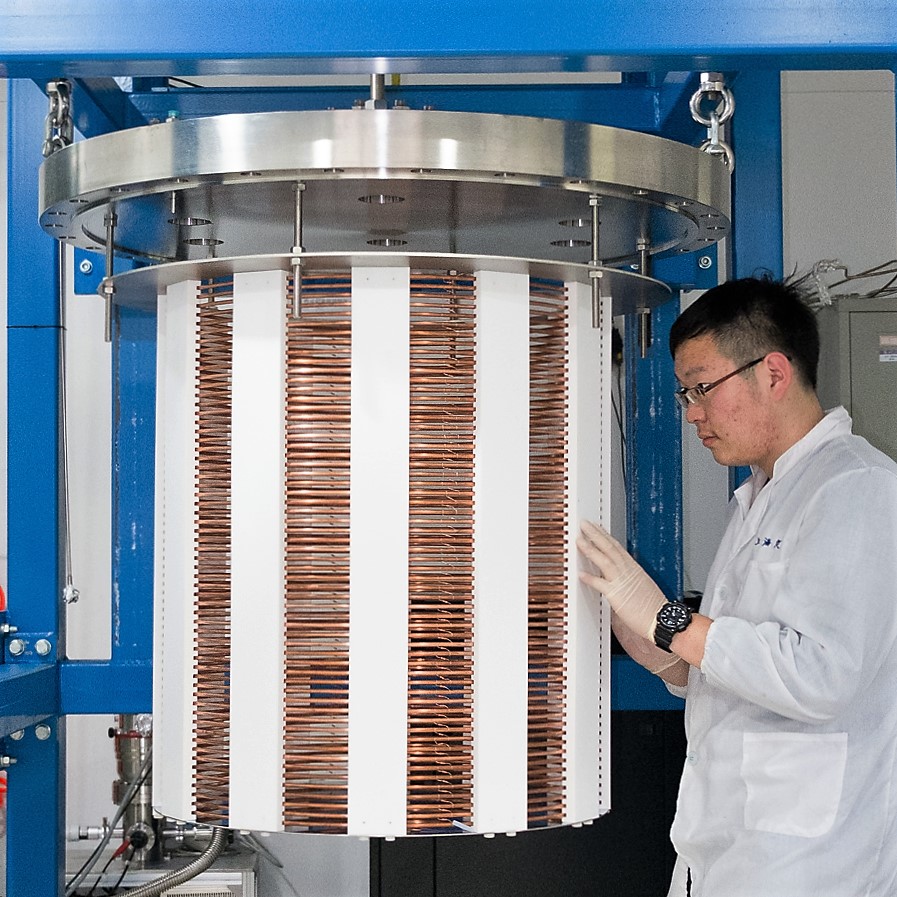}
\caption{ Pictures of the prototype stainless steel high pressure vessel (Left) and TPC (Right).
}
\label{fig:prototypePic}
\end{figure}

\begin{figure}[tbp]
\centering
\includegraphics[width=\textwidth]{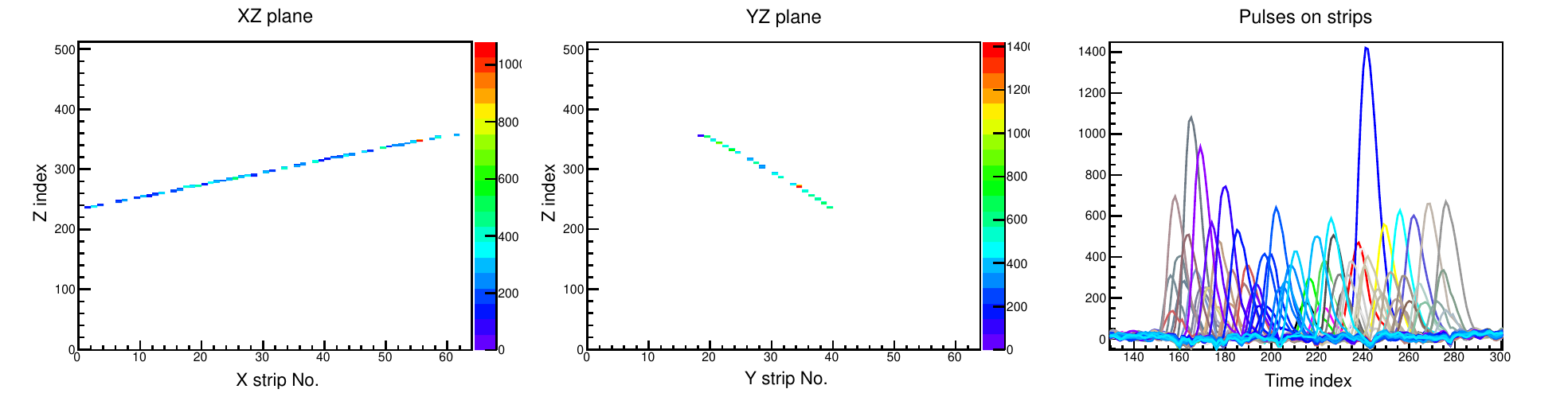}
\caption{A muon passing through the active region of our prototype TPC, as recorded by the Micromegas module. The first two plots show the moun track on the XZ and YZ planes. The X and Y coordinates are defined as the orientations of Micromegas readout strips. Z is the electron drifting direction. The figure on the right shows pulses collected on each strip.
}
\label{fig:muontrack}
\end{figure}

To test various design features of \Piii detector, we have built a prototype TPC with a accompanying stainless steel (SS) high pressure vessel.
Current setup is shown in Fig.~\ref{fig:prototype} (Left).
The inner diameter of the SS vessel HPV is 90~cm.
Its cylindrical part is 100~cm tall and the dome at the bottom adds another 22.5~cm in height.
The vessel holds about 40 kg of xenon at 10 bar operating pressure at room temperature.
The TPC inside is single-ended, with cathode at the bottom.
The active volume is 66~cm in diameter and 78~cm tall, as defined by the copper field shaping rings.
At 10 bar, 16 kg of xenon is contained within the active volume.
Secondary electrons drift up and get collected on the top readout plane with SR2Ms.
Currently one single SR2M is installed and commissioned at the center of the readout plane.
This module has 128 channel output, half of which on the X direction and the other half on the Y direction.
Each diamond-shaped readout pad has a pitch size of 3~mm.
Pictures of the vessel and TPC inside are shown in Fig.~\ref{fig:prototypePic}.
Students are shown in the pictures to illustrate the relative size of the vessel and TPC.

The prototype TPC has been commissioned for calibration runs.
Detector performance with different gas medium at different pressure were studied with alpha and gamma calibration sources.
We used pure xenon, xenon+TMA, and argon+CO$_2$ gas mixtures as the detection medium.
With argon+CO$_2$, the highest pressure we tried was 5 bar.
We have been using a \AmAlpha source with a ultra-thin window and specifically designed for vacuum use.
It emits signature 5.4~MeV alpha particles and 59~keV gamma rays, both of which are recorded by the TPC.
Passing-through muon events are also collected.
An example of such event is shown in Fig.~\ref{fig:muontrack}.

Many upgrade options are in the pipeline for the prototype TPC.
In the near future, 7 SR2Ms will be installed on the prototype to image most of the active volume of the TPC.
The coverage of the SR2M at the readout plane is shown in Fig.~\ref{fig:prototype} Right.
Those SR2Ms will have identical specifications as the one already installed in the TPC.
We will also have SR2Ms with finer pitch size (1~mm and 2~mm) to explore the impact of pixel size on tracking. with the ultimate goal to define the correct balance between tracking granularity and number of readout channels.
PMTs will be added to the area not covered by SR2Ms and used to measure scintillation light from xenon gas (without TMA).
The fast light signal will give the starting time $t_0$ of an event and help us better understand the impact of missing $t_0$ in the full experiment.

%% file: HighPressureVessel.tex
\section{High pressure vessel}
\subsection{Vessel design}
One of the key challenges of the \Piii experiment is a radio-pure vessel capable of working at 10 bars.
There are several approaches to this problem.
The classical solution is to use high-purity copper, as in the Gotthard experiment~\cite{Iqbal:1987vh, Wong:1991fg}.
For high pressure vessel, large quantity of  OFHC copper is needed for the mechanical strength.
Though copper can introduce significant background, it can shield the radio-active background from the front-end electronics.
The second approach is construct a SS or titanium vessel for their mechanical strength.
Inside the vessel wall, OFHC slabs are attached as shielding since SS and Ti usually have much more radioactive contaminations. 
The third option is to use a double-wall structure, as in EXO-200~\cite{Auger:2012gs}, in which the inner vessel is made of a thin layer of radio-pure copper, and the pressured outer vessel is situated far away and shielded.
This design requires careful control of the inner and outer pressures such that the inner vessel does not experience a large pressure difference.
We are also exploring the possibility to use high-strength synthetic fiber to reinforce a thin inner vessel, as in aerospace industry.
The challenge in our application is to accommodate various large-size feedthroughs necessary for the experiment.
To our knowledge, no one has used this method in low-background experiments.

\begin{figure}[tbp]
\centering
\includegraphics[width=0.5\textwidth]{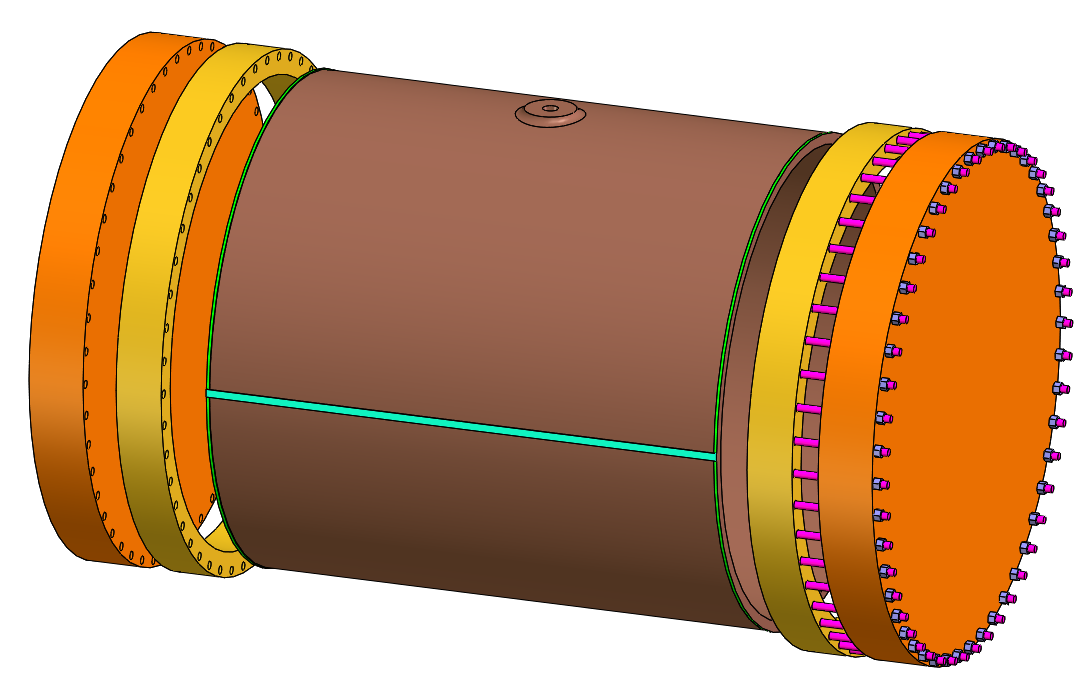}
\caption{Schematic drawing of the PandaX-III copper vessel. The green circles and cyan line highlights welding joints.}
\label{fig:CopperVessel}
\end{figure}

Our baseline design follows the classical approach, as shown in Fig.~\ref{fig:CopperVessel}.
The design requirements for the PandaX-III OFHC pressure vessel are as follows:
\begin{itemize}
\item Operating pressure: -0.1~MPa to 1.0~MPa;
\item Operating temperature: -20$^\circ$C to 30$^\circ$C;
\item Inner diameter: 1500~mm; length: 2000~mm;
\item Two flat covers;
\item Inner medium is xenon gas with about 1\% TMA;  outer medium is air or pure water;
\item Design life: $\sim$10 years;
\item Sealing: O-ring;
\item Nozzles: two 1/2 inch flanges for xenon gas inlet and outlet, one 2 inch flange for high voltage, and one 2 inch flange for vacuum pumping.
\end{itemize}
According to these requirements, the flat cover is designed to be 150~mm in thickness and weighs 2.7~ton.
The thickness of the barrel is designed to be 32~mm with flanges on both ends, whose thickness is 200~mm.
The barrel and flange weigh about 3.2~ton, and increase the total weight of the vessel to 8.6~ton.

To hold the covers and barrel together, a total of 96 sets of M24 SS bolts and nuts will be needed.
Since the background of the cleanest SS is about 1 mBq/kg level, the nuts and bolts may contribute significantly to
the total background, even though their mass is only about 190~kg.
Our simulations, as shown in a later section, confirm that SS bolts do contribute significantly to our background budget.
We will explore alternative choices, including ultra-high strength SS and high-strength titanium.

\subsection{Vessel fabrication}
Fabrication of the vessel, especially the barrel is technically challenging.
The flat covers can be routinely fabricated by forge process from a big copper ingot and then machined to required shape and precision.
For the barrel, we have investigated several processes, including forge, ring roll, extrusion, and electron beam welding (EBW).
However, the tolerance of the first three processes is all too large.

EBW is identified as the baseline approach.
The welding is performed in a vacuum chamber without soldering wires, thus no extra radioactivity is introduced when handled carefully. 
Initial welding test performed by Beijing Aeronautical Manufacturing Technology Research Institute of Aviation Industry Corporation of China (AVIC 625-Institute) was satisfactory in terms of mechanical strength.
Now we are working with AVIC 625-Institute on a prototype to test the key technologies in the vessel fabrication process.
Machine shops and welders in Germany are also contacted and quotes for the prototype are obtained. 

\subsection{Radiopurity of Copper}

Since the copper vessel is massive and very close to the TPC, it is extremely important to use radio-pure copper.
The highest grade commercial OFHC copper, TU1,  is our default choice.
However, different TU1 copper from different vendors and batches could have very different radio-purity.
The OFHC copper acquired by EXO-200 collaboration from Norddeutsche Affinerie was measured to have upper limits of 2.4~ppt in mass of \thttt and 2.9~ppt of \utte, respectively, using ICP-MS~\cite{Leonard:2007uv}.
Recently, Majorana collaboration has published the most precise radio-purity measurements of the commercial OFHC copper~\cite{Abgrall:2016cct}.
The samples from Aurubis and Mitsubishi Materials were measured to have 0.3~ppt of \thttt and 0.1~ppt of \utte, respectively in one sample, and 0.03~ppt and 0.02~ppt, respectively, in another.
These results are very encouraging.
The screening of copper samples is under way with the gamma counting facility and mass spectrometer, as described in Section~\ref{screening}. 
Contributions from copper vessel to our background budget are studied extensively with Monte Carlo simulations.

%% file: GasSystem.tex
\section{High pressure gas system}

\begin{figure}[tb]
\centering
\includegraphics[width=0.8\textwidth]{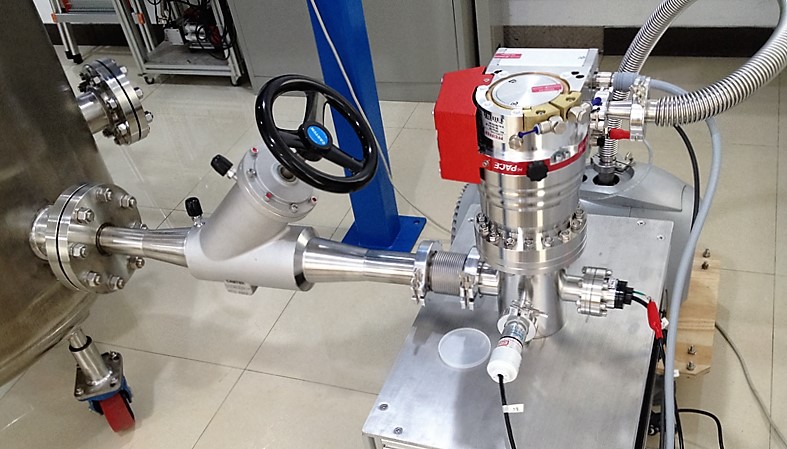}
\caption{The High pressure valve (model number: HFC2003-10PC2FSM from Carten-Fujikin), turbo-pump(HiPace300 from Pfeiffer), fore-pump (PTS300 from Agilent)  and vacuum gauges connected to our prototype TPC.}
\label{fig:pumps}
\end{figure}

\begin{figure}[!tbh]
\centering
\includegraphics[width=0.8\textwidth]{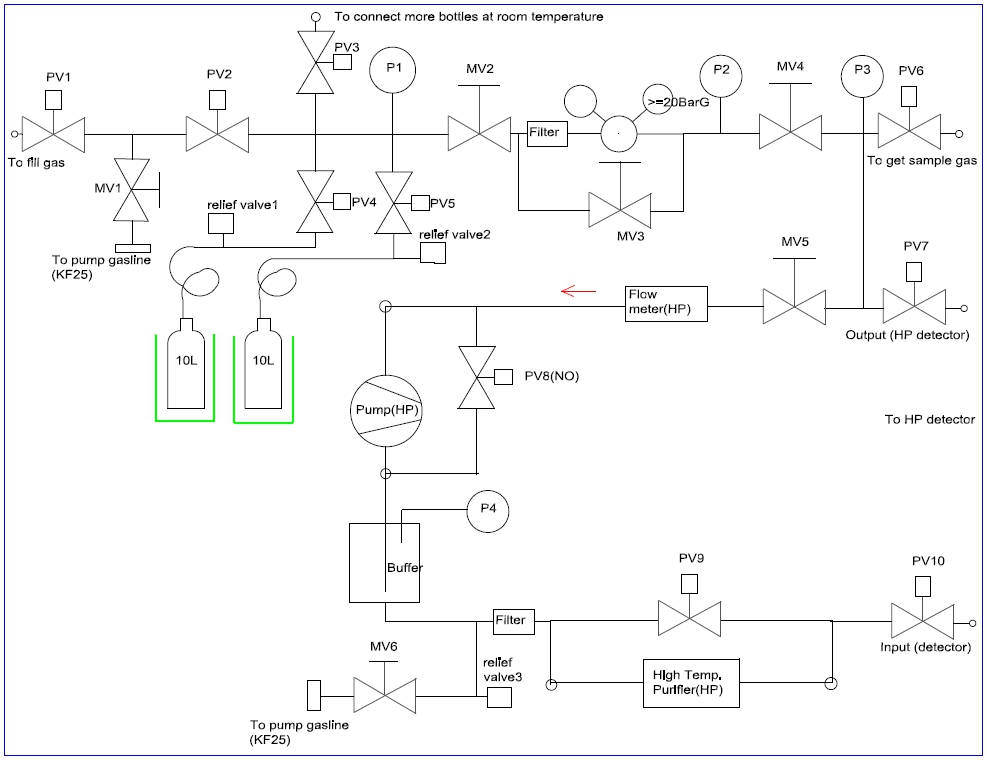}\\
\includegraphics[width=0.8\textwidth]{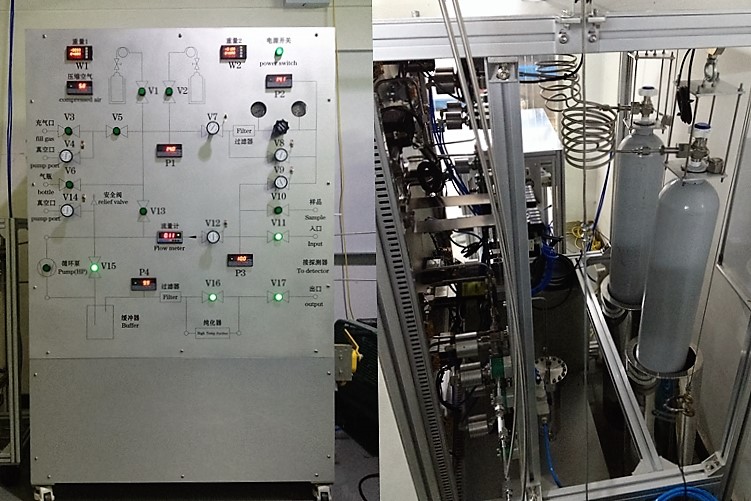}
\caption{(Top) Schematic drawing of the prototype HP gas handling system.  (Bottom) Pictures of the front control panel (Left) and two emergency storage cylinders on the back (Right).}
\label{fig:gas-handling}
\end{figure}

The xenon gas handling system prepares the vacuum of the TPC vessel, delivers gas mixtures to high pressure (HP), and purifies gas mixtures with continuous circulation.
We have built and successfully commissioned a prototype gas handling system to work with the prototype TPC at SJTU.
The full gas handling system for the PandaX-III 200~kg module will be quite similar.
In this section, we will describe this prototype system, as well as some additional components for the full system.

\subsection{Prototype HP vessel}
A prototype HP stainless steel vessel, with ports for HV feedthrough, electronics and the gas system, has been manufactured (See pictures in Fig.~\ref{fig:prototypePic}).
The vessel is made of SS with a design pressure limit of 15~bar, although the working pressure is not expected to exceed 10~bar during normal operations.
Three ports are reserved for the gas system, including a vacuum pumping port, a gas inlet port, and  a gas outlet port.
The pumping system, as shown in Fig.~\ref{fig:pumps}, includes a dry fore-pump and a turbo-pump, and vacuum gauges.
A high pressure valve with working pressure range from vacuum to 26~bar is used to connect the vessel to a vacuum pump cart.
Its large aperture with 2~inch tubing is useful for pumping down the large vessel.
The vacuum reaches around 100~Pa after half of an hour with the fore-pump only and reaches $2.5\times10^{-5}$~Pa near the turbo-pump after a day of pumping with the turbo-pump.
The gas inlet port is situated on the barrel of the HP vessel and equipped with an electronic pressure sensor.
The gas outlet port is on the small flange with a mechanical pressure meter, sitting on the top flat flange.

\subsection{Prototype gas handling system}
The schematic drawing and pictures of the prototype gas handling system are shown in Fig.~\ref{fig:gas-handling}.
It has a design pressure limit of 15~bar and includes a circulating purification system, a small storage system, and an emergency recovery system to hold up to 20~kg of xenon.
Two high pressure gas cylinders (for example, one for xenon and the other for TMA) can be connected to the system through a HP regulator, which can hold an inlet pressure of over 200~bar.
Auxiliary ports for extracting gas samples and for vacuum pumping are also included.
A high pressure relief valve (Gentec SS-RV31-TF4, 250PSI, 17.2~bar) is installed for the safety.
The modular design of the system makes it compatible not only for a small prototype system but also for the full experiment.
It is convenient to connect it to the high pressure vessels, purifiers, and circulation pumps of different capacities.
Currently,  two 10-liter aluminum gas bottles with manual valves are used for emergency gas recovery.

\subsection{Prototype circulating purification system}
\begin{figure}[tb]
\centering
\includegraphics[width=8cm, angle=90]{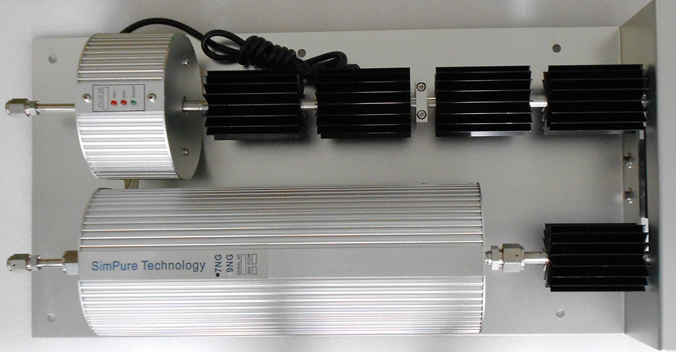} \hspace{1cm}
\includegraphics[height=8cm]{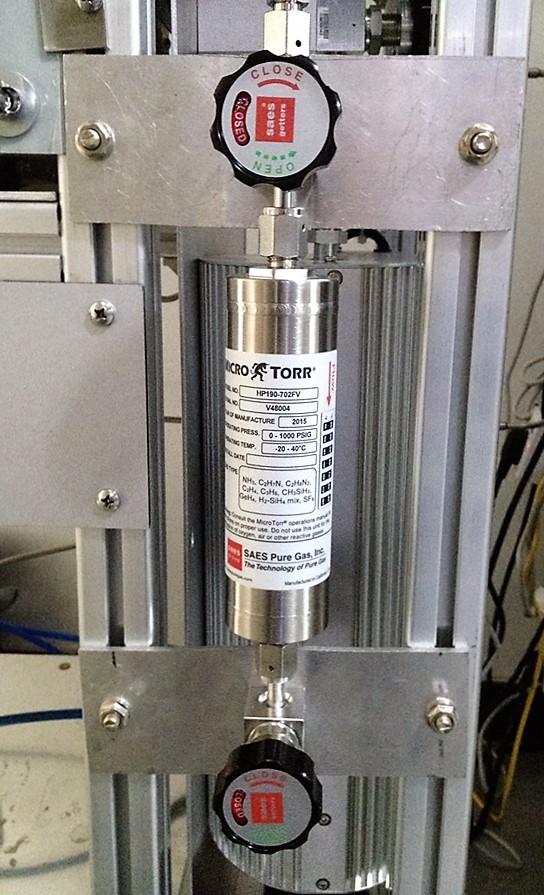}
\caption{High pressure purifier at high temperature (Left) and room temperature (Right)}
\label{fig:purifier}
\end{figure}

One of the functions of the gas handling system is to remove impurities from out-gassing of the HPV and detector in situ.
Impurities, especially the electronegative  one (H$_{2}$O, O$_{2}$, CO,etc) greatly reduce the drift length of charged tracks and degrade the performance of the detector.
In order to reduce the level of contaminants to less than a few parts per billion (ppb), continuous purification with a re-circulation loop is needed.
The high pressure hot purifier for xenon, as shown in Fig.~\ref{fig:purifier} (Left), features 3~nm filters, 160~bar pressure limit, and a nominal flow rate of 30 standard liter per minute (SLPM).
It is capable of removing electronegative impurities, including TMA, to less than 1 ppb level.
For xenon + TMA mixture gas, a room temperature purifier is needed (Fig.~\ref{fig:purifier} (Right)).
A magnetically driven and leak-proof high-pressure pump drives gas mixture through the purifier(s).
To keep the pump running at normal working temperature over a long period of time, a water and air cooling system is added.
When tested with 10~bar argon gas  circulating at a rate of 10~SLPM for 8 hours, the temperature of the pump remains constant around 17~$^{\circ}$C, well within the nominal working temperature range.

On the other hand, since the working medium is 200\,kg ($\sim$4 m$^{3}$) 90\% enriched \xeots isotope at 10~bar, the gas system must minimize accidental leakages and achieve full gas recovery in case of accidents.
To ensure the gas tightness and purity of the system, all piping and fittings have been chosen as 1/4 VCR stainless steel wherever possible, and all the valves have metal-to-metal seals.
More details about the emergency recovery is introduced in the next subsection. 

\subsection{Emergency gas recovery system}
\begin{figure}[tb]
\centering
\includegraphics[width=\textwidth]{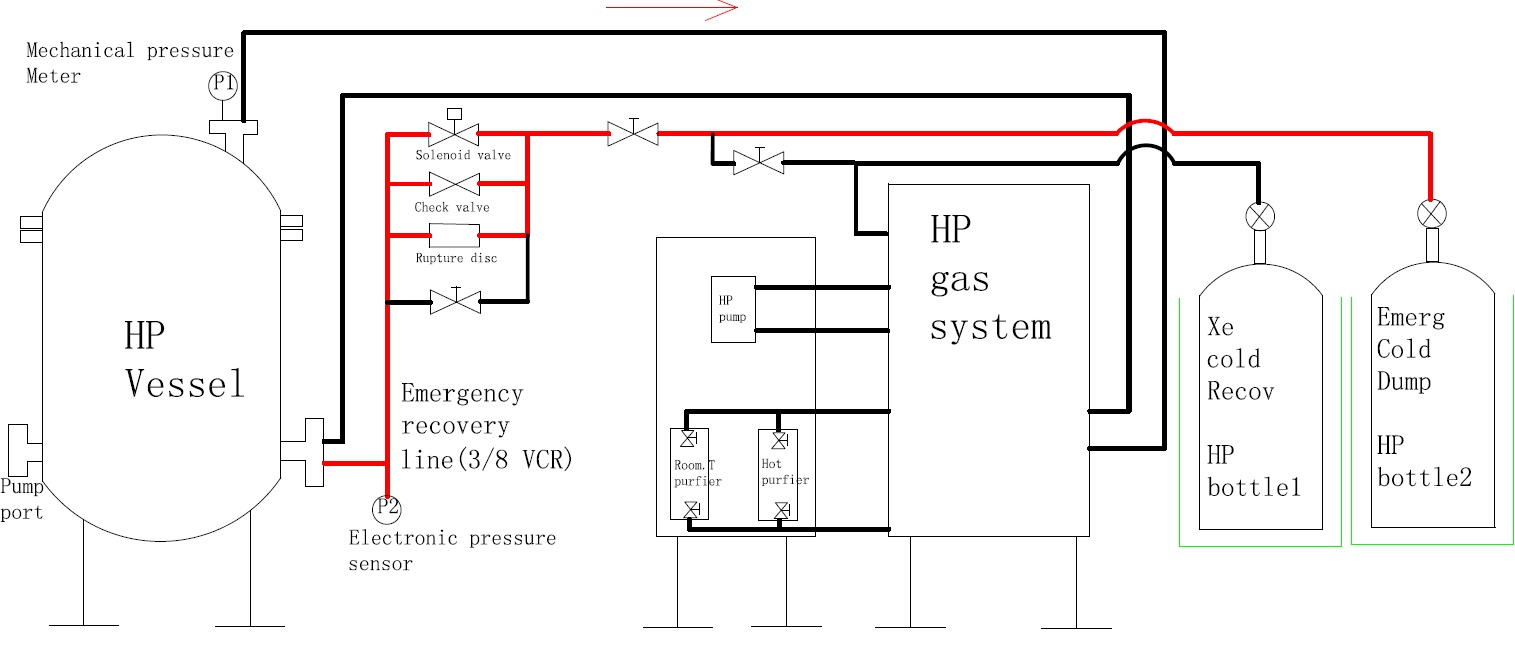}
\caption{Schematic of the emergency system. The red line is for emergency gas recovery. The black line indicates gas flow under normal operation.}
\label{fig:design-emergency}
\end{figure}

The emergency gas recovery system minimize safety hazard for operators and recovers the extremely valuable 90\% enriched \xeots in an unlikely event of an emergency condition.
Schematics of the system is shown in Fig.~\ref{fig:design-emergency}.
The recovery is achieved with a stainless steel bottle (220L), permanently kept cold with liquid nitrogen. 
In an emergency condition, a recovery valve connecting the detector vessel and the emergency reservoir will be open and the enriched Xe will be cryo-pumped into the safe container.

The system is designed to react to both overpressure and under-pressure of the HPV during normal data taking. 
Besides of the risks of losing xenon, overpressure may cause an explosion and is extremely dangerous for operators.
Since the gas system will be operated in a closed mode, there are two reasons for overpressure: mis-operation when filling gas, or laboratory fire. 
Three parallel recovery valves will be used to react to an emergency in a timely fashion. 
The first pneumatic valve would be open when the pressure is higher than the preset value, which will also trigger the Xe recovery system.
 In case of a failure of the pneumatic valve, a mechanical spring-loaded relief valve would open the pipe from the vessel to a permanently cold recovery bottle. 
A bursting disc is the third fall-back option and will burst open the connection between the vessel and the cold bottle and trigger xenon extraction.

The emergency system is also closely coupled to the slow-monitoring system.
In case of an emergency, audio and visual alarms will start in the lab and the slow-monitoring system will notify a shifter and an expert of the HP gas system.
The trained shifter needs to turn off valves of the recovery cylinders after extraction finishes.

As a final safety feature to avoid overpressure, a relief valve will be turned on to vent when all other isolating valves fail.

\subsection{Slow-Monitoring System}
The parameters of gas system, for example, pressure, flow ratio and temperature, will be recorded and monitored by a slow controller.
If there is something wrong, warning messages would be sent to on-site personnel and corresponding experts. 
This system can also be monitored from a webpage remotely.

%% file: WaterTank.tex
\section{Water Shielding}
\label{sec:water}

PandaX-III detectors will be shielded by 5.5m of ultra-clean water in all directions. 
Compared with others shielding materials, water is inexpensive and the technology for its purification is mature. 
Water shields have been used in multiple underground experiments around the world (for example,~\cite{LZ_CDR, Agostini:2016iid}). 
In this section, we will describe in details the lab space allocated for PandaX-III and the water shield. 
  
\subsection{Water pool}
 
The water pool is situated in the inner half of the Hall B4 at CJPL-II, which is about 30~m long and will be built as a class-10000 clean room. 
The pool is 27~m long, 15~m wide and 13~m deep.
Its footprint has rounded corners and the total volume is about 4800~$m^3$, as shown in Fig.~\ref{fig:detector} (Right). 
All the concrete walls are covered with 3~mm-thick SS plates.
To reduce the interference, all feedthroughs for tubes going into the water pool are located a single big flange of 400~mm diameter,  which is decoupled from the water pool liner. 
Therefore it can be re-designed and modified if necessary in future. 

The pure water plant design capacity is about 10 t/h, i.e. about 20 days needed for one time of cycling the whole pool. 
The filling tube inner diameter is about 65 mm. 
The pure water will be injected into the area around the detectors to enhance the purification performance. 
The design drain water flow rate is about 50~$m^3/h$, which requires a DN150 tube. 
The exhaust process need about 4 days. The air pressure is about 0.85~bar at CJPL, water pool depth is 13~m, so a submerge pump is necessary. 
To prevent the pump polluting the pure water, a clean grade stainless steel water pump with design flow rate of about 50~m$^3$/h will be used.

A work platform will be constructed above the water pit, at the same level as the lab floor. 
The platform is semi-permanent and can be used as auxiliary facilities.
The design static payload is 500~kg/m$^2$. 
There are rectangular openings which are 4~m long and 2~m wide to lower the detectors into the pool.
The openings are separated by removable parts to accommodate potentially larger detectors in the future. 

\subsection{Water quality requirements}

The water quality requirement is set by stringent background requirements of NLDBD experiments. 
To be effectively background-free during a 1 year run, we need to expect no more than 1 background event in the region of interest per year per module. 
Simulations of the activities in water that would produce such a rate in the detector result in estimates of the tolerable concentrations of U and Th of the order of 50--70 fg/g, with a tolerable Rn content of $\sim1$ mBq/m$^3$. 
Such low levels of contamination  have been achieved by SNO~\cite{theSNO} and Super-Kamiokande~\cite{SK},  and our design follows largely the experience from them.
We plan to employ a standard industrial ultra-pure water design~\cite{ewi} with customizations targeting at effective removal of uranium, thorium, and radon contaminations.
Given the size of the water shield (5 kton of water), the design filling and recirculation flows have both been fixed at 10 t/h, or 4.8 kton in 20 days.

\subsection{Standard ultra-pure water design}

\begin{figure}[tb]
\includegraphics[width=\linewidth]{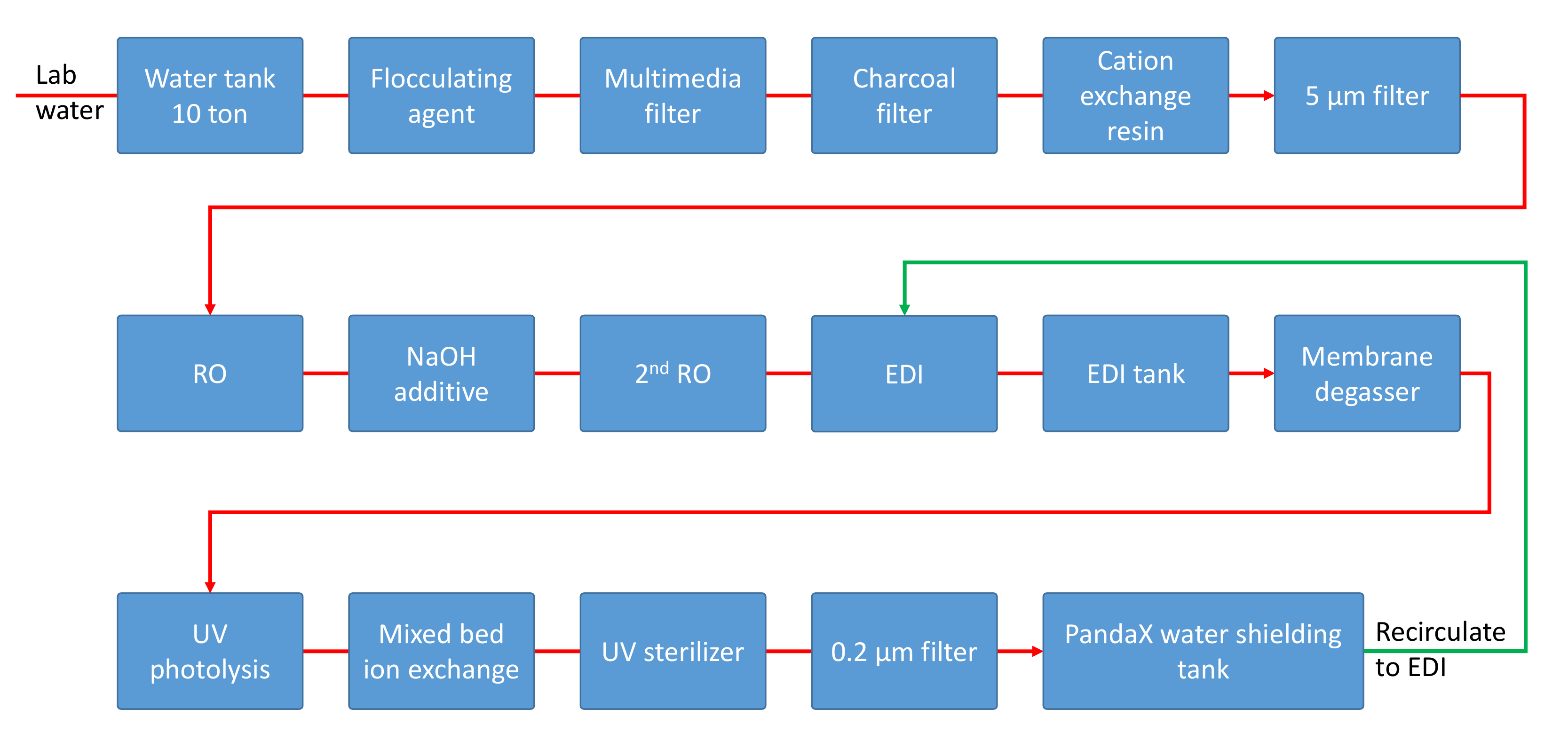}
\caption{Schema of the filtration system design.}
\label{fig:filteringSchema}
\end{figure}

The standard design is shown in Fig.~\ref{fig:filteringSchema}. 
The raw (tap) water from the laboratory infrastructure is first filtered for gross  particles through a multimedia (mainly sand and carbon) filter. 
The efficiency of this step is boosted up by the addition of a flocculating agent that causes some impurities either dissolved or in the form of fine particles to coagulate in big particles.
Then water is flowed through an activated carbon filter, which is a very good absorber. 
In particular, it is known that activated carbon can adsorb radon, so a first Rn pre-filtering is expected to occur at this stage. 
The next major step, the reverse osmosis (RO), is preceded by a softening stage. In fact, an osmotic membrane is permeable to water but not to impurities, so the permeate water is purer than the input, but the waste water (technically called ``concentrate") has a high impurity concentration. 
Unfortunately, carbonates (CaCO$_3$, MgCO$_3$, etc.) have low solubility in water, so when concentrated they tend to precipitate onto and clog the osmotic membrane of an RO. 
The softener consists of a cation exchange resin that replaces Ca$^{++}$ and Mg$^{++}$ with other ions (typically Na$^{+}$ or H$^{+}$) and, if necessary, a basic additive to re-balance the pH. 
Of course, some slow precipitation processes occur anyway, and are mitigated by running 
the RO units with a throughput of around 50\%. 
After the RO, the water resistivity is still in the tens of k$\Omega\cdot$cm, versus a UPW resistivity $>18$ M$\Omega\cdot$cm. 
This means a large amount of impurities is still present, in the form of charge carriers. 
Therefore, the following step is the EDI (ElectroDeIonization), which employs a transverse electric current, combined with semipermeable membranes, to remove the bulk of these impurities. 
It is to be noted that buffer tanks of 10~t capacity are envisioned for the raw water, and after the main purification stages.
Similar to capacitors in an electric circuit, these tanks prevent an abrupt interruption of the water flow, allowing the system to run for a full hour after a sudden water supply cutoff. 
This is plenty of time for the operators to intervene and prevent possible damage. 
Although only for the tank after the EDI N$_2$ flushing is strictly necessary, the option is kept open of flushing also the previous tanks, in case this helps with Rn control. 
The following stages are the degassing (normally employed to kill aerobic bacteria by O$_2$ removal), the UV (185 nm) photolysis of dissolved organics, and the mixed bed ion exchange resins. Further UV (254 nm) sterilization is employed to kill surviving (mainly 
anaerobic) bacteria and viruses. Of course, appropriate porosity filters are interspersed to remove particles of decreasing sizes, down to sub-micron size.

\begin{table}
\centering
\begin{tabular}{L{0.3\linewidth} L{0.15\linewidth} L{0.08\linewidth}}
\hline\hline
\textbf{Parameter} & \textbf{Value} & \textbf{Unit}\\
\hline
Output flow & 10 & m$^3$/h\\
Max input flow & 26 & m$^3$/h\\
Recirculation flow & 10 & m$^3$/h\\
Pre-RO treatment outflow & 22 & m$^3$/h\\
RO outflow & 12 & m$^3$/h\\
EDI outflow & 10 & m$^3$/h\\
EDI input pH & 4--11 & \\
EDI output H$_2$O resistivity & $>15$ & M$\Omega\cdot$cm\\
EDI tank capacity & 5 & m$^3$\\
\hline
\textbf{Degasser:} &&\\
O$_2$ removal efficiency & 99.9\% & \\
N$_2$ sweep rate & 0.8 & m$^3$/h\\\hline
\textbf{Output H$_2$O resistivity} & $>18$ & M$\Omega\cdot$cm
\\
\hline\hline
\end{tabular}
	\caption{Main parameters of the water purification system.}
\label{tab:filterPara}.
\end{table}

\subsection{Customizations}
As mentioned, customizations of mixed bed ion exchange and water degasser are introduce to remove the radioactive contaminations in water more effectively. Key parameters of the water purification system are listed in Tab.~\ref{tab:filterPara}.

The mixed bed ion exchange customization consists in measuring the U/Th removal efficiencies of several commercial resins to select those that best remove the ``chemicals" of our interest. 
Radium can be removed by a cation exchange resin that efficiently removes calcium and magnesium, because all three elements are in the same group of the periodic table. 
Using an Inductively Coupled Plasma Mass Spectrometry (ICP-MS) at Peking University, we have successfully tested two resins by measuring the U/Th content after filtration of water spiked with known amounts of these elements.

Standard degassers, like the membrane degassers we plan to employ, exploit laws of gas solubility in water, particularly Henry's law: ``the solubility of a gas in a liquid is, at equilibrium between liquid and gas, proportional to the partial pressure of that gas". 
Degassers put water in contact with highly pure N$_2$ and lowering the N$_2$ pressure to a value close to its boiling point at room temperature. 
This means we have both handles to optimize the radon removal efficiency after the system is built: we can both try to push the pressure a little lower (but monitoring the temperature more strictly) and improve the purity and flow rate of the N$_2$.

%% file: Electronics.tex
\section{Electronics and data acquisition}

\subsection{Introduction}
The readout electronics of PandaX-III TPC is one of the key subsystems of the experiment. Among many of its requirements and desired features, the readout electronics must provide low noise, high energy resolution, scalability to several tens of thousands of channels and the lowest possible level of radioactivity. The most prominent performance requirements and constraints, and the plans to solve the corresponding challenges are detailed below.
\subsection{Requirements and challenges}
\subsubsection{Channel count}
In the current baseline design, each of the two end-plates of TPC is read out by a tessellation of Micromegas detectors segmented in horizontal and vertical by 3 mm wide strips (XY readout). Assuming that each TPC end-plate is composed of 41 detectors, each segmented in 64 X and 64 Y strips, the total number of channels per end-plate is 5248 (i.e. 10.5 K channels for the complete TPC). A modular design based on the massive replication of a limited number of identical components is the most practical engineering solution to deal with systems of this scale.
\subsubsection{Tracking resolution}
A TPC allows complete 3D track reconstruction using the position of hit strips (or pixels) to determine X and Y coordinates, and electron drift time information to derive Z coordinates. The resolution along the X and Y axis are mostly determined by the segmentation pitch of the detector and various other parameters (e.g. transverse diffusion of electrons in gas). For PandaX-III, the preliminary estimates of the required track resolution is $\sim$3~mm. 
Resolution in the Z axis is primarily determined by the electron drift velocity in the TPC gas volume and the sampling rate of detector channels. In high pressure Xenon/TMA mixture at the anticipated drift field of 100 V/cm/bar, the electron velocity can be as low as ~1 mm/$\mu$s (not too strongly depending on the E field), leading to a maximum drift time of ~1 ms for a 1 m deep TPC. A spatial sampling pitch of 2 mm along the Z axis translates into a channel sampling rate of ~500 kHz which does not present any particular challenge for modern electronics. However, pre-amplifiers and shapers have to be tailored to operate with such relatively slow signals.
\subsubsection{Energy resolution}
Another critical requirement for the PandaX-III experiment, besides the precise track reconstruction, is that a very good energy resolution of 2-3\% FWMH is needed for the expected 2.458 MeV energy deposition from the NLDBD signal. This total primary energy will be shared among a variable number of channels, depending on track length and topology. Channel dynamic range, resolution and signal to noise margin have to be optimized to avoid saturation and the imprecise estimation of small amounts of deposited charges. Depending on the gain of the gas amplification stage, it is expected that the maximum charge to be measured on each individual channel after a typical Micromegas amplification of 1000 times is within a typical range from 100 fC to 10 pC. A resolution of 12 bits and a RMS noise of less than 2000 electrons for a readout channel connected to its detector via the appropriate length cable are our design goals. 
\subsubsection{Readout scheme}
The current baseline is to use a self-triggered readout using TPC itself. 
This can be achieved in two ways: a first option is to discriminate the signal of each readout channel and combine this information to build a global, per end-plate, trigger signal. The alternative is to read out and amplify the signal from the mesh of each Micromegas detector of an end-plate to elaborate the trigger signal. Either option, or a combination of both, could be used for PandaX-III. Taking the extreme case that a 30 cm long electron track is being drifted towards the detector, the overall span of the signals on one end-plate is 300 $\mu$s, an estimated length of the readout window. It should be noted that each TPC end-plate is independent from the others and from that of other TPCs. Consequently, triggering is only local to an end-plate and no particular synchronization is needed between end-plates. The time when the initial ionization happens can be determined indirectly by measuring the transverse diffusion.
\subsubsection{Event rate and data throughput}
The event rate expected is rather modest, typically a few Hz, and is dominated by background events. During dedicated calibration runs, the event rate might get significant higher, but designing a readout system capable of sustaining a maximum trigger rate of 10 Hz with less than 50\% dead-time seems to be sufficient. 
Assuming that the TPC comprises 10.5 K channel, and that each event consists of 512 sampling points per channel, each coded with a two byte amplitude, the raw event data flow at 10 Hz trigger rate is 102 MB/s, or 8.4 TB/day. Zero suppression to remove samples consistent with pedestals would provide another powerful reduction in data size. This dataflow can easily be transported by one (or a couple) of standard Gigabit Ethernet link(s). During regular data taking (non-calibration period), to cut down the data volume to a comfortable level, i.e. 1 TB/day as in the Daya Bay and PandaX experiments, some additional data reduction and compression may be needed before permanent storage.
\subsubsection{DAQ, data storage, control and on-line software}
The DAQ will be custom developed based on the firmware and embedded software of the back-end modules. An event builder builds the header containing event counter, time stamps, some trigger information, as well as all digitized samples from all channels into a structured event format, and saves the event directly into binary data files on disk. Disk arrays together with their controllers are connected to the data acquisition server via SAS (6 Gb/s) to ensure high speed and reliability data writing.
Very similar to the PandaX-II online monitoring system, as the data file is written to the distributed file system, a data copier copies the data file simultaneously to a distributed storage system via optical fiber links, likely located outside the underground lab. An online data quality production farm (likely also be above ground) launches online analyzers to produce diagnostic figures, visible from a webpage.

\subsubsection{Emplacement of the electronics}
In order to minimize noise pickup on the raw analog signals,
the readout electronics, or at least the pre-amplifier stage, must be placed close to the detectors. Fitting the electronics in the TPC gas volume would add several severe constraints, including tolerance to vacuum and high pressure, minimal outgassing for a negligible impact on gas purity, controlled impact of heating on the local temperature, etc. The solution is to place the front-end electronics of each end-plate in a waterproof vessel mounted on both sides of the TPC vessel, behind a radiopure copper shield, connected to the detector plane via custom feedthroughs.The electronics will operate in a closed environment of dry nitrogen gas, at a slight overpressure than the surrounding water. Low power dissipation is needed, however some means of cooling is still required. Access to the electronics will be difficult during the operation, therefore
particular attention must be paid to the robustness and reliability of readout electronics.
\subsubsection{Radio purity}
Commercial components and common industrial technology are developed 
without the need and awareness of having low radioactivity. This aspect is however critical for low background experiments like PandaX-III. The proposed strategy to tackle this issue is multi-fold. Firstly, the electronics will be placed behind a radiopure copper shield to relax the constraint on the radiopurity of the electronics as much as possible. Secondly, a design that minimizes the amount of material and board space will be performed. Using highly integrated components like ASICs and commercial FPGAs will help meeting this goal. Thirdly, the production of readout system will be carried out with selected PCB materials, solder paste (and components if possible), that are known to have low radioactivity or that will be qualified by our own test procedure.

\subsection{System architecture of the readout electronics}
\begin{figure}[!tbph]
\centering
\includegraphics[width=0.9\textwidth]{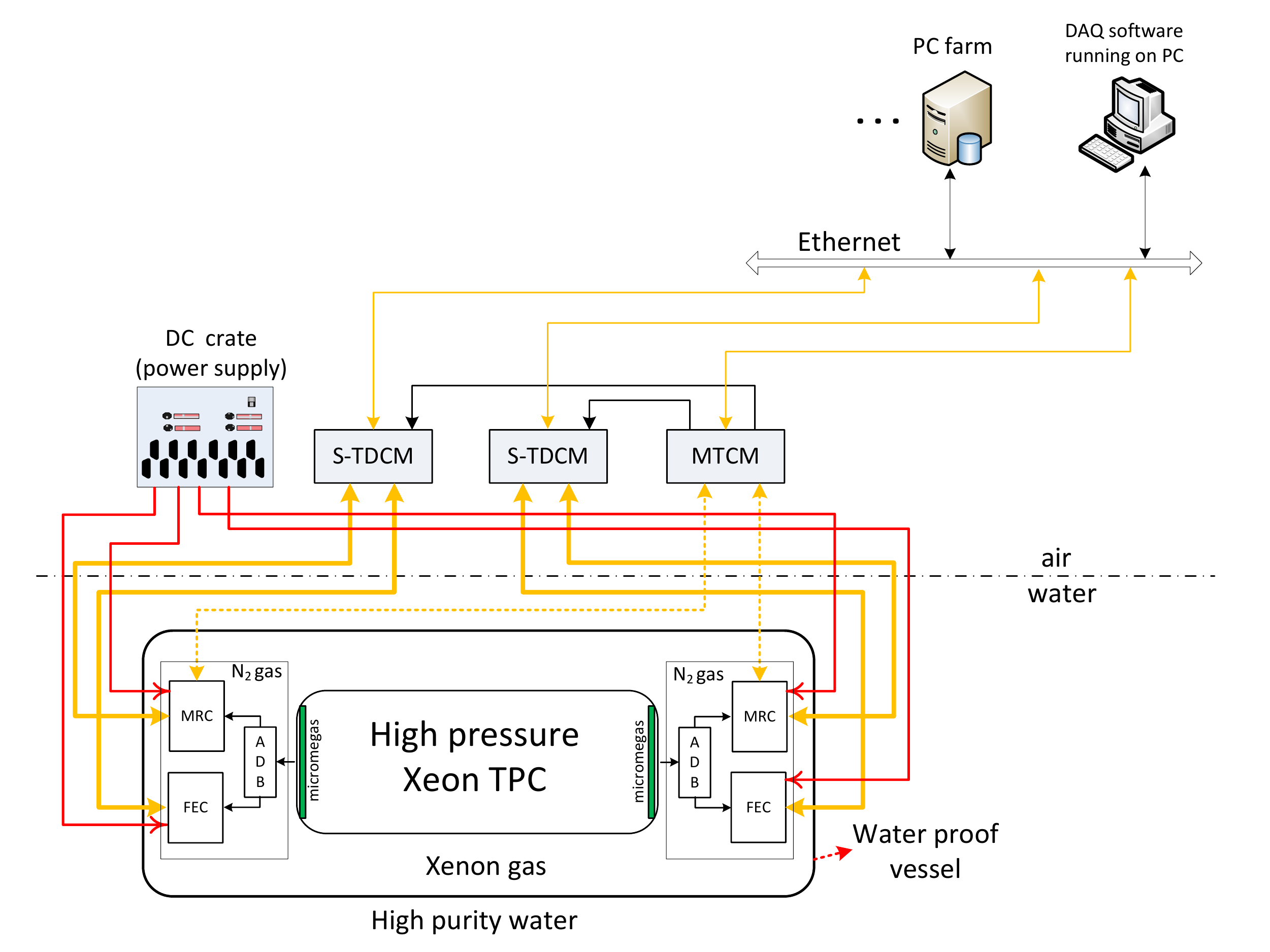}
\caption{Systematic architecture of PandaX-III readout electronics.}
\label{fig:architecture}
\end{figure}

As illustrated in Fig.~\ref{fig:architecture}, the PandaX-III readout electronics follows a modular architecture. At the top of the hierarchy, a PC farm running the DAQ software, receives event data from the Slave Trigger and Data Concentration Modules (S-TDCMs) via Gigabit Ethernet(GbE) links and user-defined protocol.

At the bottom level, there are two types of front-end modules, namely 
FEC (Front-end Card) and MRC (Mesh Readout Card) respectively. The $\sim$40 FECs, 
with 256 readout channels on each covering the anode strips of two Micromegas module, 
serve the total of about 80 Micromegas modules of the two end-caps. 
The two MRCs, with each serving one side of the end-cap, 
are designed to read out all the mesh signals and generate individual 
"Mesh-trigger" signals. The FECs and MRCs receive charge pulses from the 
Micromegas modules, integrate the charges, and digitize them under the command
of the trigger. The digitized data and status data are packed in FPGA and transmitted to S-TDCMs by a user-defined serial protocol with optic links.

\begin{figure}[!tbp]
\centering
\includegraphics[width=\textwidth]{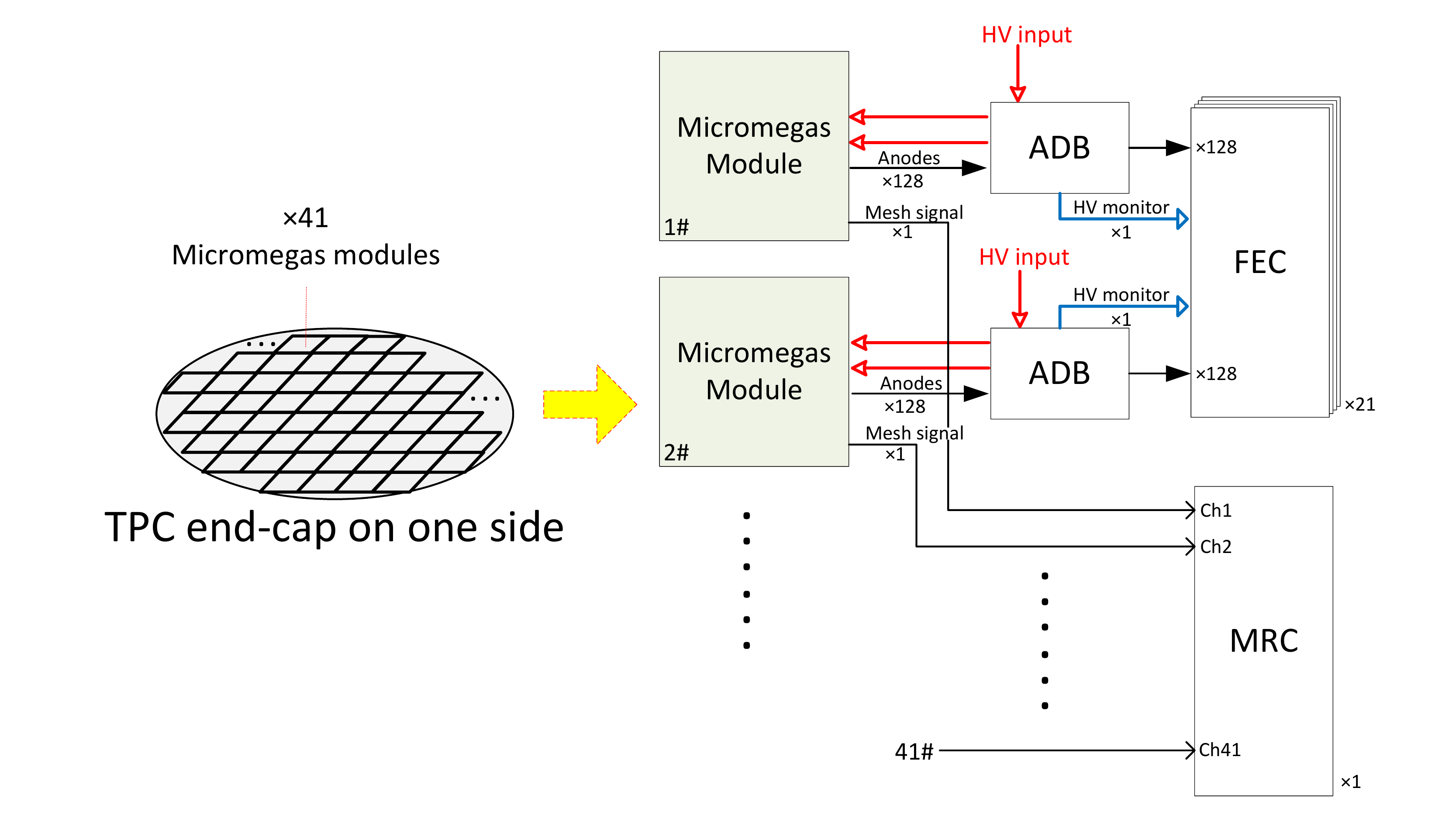}
\caption{Configuration of the electronics modules for one TPC end-cap.}
\label{fig:configuration}
\end{figure}

At the middle level, there are two S-TDCMs and one MTCM (Master Trigger and Clock Module). The S-TDCMs collect the event data packets and status information from FECs and MRCs, and send out commands and configuration data. The MTCM module receives all the 70 Mesh-trigger signals from two MRCs via optic links and to generate a Global-trigger with its FPGA logic, which then get distributed to the S-TDCMs in LVTTL or LVDS standard. Besides, a precision crystal oscillator with frequency of 25~MHz or 50~MHz is adopted in the MTCM. The clock signal and other control signals (such as Global-trigger, time stamp reset, event counter reset, etc.) are fanned out to the S-TDCMs. Control signals are then distributed from the two S-TDCMs to their corresponding FECs and MRCs. 
The configuration of the electronics for one TPC end-cap is illustrated in Fig.~\ref{fig:configuration}.

The hardware topology of the electronics is also indicated in Fig.~\ref{fig:architecture}.
The MRCs and FECs are installed inside the water-proof vessel made by high purity copper, on the side of TPC detector with optical and electrical feedthroughs for data links and 
high voltage connections. 
The optical and electrical cables are routed through the water and connected to the 
S-TDCMs and MTCM, which are installed on rack above the water level 
together with the low and high voltage supplies.

\subsection{Design of the electronics modules}
\subsubsection{Front-end Electronics Card (FEC)}
\begin{figure}[!tbp]
\centering
\includegraphics[width=\textwidth]{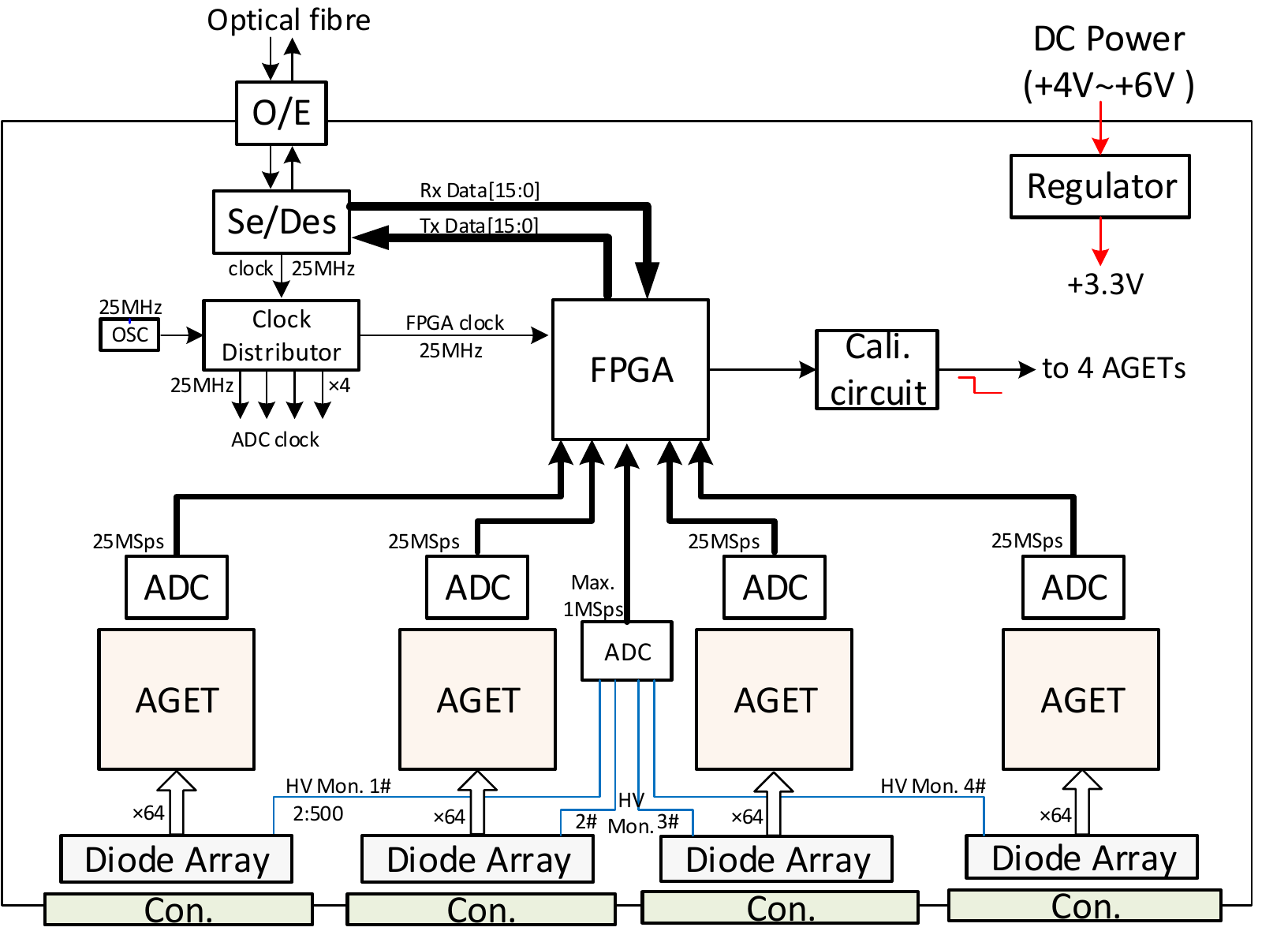}
\caption{Block diagram of FEC module.}
\label{fig:FEC}
\end{figure}

The design schematics of the FEC module is shown in Fig.~\ref{fig:FEC}. 

\begin{figure}[!tbp]
\centering
\includegraphics[width=\textwidth]{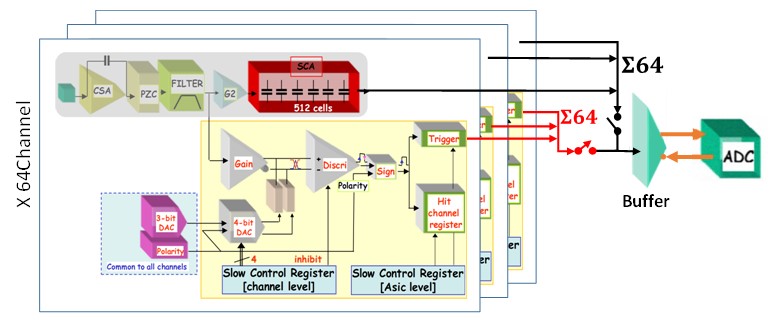}
\caption{Schematic view of AGET ASIC. Note that the ADC is not part of the AGET.}
\label{fig:AGET}
\end{figure}

It is designed to integrated the charge of Micromegas anode signals, digitize the waveform after shaping, and send the data packets to the S-TDCM. Each FEC contains 4 ASIC chips named AGET, each hosting 64 channels. 
This chip was developed at Saclay, IRFU CEA of France, as a generic 
chip used in a tracking and calorimetric TPC. 

As shown in Fig.~\ref{fig:AGET}~\cite{ref:AGET_ASIC}, each input channel includes a charge sensitive preamplifier (CSA), an analog filter (shaper), a discriminator for multiplicity building and a 512-sample analog memory with the structure of Switched Capacitor Array. As the pulse width varies from several hundred ns to possibly more than 100 $\mu$s, which is usually much greater than the shaping time of AGET, the charge of detector signal should be derived from the area rather than the peak of sampled waveform.

The AGET chip has an adjustable gain to support the dynamic range of 120 fC to 10 pC.
The peaking time of AGET is also adjustable in the range between 50~ns to 1 $\mu$s. For the application of PandaX-III TPC, the dynamic range is planned to be configured as 10 pC.
According to the simulation, the peaking time and the sampling rate should be set at
4~MS/s and 1~$\mu$s, respectively, to ensure sufficient length of the readout window
to cover the full track and to minimize aliasing error. 

The FEC connects to the pigtails of Micromegas modules via custom adapter boards (the "AdB" in Fig.~\ref{fig:architecture}). The high voltage supply to the mesh, the internal rim and external rim are feed to the 
Micromegas via the same adapter. Spark protection circuits will be implemented on the 
FEC. Attenuated high voltage on the adapter board will be fed to dedicated ADC
channels for monitoring purposes.

On each FEC board, an on-board FPGA is used to control the ADCs and send the data packets to the S-TDCM with serial optical links. The protocol will be described in the following paragraphs.

\subsubsection{Mesh Readout Card (MRC)}
\begin{figure}[!tbp]
\centering
\includegraphics[width=\textwidth]{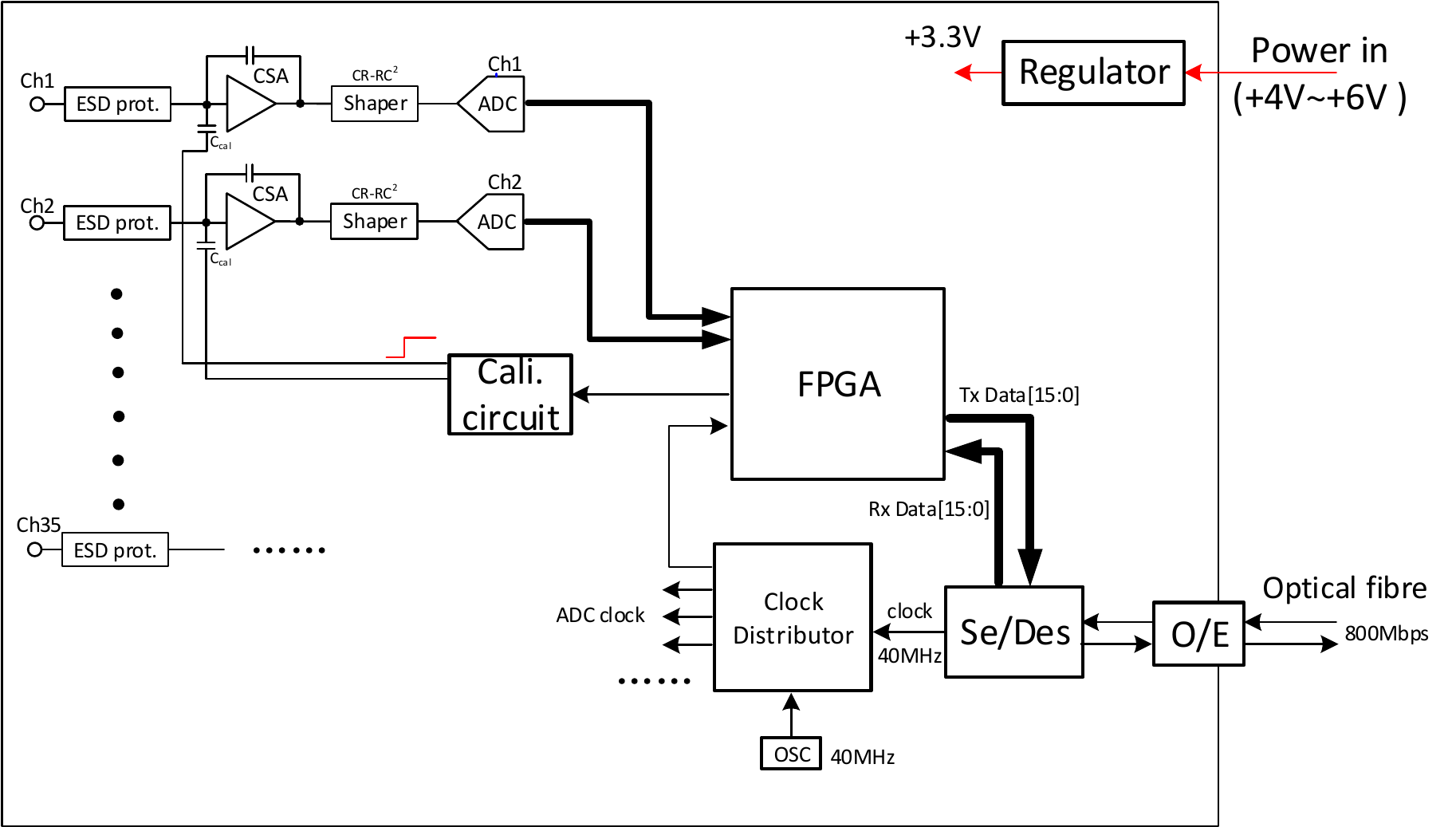}
\caption{Block diagram of MRC module.}
\label{fig:MRC}
\end{figure}

The function of MRC (Mesh Readout Card) is to process the mesh signals. 
The charge from the mesh can be used 
in combination with the anode strip signals to help the energy determination. 
The MRC uses discrete components instead of the AGET chips to implement the charge measurement circuits (Fig.~\ref{fig:MRC}).
One MRC contains 35 input channels serving all meshes from the end-cap on one side. 
For each input channel, there are a charge sensitive amplifiers and a CR-RC$^2$ shaper. 
The waveform of each shaper output is sampled by ADC channel and recorded by an FPGA 
chip. In order to assure the accuracy, a relatively high sampling rate (e.g. 10~MHz) and a slow peaking time (e.g. about 5~$\mu$s) should be applied. The generation of ``hit'' signal
from each mesh is optional on the MRC. 
The recorded data is sent to M-TCM via an optical link.
To effectively utilize the bandwidth, zero-suppression algorithm should be applied to the 
data from the MRC.

\subsubsection{Master Trigger \& Clock Module (M-TCM)}
The FECs and MRCs can operate either with the external trigger or with the internal trigger derived from AGET chips. The main function of M-TCM is to receive the waveforms and the optional multi-bit ``hit'' information from two MRCs via the optic link, and to generate the so-called Global-trigger signals. 

\begin{figure}[!tbp]
\centering
\includegraphics[width=\textwidth]{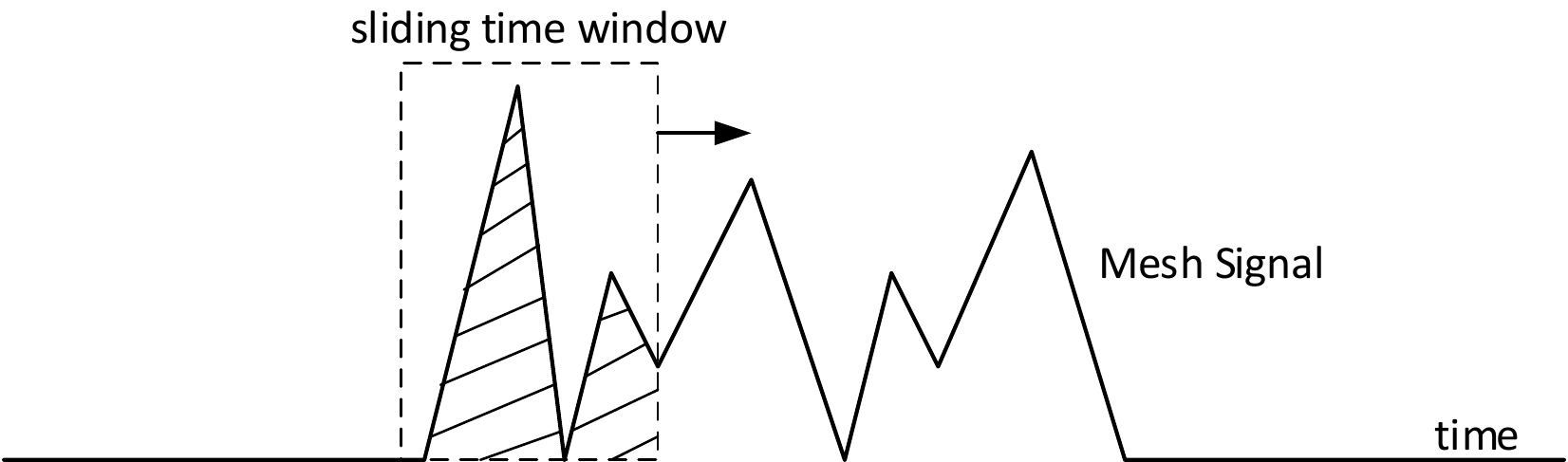}
\caption{Sliding window integration algorithm.}
\label{fig:algorithm}
\end{figure}

Two operation modes are forseen. 
In the first mode (Esum mode), a simple sliding window trigger logic can be implemented in the M-TCM  to produce the 
trigger, also illustrated in 
Fig.~\ref{fig:algorithm}. The sampled waveform from each MRC mesh channel is numerically integrated by a sliding time window. The summed signals from all meshes are 
compared to a preset threshold to generate
the trigger. In the second mode, the sliding window integration and the ``hit''
generation are carried out by the MRC and the M-TCM generates the trigger 
based on the coincidence of ``hits''.
Note that in order to fully reconstruct the track and energy for events with track 
crossing the cathode, the trigger for each MRC on each side 
should be ``ORed'' to produce the Global trigger.

In addition to the trigger, M-TCM also provides global control signals. The clock signal can be generated either by the on-board crystal oscillator or by external input. The time stamp reset and event counter reset can be generated by software commands.

\subsubsection{Slave Trigger \& Data Concentration Module(S-TDCM)}
The Slave Trigger and Data Concentration Module (S-TDCM) is an evolution of the so-called ``Trigger Clock Module'' (TCM) developed for the MINOS experiment~\cite{ref:readout_Calvet}. It is composed of three main parts, a) a commercial, off-the-shelf, System-On-Module (SOM) based on a powerful FPGA with an embedded processor, b) a custom-made carrier card that holds the SOM and the necessary interface connectors for bringing the external power, for communicating with the M-TCM, the FECs, and for the optional cascading of multiple S-TCMs, and c) up to two custom-made physical layer cards that hold the appropriate optical transceiver components to communicate with the FECs. Under this design, one S-TDCM can control up to 32 FECs via optical links, which is adequate to read out one TPC end-cap. The speed of each FEC link is limited to 800 Mbaud by the S-TDCM receivers and this card will provide a single 1 Gbit/s Ethernet interface to the DAQ and control PC, sufficient for the required data bandwidth (8~TB/day).

The main function of S-TDCM is to receive from the M-TCM the system wide reference clock, the global time-stamp reset signal, an event counter reset, and the trigger signal for the TPC end-cap which it controls. The readout of the two end-caps of one TPC takes two S-TDCM. The trigger signal sent to the S-TDCM may originate from the processing of mesh signals done by the M-TCM, or it could originate from the FECs themselves. In this case, the hit multiplicity information of the AGET chips is transferred from the FECs to the S-TDCM which processes these trigger primitives locally to elaborate a self-trigger signal that is output to the M-TCM and then looped-back to the S-TDCM.

The S-TDCM makes the fanout of the reference clock, trigger and other synchronous signals to the 18 FECs of its TPC end-cap via dedicated optical links using a proprietary protocol. The S-TDCM receives synchronously the hit multiplicity trigger primitive of the 72 AGET chips of a TPC end-plate to elaborate a self-trigger bit. The S-TDCM also combines the dead-time information of its 18 FECs to inform the M-TCM when triggers can be accepted and when triggers must be vetoed because one or several FECs are currently busy. The S-TDCM is responsible for the configuration of all the operational parameters of the FECs, front-end AGET chips, and it performs the periodical gathering of the variables monitored by the FECs and those monitored locally on-board the S-TCM: supply voltages, currents and temperatures. The S-TDCM collects event data from the 18 FECs of a TPC end-cap and assembles full events that are transferred to a dedicated control PC via a point-to-point private Gigabit Ethernet link (or some higher speed media). Assembled events may further be processed locally before they are sent to the PC farm over a globally shared Ethernet network. The PC farm runs on-line analysis programs and performs the final stage of data reduction and compression before mass storage.

The S-TDCMs are placed above the water level, therefore there is no constraints on the radiopurity of these cards and standard industrial components can be used.

\subsection{Communication protocols}
\subsubsection{Between S-TDCM and FECs}
The S-TDCM needs to deliver synchronous information to all FECs (clock, trigger, time stamp reset, event counter reset, etc.) and also messages that have no specific constraints in terms of latency and synchronization: configuration parameters, data requests, etc. The FECs need to send to the S-TDCM synchronous and time coherent multiplicity trigger primitives, as well as asynchronous messages: configuration responses, monitored variables, and event data. The network linking the S-TDCM and the FECs is fully synchronous. It features a deterministic latency in both transfer directions but uses a different structure and encoding in each direction. In the downstream direction, i.e. from the S-TDCM to the FECs, a broadcast network is used: the same serial stream is duplicated and sent to all FECs. The duplication may be done by an electrical hardwired fanout placed on the S-TDCM before the optical transceivers. 
In the upstream direction, i.e. from the FECs to the S-TDCM, each FEC has a dedicated point-to-point optical communication link with the S-TDCM. All links in both directions are fully synchronous to the clock originating from the M-TCM and feature an equal and predictable latency (within one unit interval of the link).

In order to accommodate the various types of traffic carried by these links, the available bandwidth is divided into several virtual channels that are time-multiplexed over the same physical link. 

\subsubsection{Between M-TCM and MRC}
The optical links deliver
synchronous information (clock, trigger, time stamp reset, event counter reset, etc.) from M-TCM to two MRCs (downstream), and to transmit mesh waveform data from MRC to 
M-TCM (upstream). Such links can be achieved via commerical transceiver with a Gbps 
speed.  
Similar to the S-TDCM \& FEC protocol, the optic links are divided into several time-multiplexed virtual channels.

%% file: RadScreening.tex

\section{Material screening}
\label{screening}
\begin{figure}[tbp]
\centering
\includegraphics[height=0.34\textwidth]{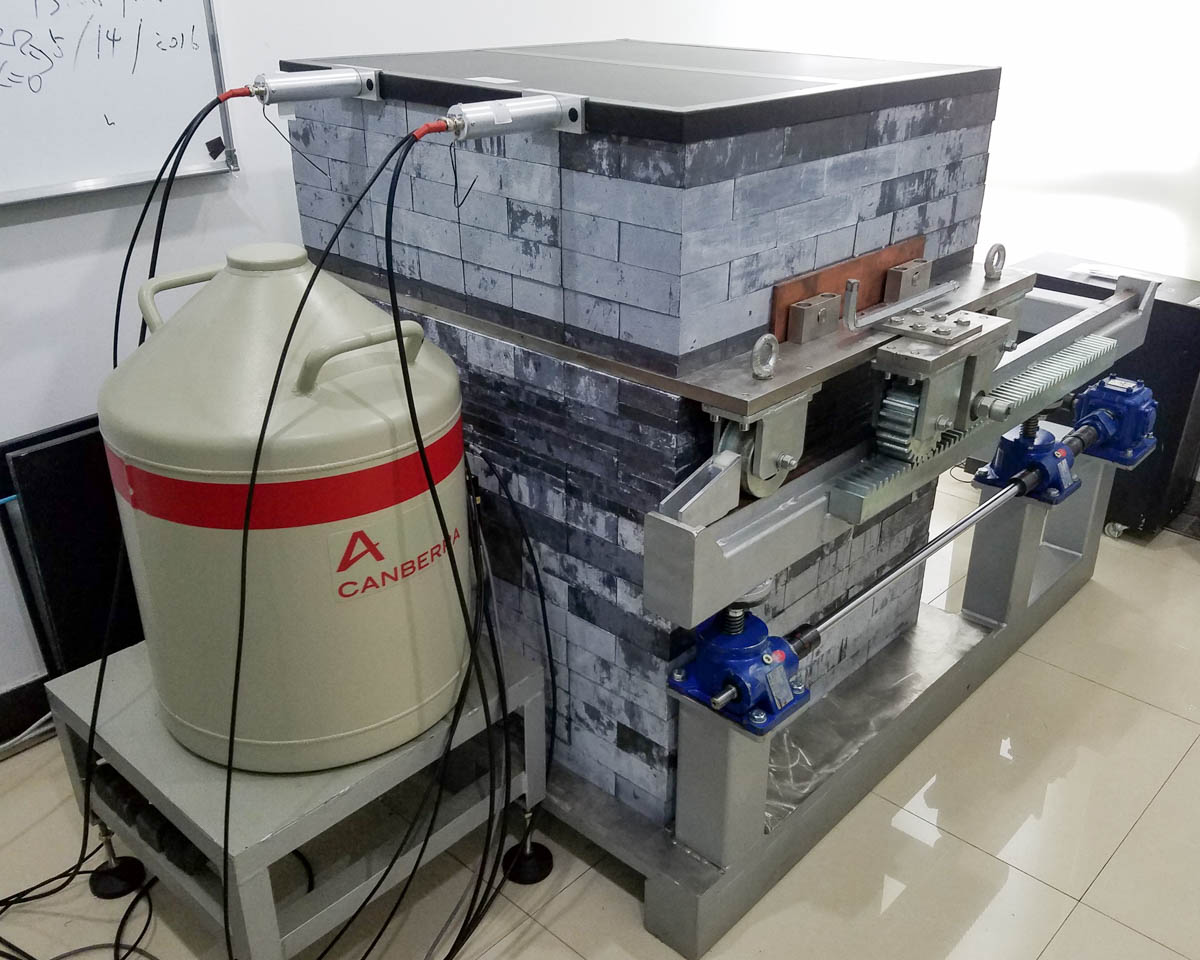}
\hfill
\includegraphics[height=0.36\textwidth]{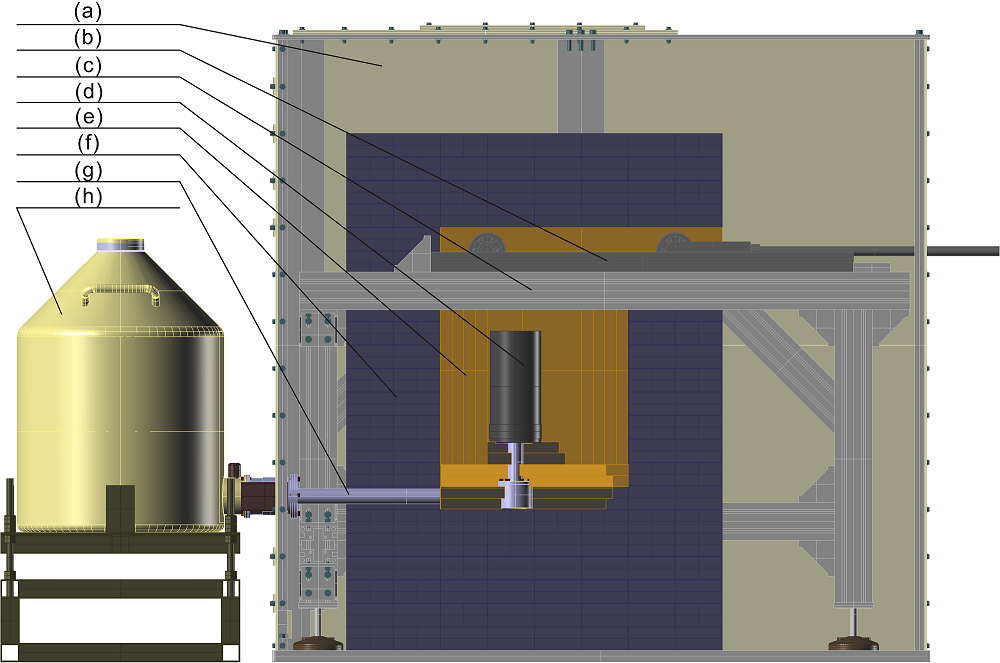}
\caption{(Left) Picture of the counting station at SJTU with scintillator panels for muon veto on top.
 (Right) Schematic of the PandaX $\gamma$-ray counting station. 
The HPGe and its carbon fiber cover (d) is placed in the center. 
The copper chamber (e) surround the HPGe crystal. 
The lead bricks layer (f) and outer acrylic vacuum shell (a) are also shown in the figure. 
The supporting structure (c) can hold the air tight acrylic shell and the rails (b) where the shielding's top cover can move on it. 
Liquid nitrogen is stored in the dewar (h). 
The electronic cables as well as cold transportation rod are sealed in the cooling finger (g). 
The nitrogen blowing equipment is omitted in this schematic.
}
\label{fig:hpge} 
\end{figure}

The low-radioactive background controlling is one of the most critical issues for rare event search experiments. 
Various techniques have been developed for measuring extremely low radioactive background, including High Purity Germanium (HPGe) $\gamma$-ray counting, Inductively Coupled Plasma Mass Spectrometry (ICP-MS), Neutron Activation Analysis (NAA), etc. 
Multiple techniques are often desired to measure the radioactive contaminations of different materials, due to different chemical and physics properties. 
Different techniques also have different intrinsic sensitivity to different radioactive isotopes. 
The \Piii collaboration benefits from experience and infrastructure accumulated during the previous generation of PandaX dark matter experiments.
Currently we have two HPGe $\gamma$-ray counting stations and one ICP-MS available for our use.

\subsection{$\gamma$ Counting Station}

The PandaX group at SJTU runs one HPGe counting station in CJPL-I underground lab and another one on the surface at Shanghai.
The CJPL detector was fabricated by Ortec~\cite{GEMMX-specifications}, customarily designed for rare radiation counting purpose. 
It has a P-type coaxial HPGe crystal with the sensitive mass of 3.69 kg, leading to a 175\% relative detection efficiency. 
The large crystal makes it highly effective in detecting photons with energy range from 10~keV to 20~MeV. 
The 630 mm long cooling finger allows us to fill sufficient shielding materials between the crystal and its liquid nitrogen (LN$_2$) dewar. The dewar can hold up to 30 liter LN$_2$. 
The Shanghai counting station has a smaller HPGe from Canberra, which weighs at 0.63~kg and has a relative efficiency of 34\%.
High energy cosmic radiation can introduce various types of radioactive events, and thus ruin the experiment. 
For the CJPL, there is about 2400 m rocks (or 6720 m water-equivalent) over head, which makes it the deepest underground lab in the world to date~\cite{CJPL}. 
The muon flux there was measured to be (2.0$\pm$0.4)$\times$10$^{-10}$cm$^{-2}$s$^{-1}$, a factor of $10^{8}$ reduction compared to that at sea level~\cite{CJPL-muon}. 
Therefore, practically, this counting station was in a cosmic-ray-free environment. 
The Shanghai station has effectively no overburden. 
A pair of scintillation panels (Shown in Fig.~\ref{fig:hpge} (Left)) are added on top of the counting station to effectively veto muons passing through the HPGe. 

Both counting stations share similar, extensive passive shielding to further reduce the $\gamma$-ray background from the environment. 
The shielding structure is visible in the picture in Fig.~\ref{fig:hpge} (Left) and the schematic is shown in Fig.~\ref{fig:hpge} (Right). 
The innermost layer of shield around the HPGe crystal is a chamber made of OFHC copper. 
the space to hold samples inside the chamber is $20\times20\times35$ cm$^{3}$.
On each side, the copper chamber wall is 10-cm thick. 
Due to its high purity and relatively high Z, the copper can shield the radioactivity from outside without emitting too much $\gamma$-ray from itself.
The OFHC copper placed at the bottom of this chamber were  designed to match the curvature of the HPGe cryostat cold finger. 
In this way, the cooling finger can be protected well, while still keep good $\gamma$-ray shielding efficiency. 

Outside the copper layer, a 20~cm layer of lead provides further shielding. 
This layer was made of over 500 pieces of lead bricks, and each is  $20\times10\times5$ cm$^{3}$ large. 
The top cover of the shielding  can move along sliding rails on the sides. which makes it easier to access the copper chamber and handle test samples.

The CJPL station has an air-tight acrylic shell covering the whole system. 
To further reduce the Rn background, boiling nitrogen gas was blown into the air-tight acrylic shell continuously. 
Since the boiling nitrogen gas itself is almost Rn free, therefore, it can flush out the Rn in the shielding system, especially that in the volume of the copper chamber.
When open and close the top shield to replace test samples, the Rn gas could be trapped there and hard to be blown out. 
Normally, it takes about 12 hours before the Rn level drops from $\sim$200 Bq/m$^3$ to $\sim$1.5 Bq/m$^3$. 

Samples are usually counted at Shanghai for preliminary evaluation.
Should the events from characteristic gamma lines from a sample is too few to distinguish from background, the sample will be sent to CJPL for further evaluation. 
For each sample, a dedicated Geant4-based simulation will be performed to evaluate the efficiencies of each $\gamma$ line. 
Combined with measured background and signal spectra, a measurement or upper limit is given for the sample.
Currently, we are counting samples of OFHC copper, acrylic and Teflon.

\subsection{Inductively Coupled Plasma Mass Spectrometry}

\begin{figure}[tb]
  \centering
    \includegraphics[width=0.4\textwidth]{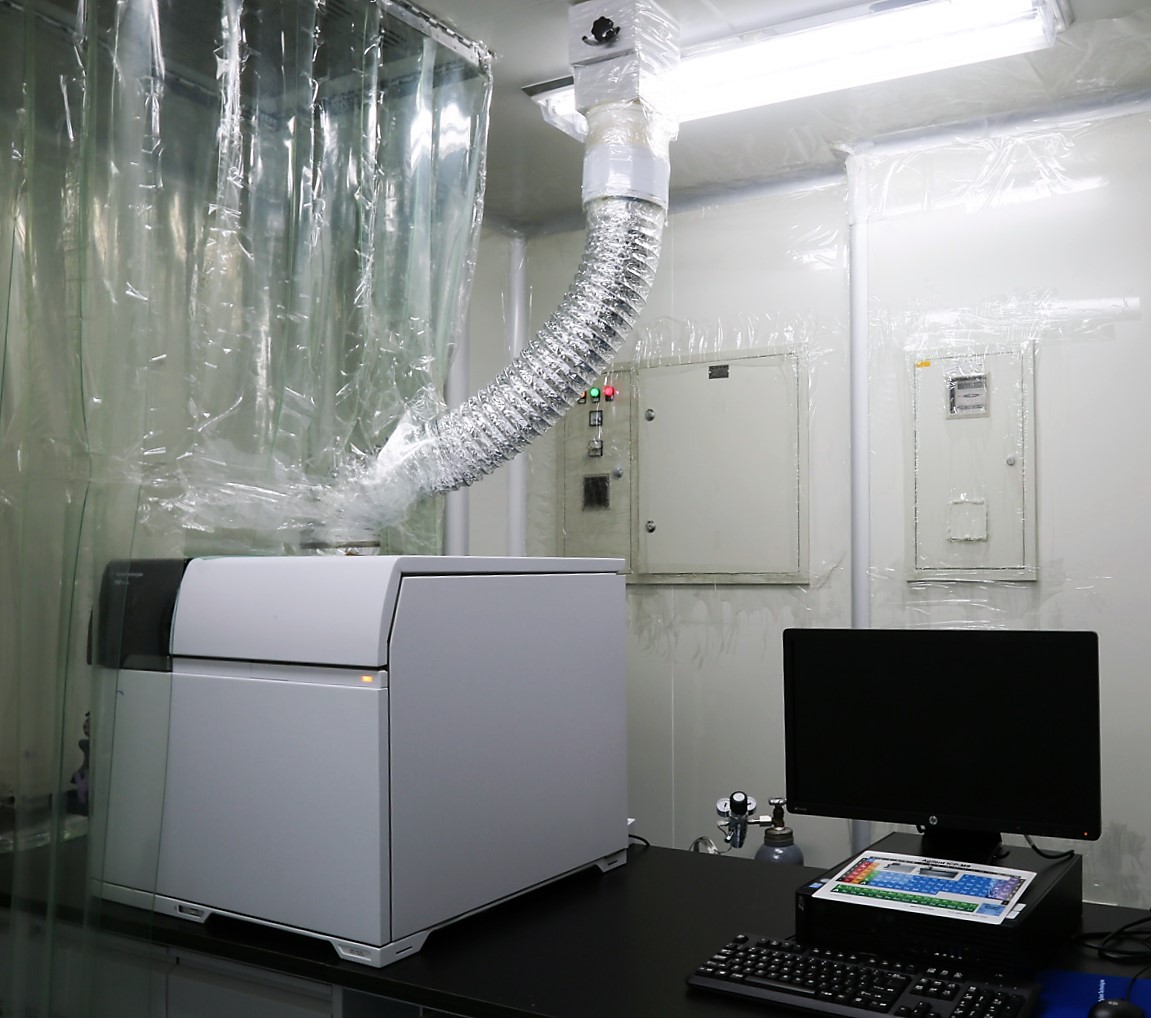}
    \hfill
    \includegraphics[width=0.52\textwidth]{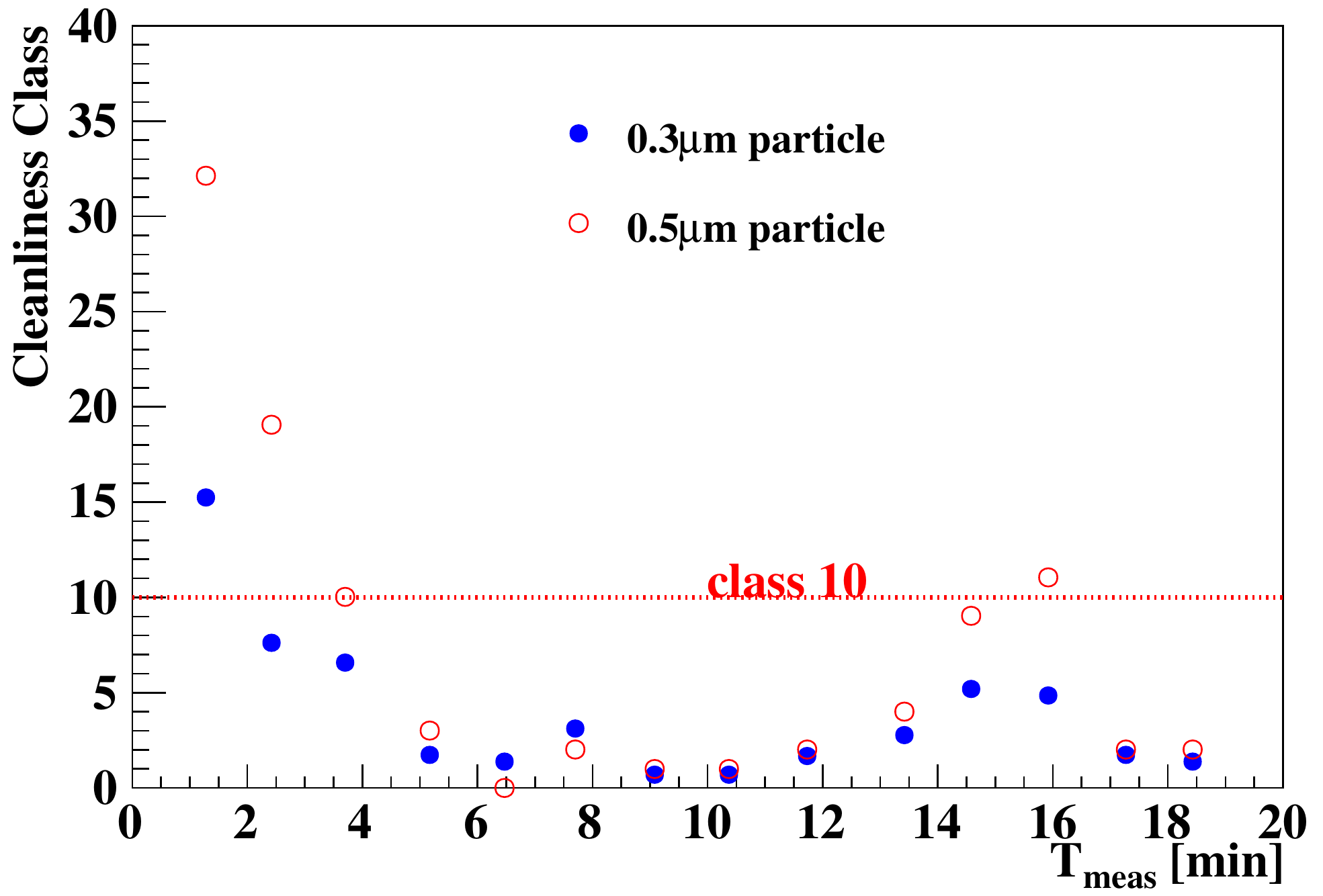}
  \caption{
(Left) Picture of the Agilent 7900 ICP-MS in the cleanroom of ULRML at Peking University. (Right) The cleanliness of the cleanroom changed with time. The measured point is in the sample preparation room and outside the clean workbench where the sample will be prepared in the bench. Due to the presence of operator at the beginning, the value was worse at $T_{meas}=0$. After the operator left, the value was dropping step by step until $T_{meas}=12$~minitues when the operator came and started to exchange the cleanroom suit outside its air-shower door.
  }
  \label{fig:icpms}
\end{figure}

\begin{table*}[tbh]
\centering 
\begin{tabular}{ c c c c c c }
\hline
\hline
 \multirow{2}{*}{\textbf{Sample}} & \multirow{2}{*}{\textbf{Time}} & \multicolumn{2}{c}{\textbf{\thttt}}  & \multicolumn{2}{ |c }{\textbf{\utte}}\\
 \cline{3-6}
   & & [CPS] & [ppt]& [CPS]& [ppt]\\
\hline UPW & 2016/5/9 0:26 & $13.1\pm1.3$ & $0.0104\pm0.0010$ & $8.8\pm1.6 $& $0.0088\pm0.0016$\\
\hline UPW & 2016/5/9 0:28 & $11.9\pm1.6$ & $0.0094\pm0.0013$ & $7.8\pm2.6 $& $0.0078\pm0.0026$\\
\hline Watsons water & 2016/5/9 0:29 & $13.9\pm2.5$ & $0.0110\pm0.0020$ & $310.7\pm6.1 $& $0.309\pm0.006$\\
\hline Watsons water & 2016/5/9 0:31 & $17.3\pm10.5$ & $0.014\pm0.008$ & $315.7\pm7.5 $& $0.314\pm0.007$\\
\hline UPW & 2016/5/9 0:34 & $9.4\pm4.0$ & $0.0075\pm0.0032$ & $12.8\pm2.4 $& $0.0127\pm0.0024$\\
\hline UPW & 2016/5/9 0:35 & $10.9\pm2.9$ & $0.0086\pm0.0023$ & $13.5\pm2.1 $& $0.0135\pm0.0021$\\
\hline No Sample & 2016/5/9 0:38 & $1.6\pm0.7$ & $0.0013\pm0.0006$ & $1.0\pm0.9 $& $0.0010\pm0.009$\\
\hline Calibration & 2016/5/9 0:41 & $118450\pm516$ & $94.3\pm0.6$ & $100307\pm655 $& $99.8\pm0.7$\\
\hline
\hline
\end{tabular}
\caption{U and Th in Ultra Pure Water(UPW) and the background counts of ICP-MS. The UPW is Watsons water filtered by Milli-Q system.
The background of the machine can be obtained by no sample (just leaving the sample input tube in the air). CPS (Counts Per Second) stands for the count rate and ppt (Parts Per Trillion) for the concentration of U or Th in solution.
Listed errors of CPS and concentration are statistics errors except those of the calibration solution are provided by their dealer.
Each number is the averaged value among three measurements and each one was integrated five seconds.}\label{tab:icpms}
\end{table*}

In Peking University, an Ultra Low Radioactivity Measurement Laboratory (ULRML) is established and equipped with an Agilent 7900 ICP-MS (Inductively Coupled Plasma Mass Spectrometry) and a 45~$m^2$ cleanroom (Fig.~\ref{fig:icpms} (Left)).
The ULRML is currently assaying the detector materials. 

Since the first generation of commercial ICP-MS appeared on the market in the 1980s, the technique has been adopted for trace elemental analysis across a wide range of industries and applications.
In comparison to the traditional methods, ICP-MS can assay a wide range of analytic elements and provide isotopic information with fast speed and high-precision. 
However, it is limited in the amount of total dissolved solids in one assay. 
Generally, it is recommended that samples have no more than $0.2\%$ total dissolved solids for best instrument performance and stability. 
The limitation requires that solid sample must be diluted before running through ICP-MS which can make the detection limits worse.
One way to improve the detection limits is to undertake prior separation of the analytes from the matrix. 
Methods, such as co-precipitation, liquid-liquid extraction, distillation, ion-exchange can be used. 
These techniques often require longer analysis times and give rise to additional analytical problems, including contamination during sample pretreatment and increased blank levels, which must be carefully controlled~\cite{EXO08}.

To suppress contamination from dust in air during sample preparation and measurement, a cleanroom is built. 
The room is mainly divided into four regions: utility room, chemical room, sample preparation room and instrument room. 
The gas cabinets are set in the utility room, where the gas supply bottles for ICP-MS are stored. 
A chemical hood is seated in the chemical room, all the operation involving concentrated acid, such as purification for acid and pressurized sample digestion, will be done there. 
The ICP-MS is installed in the instrument room. 
Measurement done in the sample preparation room demonstrates the  cleanroom is rated better than class 10, as shown in Fig.~\ref{fig:icpms} (Right). 
To obtain a better clean region for sample preparation, a clean workbench is placed inside the sample preparation room and the cleanliness inside the workbench is better than class 10.

The performance of the ICP-MS are tested with Ultra Pure Water (UPW) and the results are shown in Table~\ref{tab:icpms}. 
Before measurement, the ICP-MS is optimized to favor high mass range by tuning its setting parameters, which includes torch tube positions, gas speed and voltages of lens.
Then its sample input system has been carefully washed. 
UPW are measured twice to check the stability between measurements.
Then the commercial distilled water (called Watsons water by the brand name) water are measured. 
After this, UPW are measured again to check the contamination from the Watsons water. 
Then we leave the sample input tube in the clean air (tagged as no sample in the table) to measure the background counts of the ICP-MS.
A standard solution which are diluted from a 100~ppm Uranium and a 100~ppm Thorium standard solutions are measured finally as the system calibration. 
The relative precision of the standard solution is $\pm0.6\%$ according to the vendor. 
Detection limits for U/Th are somehow measurement period dependent and one measurement shows it is 0.005 ppt for Uranium using the 3~$\sigma$ of ten repeated measurements. 
Their level can be deduced from the background levels shown in the table ( three times of background level $0.0010\pm0.009$~ppt for Uranium and $0.0013\pm0.0006$~ppt for Thorium). 
Calibration results indicate its sensitivity is $1005\pm9$~cps/ppt for \utte and $1256\pm10$~cps/ppt for \thttt.
We are measuring the contamination levels of resin samples used in the water purification system.

%% file: Simulation.tex

\section{Monte Carlo background simulation and signal detection efficiency}
\label{sec:Montecarlo}

Given the low decay probability of NLDBD events, a low background level is critical for PandaX-III. The expected background level should be studied carefully before the construction of the detector. We have developed a Monte Carlo (MC) background model to assess the final sensitivity achievable by our experiment. This model will allow us to estimate beforehand the expected background contribution of each of the detector components, and the surrounding environment. The study of those independent background contributions will serve to define constrains on the maximum radiopurity budget, allowed by each of those components, to be considered during the detector design and construction. The results obtained with this model will also serve as the basis for the projected PandaX-III sensitivity.

\begin{figure}[tb]
  \centering
  \includegraphics[width=\textwidth]{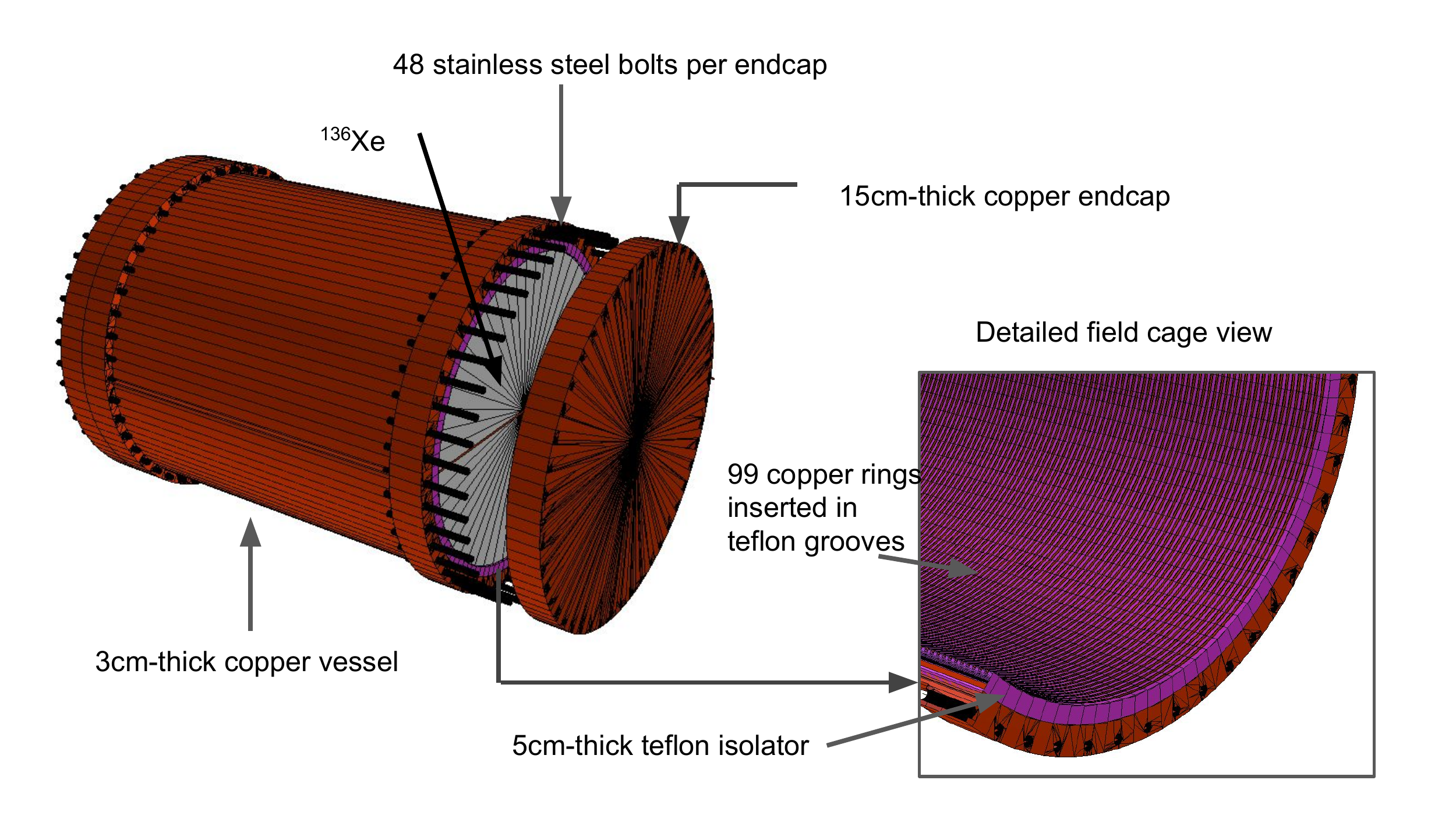}
  \caption{A schematic view of the detector geometry used in MC simulation. One of the end-caps is open to show the vessel contents. The water shielding and the laboratory concrete walls implemented in our geometry are not shown in this plot.   }
  \label{fig:mc_overview}
\end{figure}

The detector geometry used in the simulation, shown in Fig.~\ref{fig:mc_overview}, is defined based on the current design.
The high pressure copper vessel is of a cylindrical shape with flanges and end-caps in each side. The thicknesses of the vessel barrel and the end-caps are 3~cm and 15~cm, respectively. Each end-cap is connected with the barrel by 48 stainless steel bolts. The main components of the TPC are the field cage, the central cathode and the Micromegas readout planes. The field cage is made of a 5~cm thick PTFE wall and 99 copper shaping rings, which helps to generate a uniform drift field. The two Micromegas readout planes are placed at each end of the detector. The copper made central cathode splits the inner space into two independent drift chambers. A summary of the main components and the main parameters considered in the geometry are shown at Table~\ref{tab:parameters_geometry}.

The simulation setup consists of the detector vessel filled with 200\,kg of xenon ( +1\%TMA), corresponding to a total pressure of 10\,bar. The vessel is immersed on a large volume water pool (as described in section~\ref{sec:water}) that protects from gamma radiation produced within the laboratory walls. 

The simulation results will be presented in three stages. First, we report the total energy deposited in the gas. Second, we include the detector response related to the event acquisition. Third, we describe the topological rejection advantages of our particular detector readout.

\begin{table}
  \centering
  \begin{tabular}{L{0.15\textwidth}L{0.15\textwidth}ccc}
    \hline
    \hline
    \textbf{Component} & \textbf{Parameter} & \textbf{Value} & \textbf{Material} & \textbf{Weight} \\ \hline
    \multirow{3}{*}{Vessel} & inner diameter & 80\,cm & \multirow{3}{*}{Copper} & \multirow{3}{*}{3438 kg} \\
              & inner height & 200\,cm &  &    \\
              & wall thickness & 3\,cm &  &    \\\hline
    \multirow{2}{*}{End-cap} & diameter & 88\,cm & \multirow{2}{*}{Copper} & \multirow{2}{*}{3320 kg} \\
              & thickness & 15\,cm &  &    \\\hline
    \multirow{2}{*}{Bolts} & diameter & 1.4\,cm & \multirow{2}{*}{Stainless steel} & \multirow{2}{*}{230.1 kg} \\
              & height & 40\,cm &  &    \\\hline
    \multirow{4}{*}{Field cage} & inner radius & 75\,cm & \multirow{3}{*}{PTFE} & \multirow{3}{*}{1042 kg} \\
              & height & 200\,cm &  & \\
              & thickness & 5\,cm &  & \\
              & rings & 99 &    \multirow{1}{*}{Copper}  & 118.2 kg \\\hline
    Cathode & thickness   &   50\,$\mu$m     &   \multirow{1}{*}{Copper}  &    0.79 kg   \\
    \hline    
    \hline
  \end{tabular}
  \caption{Main parameters of the detector components defined in the geometry used for MC simulation.}
  \label{tab:parameters_geometry}
\end{table}

\subsection{Raw background contributions}
\label{sec:rawBackground}
The radioactive isotopes, especially the descendants from the \utte and \thttt, inside the laboratory and detectors will produce high energy gamma rays. The energy deposited by these gamma rays fall inside the Region Of Interest (ROI) for NLDBD.
The raw background contribution of an isotope is defined as the expected energy within the ROI. The study of the raw background contributions provide a straightforward way to identify the most critical parts of the detector setup. The long-lived radioactive isotopes of \utte and \thttt and their descendants are assumed to be in secular equilibrium. The full decay chain is included in our simulation. The input activities considered for different materials are given in Table~\ref{tab:activities}.

\begin{table}[tbp]
  \centering
  \begin{tabular}{lccc}
    \hline
     \hline
    \multirow{2}{*}{\textbf{Material}} & \multicolumn{3}{c}{\textbf{Activity ($\mu$Bq/kg)}}
    \\      
                                 & $^{232}$Th & $^{238}$U  & $^{60}$Co \\ \hline
    Copper                       & 0.2        &   0.75     &     100     \\
    PTFE                         & 0.1        &   4.94      &    -      \\
    Stainless Steel              & 0.32$\times$10$^3$          &    0.5$\times$10$^3$      &     2.6$\times$10$^3$     \\
    UPW                          & 0.04          &     0.12      &     -     \\
    Concrete                     & 9.9$\times$10$^6$          &    4.4$\times$10$^6$   &    -    \\
      \hline
      \hline
  \end{tabular}
  \caption{Input activities of isotopes considered in the simulation for different materials. Copper and PTFE activities are extracted from Ref.~\cite{Abgrall:2016cct}, stainless steel from Ref.~\cite{LZ_CDR}. Activities in ultra pure water (UPW) are also given in Table~\ref{tab:icpms}. The activities in concrete measured at CJPL-I are used~\cite{Zeng2014}. }
  \label{tab:activities}
\end{table}

The following list describes the main background contributions considered in the simulation.

\begin{description}

\item[Copper Vessel] The high pressure vessel is the heaviest component close to the detection volume. Therefore, a non-negligible contribution is expected even if the highest radiopure copper is available.

\item[Electronics] Low radioactive electronics is a challenge for the PandaX-III experiment. Some electronics components may not be made radiopure. The end-caps of the vessel are thick enough to reduce the impact of electronics to the detector background level. The total activity is 0.26 Bq for \utte and 0.07 Bq for \thttt, respectively, by using the measured value of electronics in previous PandaX experiments~\cite{PCB-PandaX-II}.

\item[Field Cage] The field cage is made of a 5\,cm thick resistive PTFE wall, with equally separated copper rings interconnected by resistors. The weight of the field cage, dominated by the isolating material, is the second heavier element related to the construction of the TPC. Therefore, the choice of ultra pure materials to be used on the field cage construction is a major concern.

\item[Steel Bolts] The stainless steel bolts are used to connect the barrel and the end-caps together. There are 48 bolts connecting each end-cap. Because of the relative higher radioactive contamination in steel, bolts are also important sources of background. 

\item[Water Shielding] We have simulated the background contribution coming from $^{238}$U and $^{232}$Th isotopes diluted in the water shielding. Due to self-shielding effects we launch decays only from a reduced region of water, of at least 125\,m$^3$.

\item[Laboratory Walls] The walls of the laboratory are made of concrete, which would be the most important contribution in the absence of a shielding. We must assure that the size of the water shielding will be large enough to reduce its contribution to negligible levels. The volume of water pool in our simulation is smaller than the real design. In our geometry model the minimum distance from the vessel to the surface of water is larger than 5\,m, which will be ensured in the real experiment. We have simulated the contribution from the concrete (or laboratory walls) in three steps.

\begin{itemize}
\item \emph{First}, we have simulated the gamma flux radiating from the concrete walls. We register in this step the energy and angular gamma flux, produced by $^{238}$U and $^{232}$Th full decay chains, through the walls of a large block of concrete (see Figure~\ref{fig:sim_lab_gamma_flux}, left).

\item \emph{Second}, we use the gamma flux produced in the previous step to simulate the gammas transferred through the water using a biasing technique. The basic idea of the biasing technique is to divide the water into many layers. We iterate to obtain the flux passing through the layers from outside to inside. Because our desired events should have an energy around 2457.83~keV, only the gammas with energy larger than 2.2\,MeV are counted. The original events are generated from the energy and angular distribution calculated in the previous step. Then, the energy and angular distribution of flux passing through the first layer is recorded for the second iteration. For the $i$-th iteration, the number of simulated events is $N_{i}$ and the number of gamma rays recorded is $n_{i}$. We define an amplification factor $a_{i}$ used to scale the number of simulated events in the next iteration, or, $N_{i+1} = n_i \cdot a_i$. In the last layer (the closest to the detector) the energy and angular spectra of gammas are registered (see Figure\,\ref{fig:sim_lab_gamma_flux}, rigth) and used in the following step.

The equivalent number of simulated events, $N_{eq}$, is given by
\begin{equation}
  \label{eq:n_equiv_biasing}
  N_{eq} = N_0 \prod_{i=1}^{t} a_i,
\end{equation}
\noindent where $N_0$ is the number of initial simulated events, and $t$ is the number of iterations.

\item \emph{Third}, we launch a final simulation from the last biasing layer (sized 2.5$\times$2.5$\times$2.5\,m$^3$) taking into account the energy and angular distributions obtainned from the previous step. We register at this step the energy deposits in the gas volume.
\end{itemize}
    
\begin{figure}[tb]
  \centering
  \includegraphics[width=0.49\textwidth]{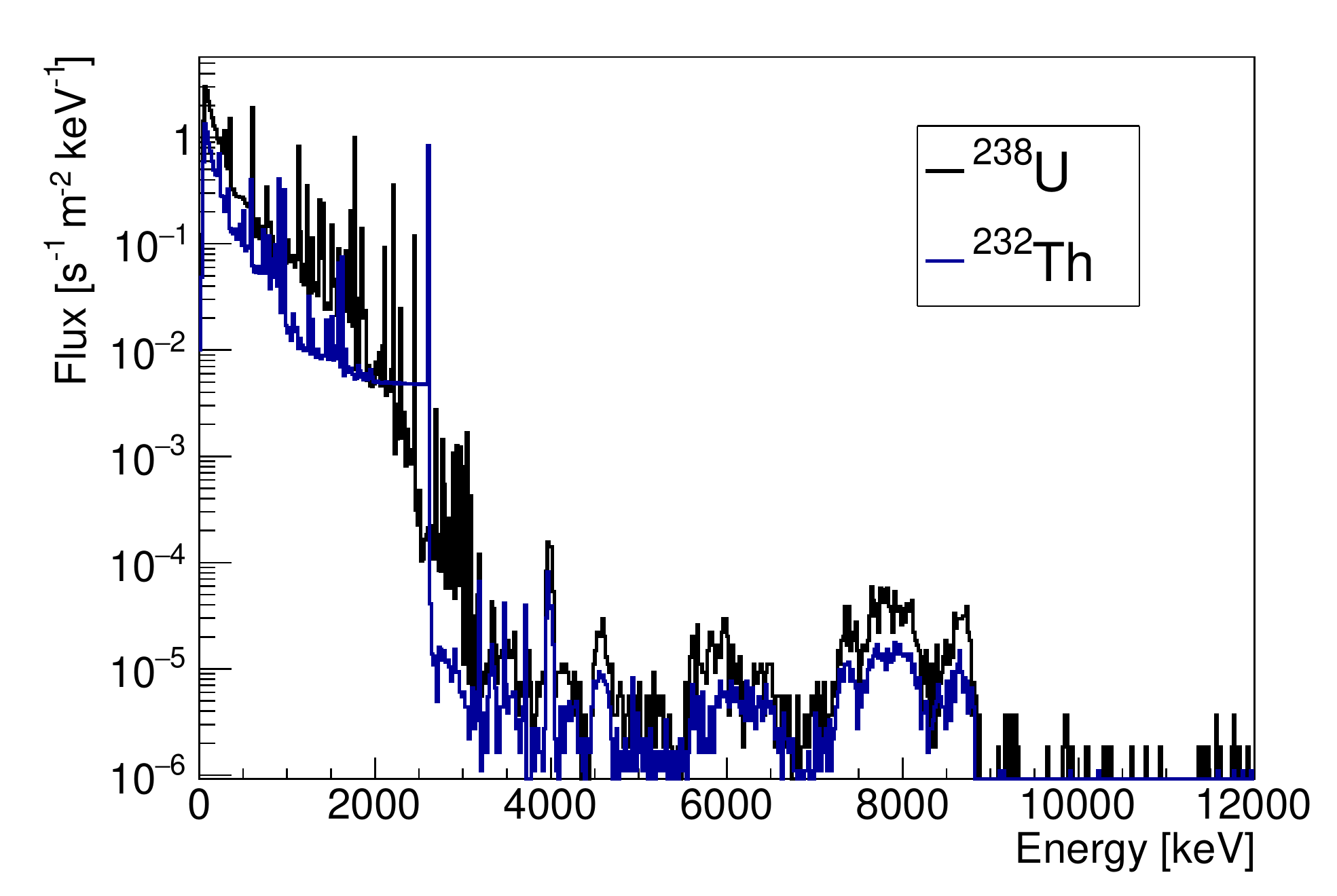}
  \includegraphics[width=0.475\textwidth]{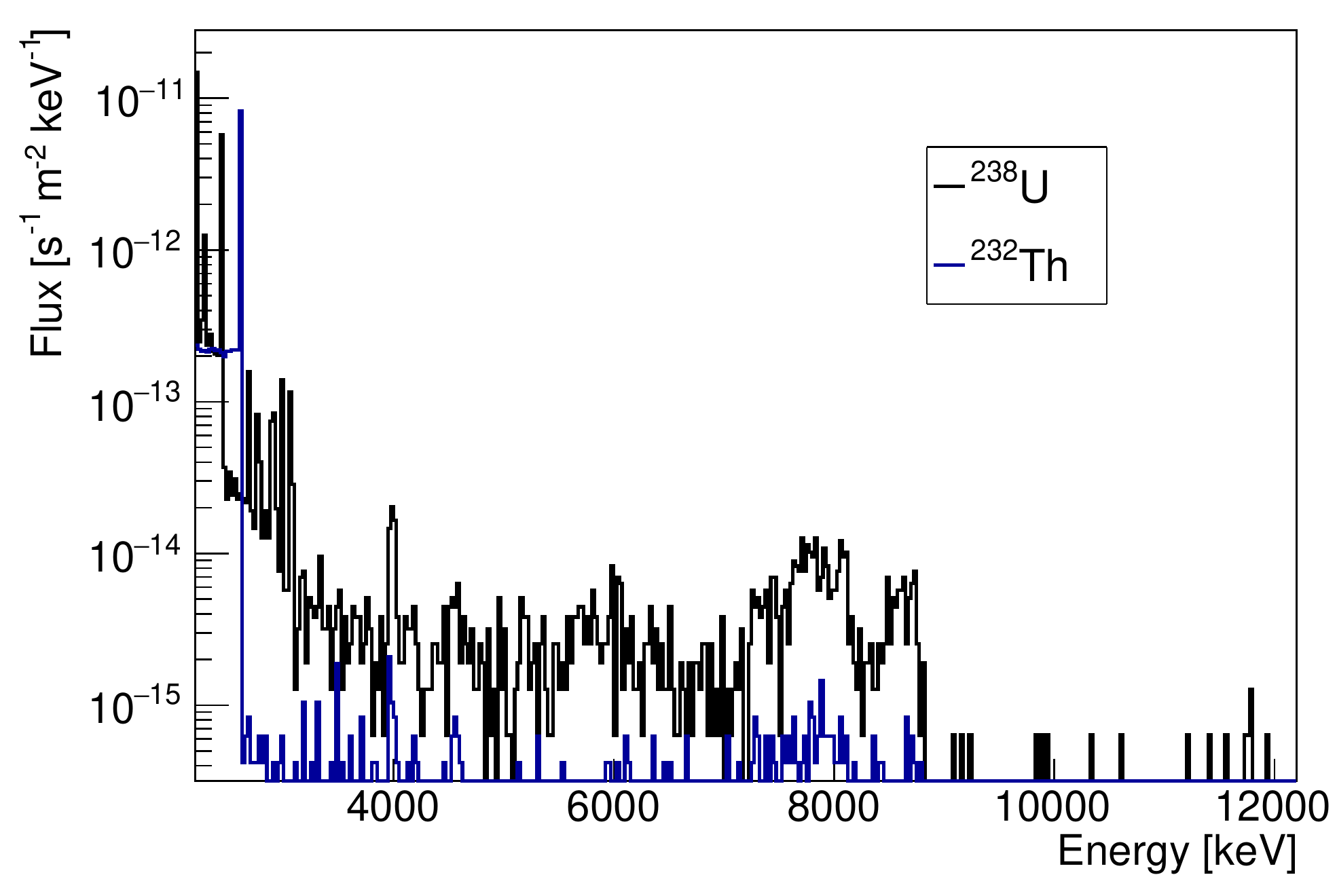}
  \caption{On the left, the simulated gamma energy flux from the \utte and \thttt decay chains within the concrete wall of the CJPL-I laboratory. The total flux rates for \utte and \thttt are $2.23\times10^3$\,Hz/m$^2$ and $6.64\times10^2$\,Hz/m$^2$, respectively. On the right, the resulting gamma energy flux at a box of size 2.5$\times$2.5$\times$2.5\,m$^3$ surrounding the detector, after the gamma transfer simulation using the biasing method described. Only gammas above 2.2\,MeV are considered in the gamma transfer. }
  \label{fig:sim_lab_gamma_flux}
\end{figure}

\item[Readout Planes] The readout of the first TPC module will be constructed with Micromegas technology. The Microbulk type Micromegas are made of typically low radioactivity materials, as copper and Kapton. A low mass budget (<1 kg) is required to construct a readout covering a large detection area. Recent material radiopurity assay measurements show that the $^{238}$U and the $^{232}$Th levels are below 45\,nBq/cm$^2$ and 14\,nBq/cm$^2$, respectively~\cite{MM-BiPo3}.

\item[Cathode] The cathode is made of radiopure copper. The background from the material is negligible due to its small mass. But $^{222}$Rn contamination in xenon gas will produce $^{214}$Bi, which may be driven by the electric field and attached to the cathode. It is difficult to estimate beforehand the final radon contamination we will have in the gas volume. We will assume a reasonably high activity of 1\,mBq/m$^{3}$, leading to a $^{214}$Bi activities of about 2~nBq/cm$^2$ at the surface of the cathode.

\end{description}

Two independent Geant4~\cite{Agostinelli:2002hh} based programs, \emph{BambooMC}\footnote{BambooMC is a modularized Geant4 based simulation program, developed by the SJTU group. It has been used extensively in the PandaX-II dark matter project for the prediction of backgrounds.} and \emph{RestG4} (a package integrated in REST~\cite{Tomas:1557181}), have been used to produce simulation data. Consistent results were obtained previously with the two codes on a set of benchmark results. Independent geometry implementations, based on the geometry model described, have been used by each package.

We have defined the ROI around $Q_{\beta\beta}$ considering the energy resolution, $\sigma$, of our detector. We provide our results in the energy range ($Q_{\beta\beta}$-2$\sigma$,$Q_{\beta\beta}$+2$\sigma$). The initial Monte Carlo energy depositions of each event have been smeared using a Gaussian convolution defined by the energy resolution of our detector. Therefore, we are taking into account energy deposits originally falling outside the ROI, but that would contribute to it due to the effect the energy resolution of the detector. 

\begin{figure}[h]
  \centering
  \includegraphics[width=0.49\textwidth]{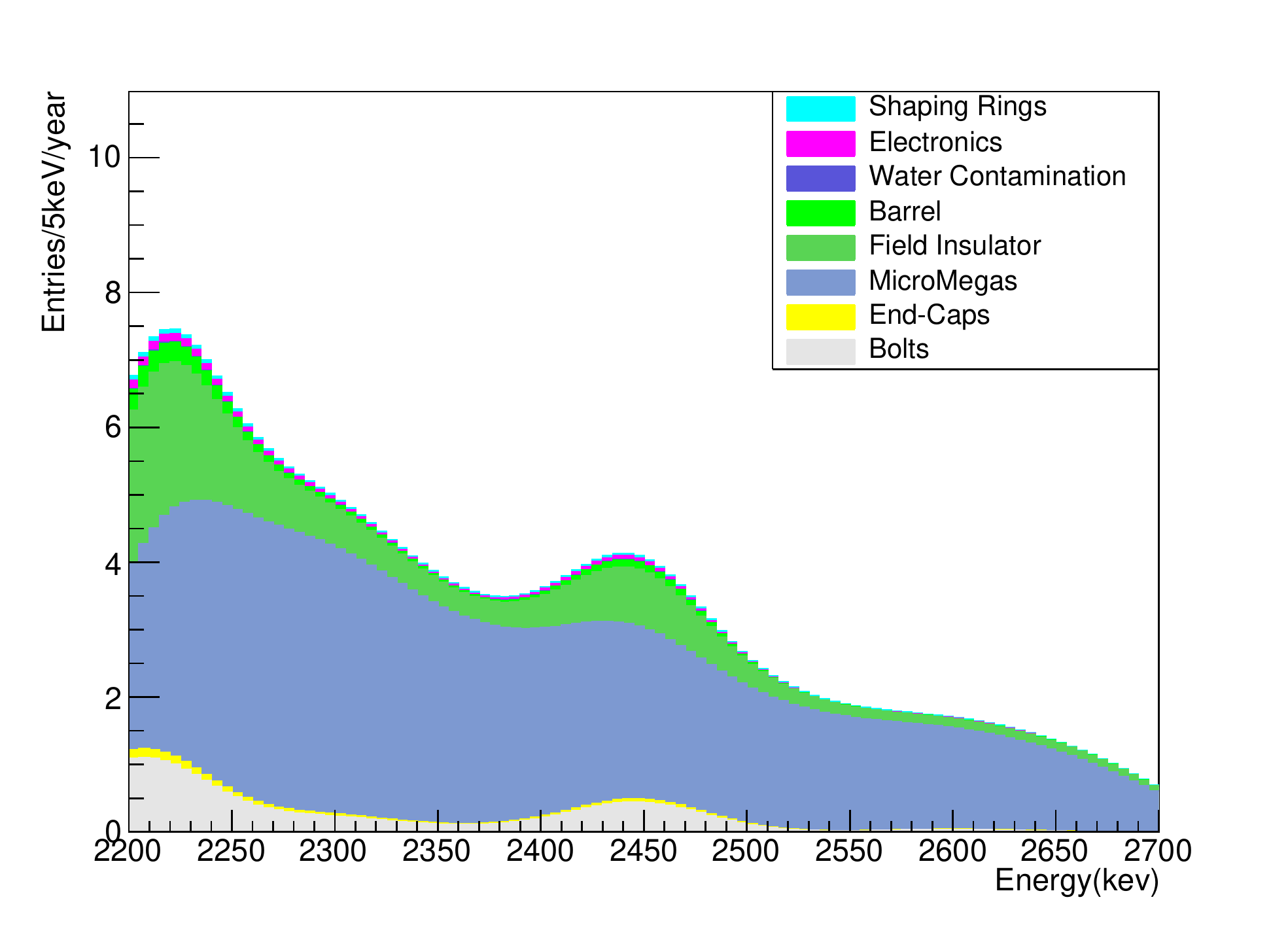}
  \includegraphics[width=0.49\textwidth]{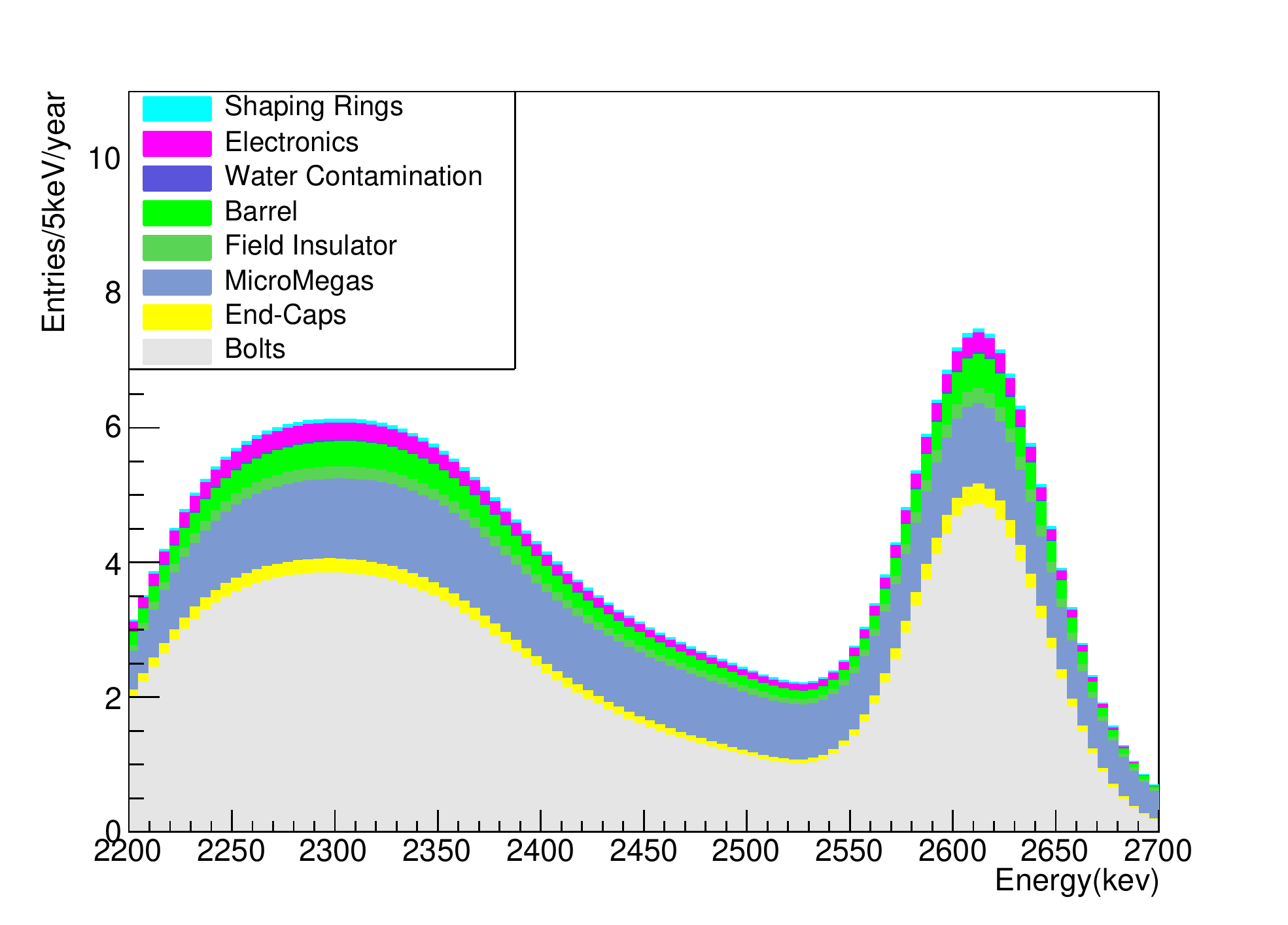}
  \caption{The raw background spectrum (stacked) from different parts of the detector and the water shielding system when the energy resolution of the detector is of $3\%$ FWHM at $Q_{\beta\beta}$. The ROI is from 2395 keV to 2520 keV. Left: backgrounds from \utte; Right: backgrounds from \thttt.}
  \label{fig:raw_bkg_summary}
\end{figure}

The raw background contributions are summarized in Table~\ref{tab:rawBck}. Fig.~\ref{fig:raw_bkg_summary} shows the background spectrum of each of the components in the detector for a resolution of 3\% FWHM at $Q_{\beta\beta}$.

\begin{table}
  \centering
  \begin{tabular}{ccccccc}
    \hline
    \hline
    \multirow{2}{4em}{} & \multirow{2}{4em}{\textbf{Isotope}} & \multirow{2}{4em}{\textbf{Activity}} & \multicolumn{2}{c}{\textbf{Background (cpy)}} &  \multicolumn{2}{c}{\textbf{BI (10$^{-5}$\ckky)}} \\ 
                                 &              &     & BambooMC & RestG4  & BambooMC & RestG4 \\ \hline
    \multirow{2}{4em}{Laboratory walls}   & $^{238}$U  &  9.9 Bq/kg & $<0.40\pm0.03$ & $<0.09 \pm$ 0.01 & - & $<$0.4  \\ 
                                           & $^{232}$Th &  4.4 Bq/kg &  $<0.22\pm0.02$  & $<0.15 \pm$ 0.01   & - & $<$0.6 \\ \hline

    \multirow{2}{4em}{Water} & $^{238}$U  & 0.12 $\mu$Bq/kg & 0.20 $\pm$ 0.1 & 0.22 $\pm$ 0.03 & 0.74 & 0.86 \\ 
                                           & $^{232}$Th & 0.04 $\mu$Bq/kg & 0.24  $\pm$ 0.06 & 0.55 $\pm$ 0.03 & 0.96 &  2.21 \\ \hline
    \multirow{3}{4em}{Barrel}              & $^{238}$U  &  0.75 $\mu$Bq/kg & 1.73  $\pm$ 0.12 & 1.77 $\pm$ 0.1 & 6.9 & 7.05 \\ 
                                           & $^{232}$Th & 0.2  $\mu$Bq/kg & 4.63  $\pm$ 0.18 & 4.55 $\pm$ 0.05 & 18.5 &  18.2 \\ 
                                           & $^{60}$Co  & 10 $\mu$Bq/kg & 9.8  $\pm$ 1.0 & 9.9 $\pm$ 0.9 &  39.0  & 39.7 \\ \hline

    \multirow{3}{4em}{End-caps}            & $^{238}$U  & 0.75 $\mu$Bq/kg  & 0.83  $\pm$ 0.11 & 0.90 $\pm$ 0.11 & 3.3 & 3.6 \\ 
                                           & $^{232}$Th & 0.2 $\mu$Bq/kg & 2.4  $\pm$ 0.1 & 2.2 $\pm$ 0.1 & 9.8 & 9.0 \\ 
                                           & $^{60}$Co  & 10 $\mu$Bq/kg & 4.4  $\pm$ 1.0 & 4.2 $\pm$ 0.9 & 17.8 & 16.7 \\ \hline

    \multirow{2}{4em}{Bolts}              & $^{238}$U   &  0.5 mBq/kg & 7.5 $\pm$ 1.5 & 7.3 $\pm$ 0.9 & 30.1 & 29.2 \\ 
                                          & $^{232}$Th  & 0.32 mBq/kg & 39.8 $\pm$ 2.7 & 46.7 $\pm$ 1.9 & 159 & 186.3 \\ \hline
    \multirow{2}{4em}{Field insulator}    & $^{238}$U   & 4.94 $\mu$Bq/kg  & 15.0  $\pm$ 0.5 & 15.7 $\pm$ 0.3 & 59.9 & 62.6 \\ 
                                          & $^{232}$Th  & 0.1 $\mu$Bq/kg & 2.69 $\pm$ 0.03 & 2.61 $\pm$ 0.1 & 10.7 & 10.4 \\ 
    \multirow{2}{4em}{and rings}          & $^{238}$U  & 0.75 $\mu$Bq/kg  &  0.67 $\pm$ 0.01  & 0.72$\pm$0.05 & 2.7 & 2.9 \\ 
                                          & $^{232}$Th  & 0.2 $\mu$Bq/kg & 0.95 $\pm$ 0.01 & 0.92 $\pm$ 0.03 & 3.8 & 3.7 \\ \hline
    \multirow{2}{4em}{Electronics}        & $^{238}$U  & 0.26 Bq & 1.0 $\pm$ 0.3 & 2.4 $\pm$ 0.5 & 4.2 & 9.5 \\ 
                                          & $^{232}$Th  & 0.07 Bq & 2.8 $\pm$ 0.2 & 4.1 $\pm$ 0.5 & 11.3 & 16.3 \\ \hline
    \multirow{2}{4em}{Micromegas}         & $^{238}$U  & 45 nBq/cm$^2$ & 60.5 $\pm$ 1.7 & 63.7 $\pm$ 1.8 & 241.6 & 254.4 \\ 
                                          & $^{232}$Th  & 14 nBq/cm$^2$ & 23.5 $\pm$ 0.6 & 25.3 $\pm$ 0.6 & 93.9 & 101  \\ \hline
    \multirow{1}{4em}{Cathode}            & $^{214}$Bi  & 2 nBq/cm$^2$ & 4.1  $\pm$ 0.2 & 3.3 $\pm$ 0.1 & 16.5 & 13.2 \\ 
    \hline
    \hline
  \end{tabular}
  \caption{The raw background contribution from the different parts in the laboratory and the detector by taking the $3\%$ FWHM detector resolution into account. BI stands for \emph{ Background Index}.}
  \label{tab:rawBck}
\end{table}

\subsection{The detector response}
\label{sec:fiducialization}

The TPC module is divided in two active volumes by a central cathode. The charge produced by interacting particles inside these active volumes is drifted and collected by two independent Micromegas planes placed on both ends of the TPC. The optimum drift field in terms of Micromegas transparency, drift velocity and electron diffusion is about 100\,V/cm/bar. For that field intensity the electron drift velocity in xenon + 1\%TMA is 1.87\,mm/$\mu$s\footnote{Drift velocity values and gas properties have been calculated using \emph{Magboltz}}. Therefore, the total drift time in each of the TPC sides is slightly above 500\,$\mu$s. 

The electronic acquisition cards are able to register up to 512 time bins per event, at a tunable sampling rate. If we would like to scan the event signature in the whole TPC volume we would require to use a sampling rate of about 1\,MHz. However, we are interested in having the highest sampling rate possible in order to have a better definition of the NLDBD decay event for two main reasons. \emph{First}, a low sampling rate will have a negative effect on the energy resolution of the event. \emph{Second}, the better time resolution of the event will enhance the track image definition, hence improving the topological event recognition potential.

The length of electron tracks for NLDBD events are of the order of 10\,cm. Although, few times those electrons will loss energy emitting low energy gammas by bremsstrahlung. We must assure that most of those gammas are found within the acquisition window that we will define in our setup. We have studied the response of our detector to NLDBD events from $^{136}$Xe. We use the DECAY0 package to generate electrons from the double beta decay process~\cite{Ponkratenko:2000um}, and simulate the resulting electron tracks with the packages described in section~\ref{sec:rawBackground}.

\begin{figure}[tb]
  \centering
  \includegraphics[width=\textwidth]{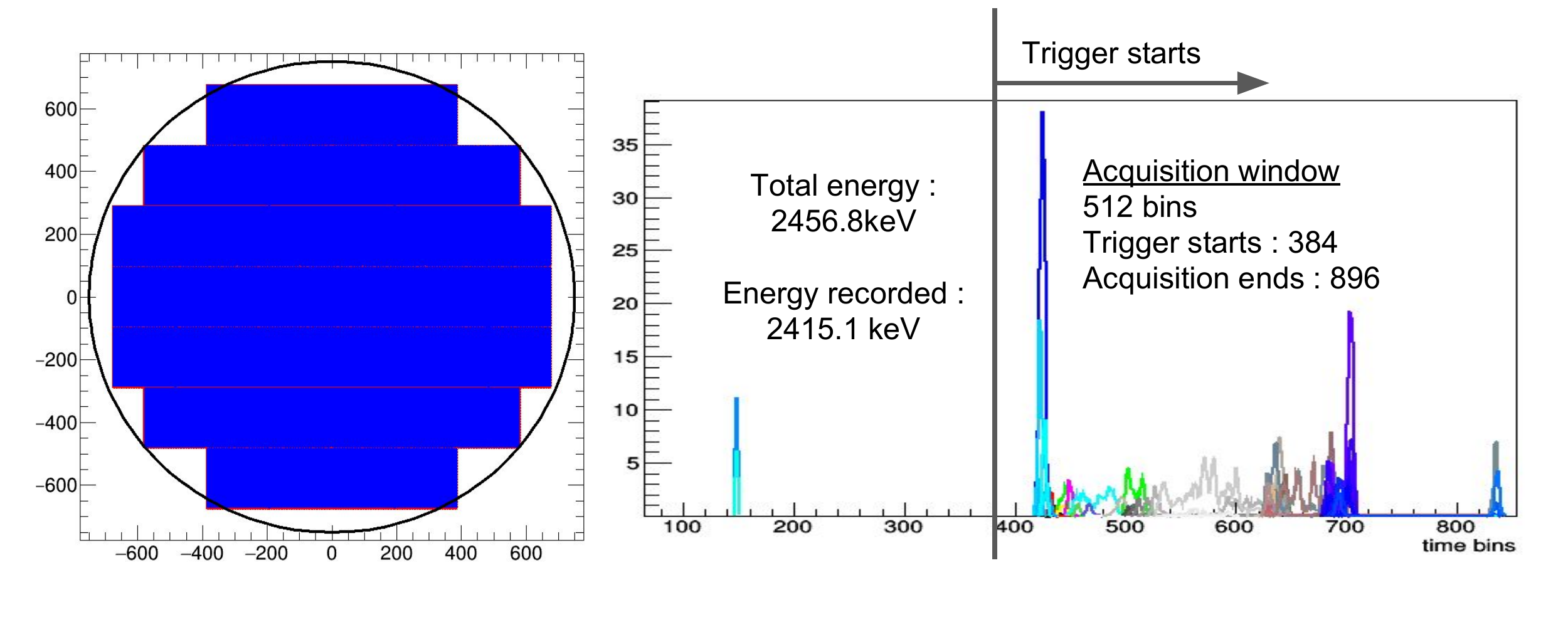}
  \caption{ (Left) The readout plane definition composed by 41 Micromegas modules of 20$\times$20\,cm$^2$. (Right) The signal response corresponding to our readout definition. Each color corresponds to different readout channels. The definition of the acquisition window for this particular NLDBD event produces a small energy loss. }
  \label{fig:response}
\end{figure}

We define the time acquisition window using a trigger time bin, $t_o$, such that the integral of our readout signal, $S(t)$, between $t_o$ and $t_o+256$ contains half of the energy of a NLDBD event,  $Q_{\beta\beta}/2$, being $t_o$ the lowest value that satisfies this condition for each event. The acquisition window is defined between $t_o$ and $t_o+512$ time bins. We have simulated the electron diffusion using the gas properties, generating readout signals equivalent to those that would be obtained with the real readout channel segmentation. We have studied different sampling rates, being 10\,MHz the highest sampling rate allowing to keep a reasonable signal efficiency. We have reproduced the readout shown in Figure~\ref{fig:detector} to fiducialize our events. Only energy deposits falling inside the detector module boundaries are considered, with the consequent loss of efficiency due to non active detection regions. Figure~\ref{fig:response} shows the description of one the readout planes used in our detector response simulation, together with a NLDBD event with produced charges distributed along the time and at different readout channels, the charge spread in this particular case is only due to the simulation of the electron diffusion, effects from shaping time or electronic noise are not considered here.

We have obtained the efficiency of detection of NLDBD events at different steps. We start by determining the probability of having a full NLDBD event completely contained in the gas, then we consider the effect of the energy resolution of the detector, the time acquisition window and the active readout area of the detector. Table~\ref{tab:signalResponse} summarizes the progressive loss of efficiency due to event acquisition.

\begin{table}[tb]
  \centering
  \begin{tabular}{cccc}
  \hline
  \hline
      \textbf{Full event} & \textbf{Smearing (3\% FWHM)} & \textbf{Time window} & \textbf{Active area} \\ \hline
      74.2\%     & 70.8\%          & 64.9\%      & 54.2\%      \\
  \hline
  \hline
  \end{tabular}
  \caption{NLDBD signal detection efficiency after different hardware cuts. \emph{Full event} only considers that all the energy of the event has been deposited in the gas volume. \emph{Smear} considers the energy resolution of the detector. \emph{Time window} adds the hardware cut imposed by the electronics acquisition window. And \emph{Active area} is defined by the readout module area coverage.}
  \label{tab:signalResponse}
\end{table}

This loss on signal efficiency due to the hardware cuts applied to the event readout will be also affecting to the background contributions studied in section~\ref{sec:rawBackground}. We have applied the same event processing to the simulated background data. We provide results for the most important contributions at Table~\ref{tab:results_trigger}.

\begin{table}[tb]
  \centering
  \begin{tabular}{lccccc}
    \hline
  \hline
      \multirow{2}{*}{\textbf{Component}} & & \multirow{2}{2em}{\textbf{Isotope}} & & \multicolumn{2}{c}{\textbf{Background (10$^{-5}$\,\ckky)}}\\
                                 &  &                       &   &   BambooMC & RestG4  \\\hline
      \multirow{2}{*}{Water}  & & \utte     & & -   & 0.23 \\ 
                              & & \thttt    & & 0.56  & 0.63 \\\hline 
      \multirow{3}{*}{Barrel} & & \utte     & & 1.07  & 2.41 \\ 
                              & & \thttt    & & 7.54  & 7.86 \\ 
                              & & \cose     & & 3.02  & 2.11 \\ \hline
      \multirow{3}{*}{End-caps } & & \utte  &   & 0.30 & 1.26    \\ 
                                & & \thttt  &  & 3.89 & 4.16    \\ 
                                & & \cose    & & 2.98 & 0.76    \\ \hline
      \multirow{2}{*}{Bolts}  & & \utte     & & 3.50   & 11.9 \\ 
                              & & \thttt    & & 73.8  & 78.5 \\\hline
      \multirow{2}{*}{Field insulator}  & & \utte     & & 19.5   & 16.5 \\ 
                              & & \thttt    & & 3.80  & 3.86 \\
      \multirow{2}{*}{and rings}  & & \utte     & & 1.52   & 0.45 \\ 
                              & & \thttt    & & 1.41  & 1.17 \\\hline 
      \multirow{2}{*}{Electronics} & & \utte     & & -  & 1.42  \\ 
                              & & \thttt    & & 5.02  & 8.69 \\ \hline
      \multirow{2}{*}{Micromegas} & & \utte     & & 144  & 158  \\ 
                              & & \thttt    & & 36.9  & 44.5 \\ \hline \hline
      \multirow{1}{*}{Total}  & &      & & 308.8   & 344.4 \\ 
        \hline
  \hline
  \end{tabular}
  \caption{ Summary of the most relevant background contributions taking into account the detector response.}
  \label{tab:results_trigger}
\end{table}

\subsection{Topological signal of NLDBD events in high pressure gas}
\label{sec:topology}

A unique feature of a HPXe gas TPC for NLDBD searches is the possibility to access the topological information of the electron ionization tracks in the gas, which can be used to discriminate (with some efficiency) background from signal events and thus further reduce the effective background of the experiment.

\begin{figure}[tb]
  \centering
  \includegraphics[width=\textwidth]{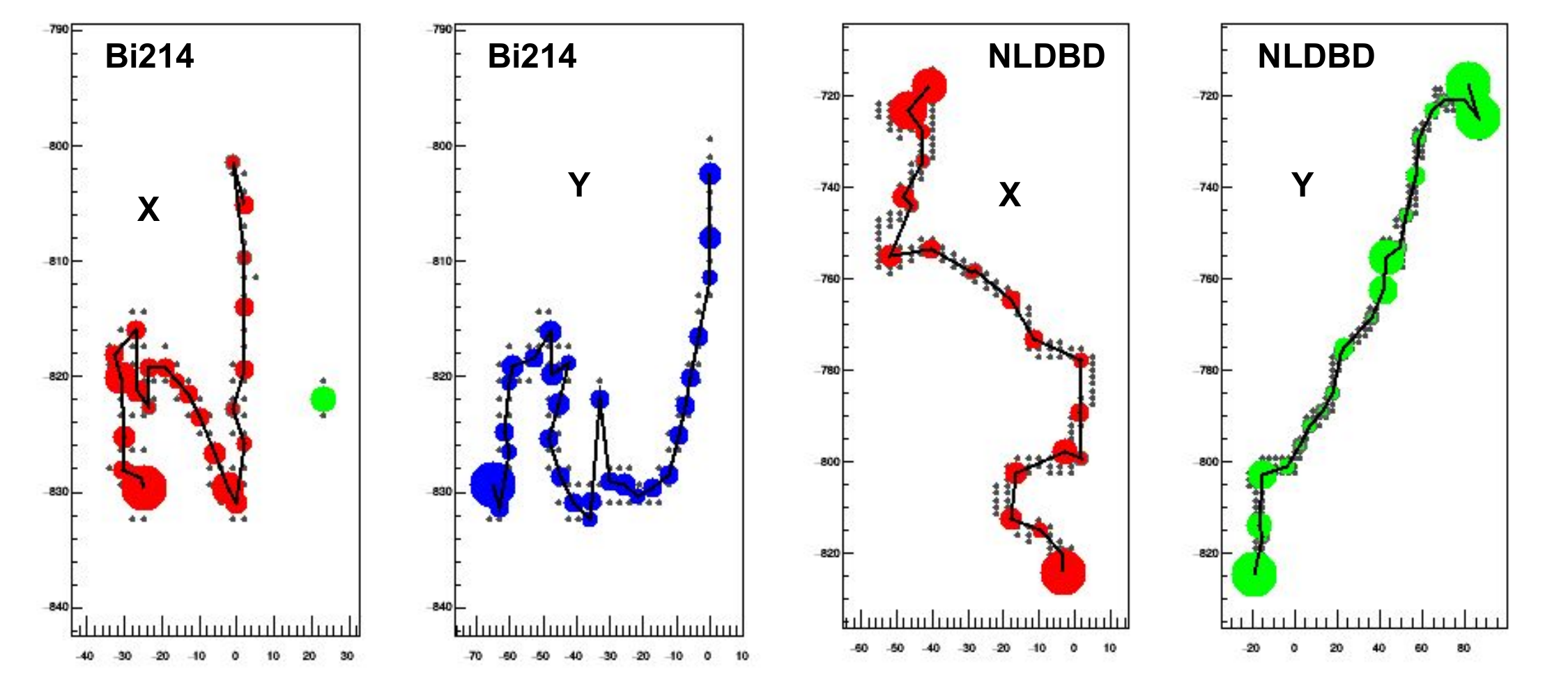}
  \caption{Reconstruction of the projected electron tracks (XZ and YZ) produced by two Monte Carlo simulated events, a typical background event coming from a $^{214}$Bi decay (Left), and a NLDBD event (Right). The reconstruction has gone through different processing stages; Geant4 deposit production, electron diffusion, readout charge collection, time signal generation and re-construction of tracks with the same algorithms that we will use for reconstruction from real data. Track path minimization algorithms implemented in the REST framework are used to find the most likely trajectory of the reconstructed hits. These events have been processed using a Micromegas readout module 3mm-pitch, using a 1~$\mu$s sampling rate acquisition, and applying the diffusion properties of Xe+1\%TMA at 1~kV/cm. A hit reduction algorithm has been applied resulting in a minimalist reproduction of the track (coloured circles). The size of the circles is related to the charge density allowing to identify visually the blobs at the end of the tracks. }
  \label{fig:tracks_REST}
\end{figure}

The typical topology of NLDBD events in 10\,bar Xe gas consists of a straggling ionization track of around 15\,cm finishing, at both ends of the track, in higher energy depositions (or blobs) due to the Bragg peak of the electrons. This is illustrated in Fig.~\ref{fig:tracks_REST}. On the other hand, the typical background events in the energy window around $Q_{\beta\beta}$ come from energetic photons from environmental natural radioactive contaminations (mainly 2614.5\,keV photons from the decay of $^{208}$Tl, and 2447.8\,keV photons from the $^{214}$Bi decay). These photons produce (via Compton or photoelectric interactions) one or more single-electron ionization tracks, and their prototype topology will be that of a ionization track with only one higher energy blob at one of the track's ends.

These two conceptual topologies offer a clear distinction between signal and background events and are the basis for the design of discrimination criteria in gaseous xenon experiments. The pioneer work in this respect was carried out by the Gothard group in the 90's~\cite{Wong:1993uq}, introducing some key ideas like the presence/absence of the two blobs at the end of the tracks. They successfully demonstrated the application of these concepts to real data (in a 5.3 kg gas Xe TPC), although the discrimination was done through visual inspection. More recently the group at the University of Zaragoza, as part of the aforementioned T-REX project, has superseded these results~\cite{Irastorza:2015dcb}, introducing new ideas for reconstruction and analysis, performing full simulation and analysis of the expected signals and background in a high pressure Xenon TPC, and defining the basis of automated algorithms for recognition and discrimination. The status of these studies is briefly described here followed by the needed next steps.

The referred work proved that a rejection factor of about 3$\times$10$^3$ on the events falling in the energy region of interest around $Q_{\beta\beta}$ is feasible, while keeping a signal efficiency of 40\%. For comparison, this number corresponds to more than one order of magnitude better rejection factor than experiments counting with only single site / multiple site discrimination (like liquid Xe). The effect of a low diffusion gas versus a high diffusion gas (like pure Xe) in the topological quality was quantified to be a minimum of a factor of 3. 

Combined with state-of-the-art radiopurity specifications, these numbers already prove that HPXe TPCs like the one here proposing have very promising sensitivity prospects to the NLDBD signal. Moreover, all these numbers must be considered conservative, (and indeed there is evidence -unpublished- for better discrimination power) due to the limited scope of the study in a number of points:

\begin{itemize}
    \item First, this study was performed with a relatively coarse pixelization of 1\,cm. This was the case because it was originally targeted to the case of operation in a high diffusion gas like pure Xe, and certainly was not adequate for a low diffusion mixture like Xe-TMA as argued before. Even with this limitation, the effect of low diffusion still reflects on an improvement of a factor of 3 in the one blob/two blob identification. Presumably better factors are achievable with better pattern granularity.

    \item Although the discrimination criteria have been automated and quantified, a number of parameter optimizations remained for future work.

    \item A number of features of surviving events were identified that can be used as additional discrimination criteria but not actually used in the algorithm, and were left as the basis for future improvements.
\end{itemize}

Therefore this result must be viewed as preliminary and certainly can be improved by future studies. Work is ongoing to explore these lines for improvement. In addition, this result supports the design choice made in section~\ref{sec:TPC} to use thin strips pattern instead of fully independently read pads in order to reduce number of channels and simplify design. Going to a strip-pattern must penalize somehow the topological discrimination ability (due to the fact that now we are dealing with two "2D-projected" images XZ and YZ of the event (see Fig.~\ref{fig:tracks_REST}), instead of the real 3D topology). To quantify this penalty is one of the most urgent simulation tasks in the short term. However, we anticipate that this penalty will be small and probably over-compensated by the improvement of topological quality given by the better pattern granularities (3 mm strip pitch, versus the 1 cm pitch of the pixel readout simulated in study of~\cite{Iguaz:2016hlu}).

%% file: Sensitivity.tex
\section{Sensitivity Projection}

\begin{figure}[tb]
\centering
\includegraphics[width=0.7\textwidth]{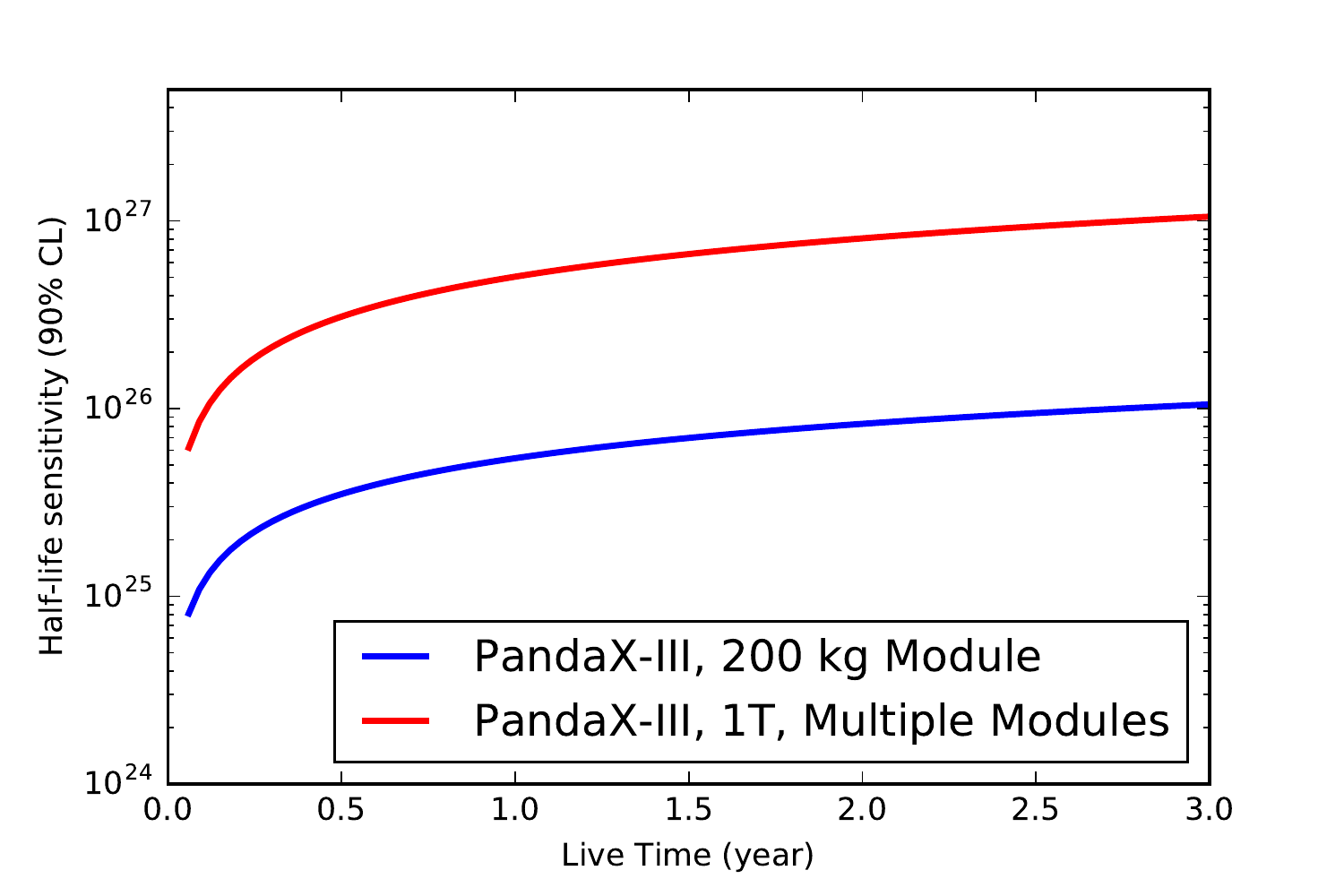}
\caption{Projected PandaX-III sensitivity as a function of live time for the first 200 kg module and for the future ton-scale experiment. Parameters used in the calculation are described in the text.}
\label{fig:sensitivity}
\end{figure}

\begin{figure}[tb]
\centering
\includegraphics[width=0.85\textwidth]{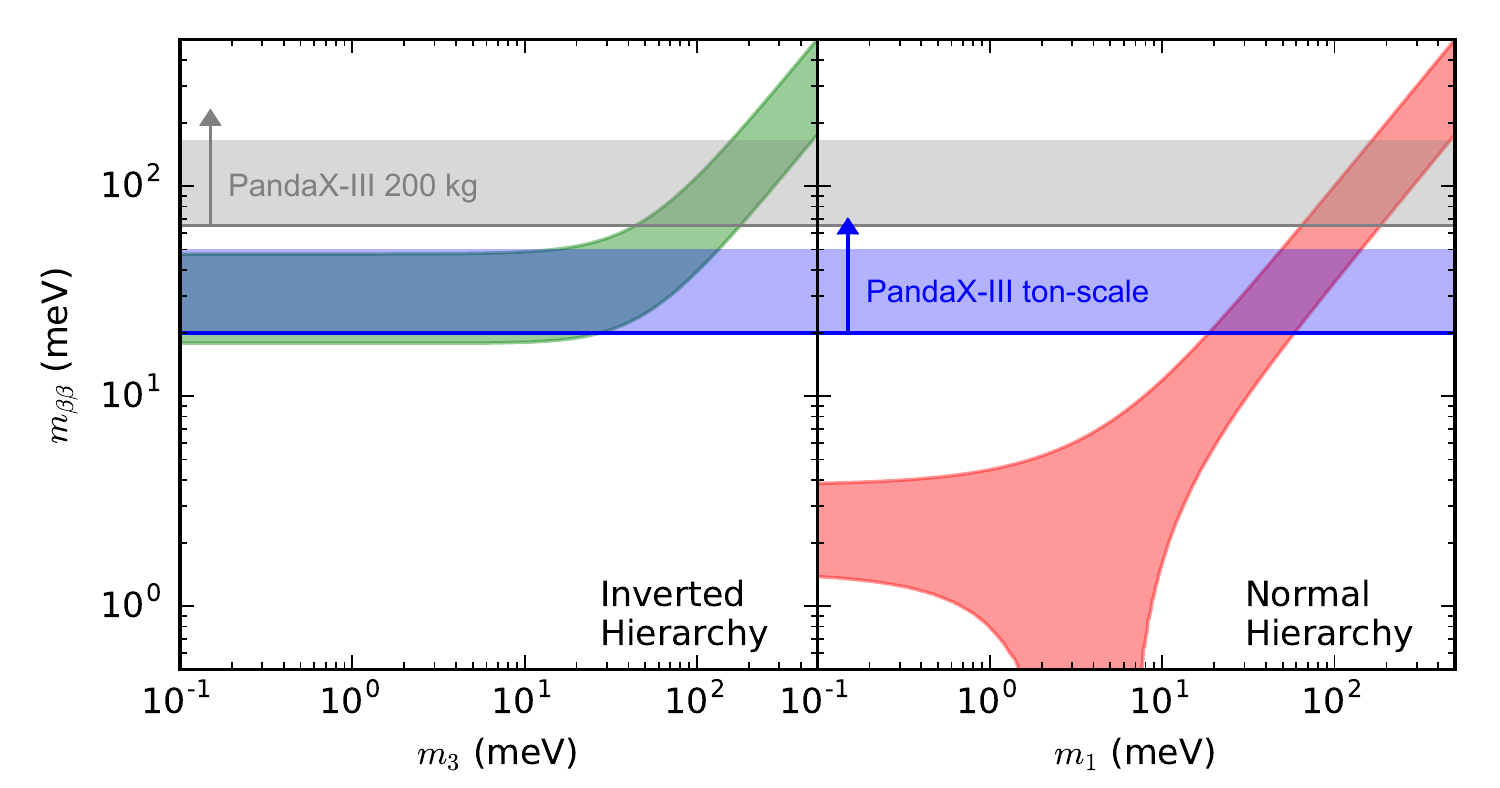}
\caption{Connectivity plot between the effective Majorana mass \mbb and the lightest neutrino mass, $m_{lightest}$ for the two mass orderings. 
The best-fit values of the oscillation parameters are used to generate the allowed regions. 
The horizontal gray and blue bands are the projected sensitivities for the PandaX-III first module and full ton-scale experiment respectively.
The live time is assumed to be 3 years and all other parameters are as described in the text.
The spread of the sensitivity bands are due to spread of NME calculations.
}
\label{fig:connectivity_plot}
\end{figure}

For NLDBD experiments with non-negligible background events, a commonly used figure-of-merit sensitivity formula is 
\begin{equation}
  T_{1/2}\propto  \eta \cdot \varepsilon \sqrt{\frac{M \cdot T}{r\cdot\delta E}}, 
\label{eq:sens-exp}
\end{equation}
where the following parameters contribute to final half-life sensitivity:
\begin{description}
\item[Detector efficiency $\eta$] determines the percentage of  NLDBD events are recorded  correctly by the detector. 
The efficiency includes detector containment efficiency, fiducialization, and event reconstruction efficiency. 
For PandaX-III gaseous TPC, the expected detector efficiency depends where we put the topological cuts. 
The overall efficiency is estimated at 35\%.
\item[Isotopic abundance $\varepsilon$] of the candidate isotope in the detector. 
PandaX-III uses enriched $^{136}$Xe and the nominal abundance is 90\%.
\item[Detector mass $M$] determines the total number of candidate isotope nucleus and thus the possible number of NLDBD events, given a certain half-life $T_{1/2}$. 
PandaX-III has 200 kg of enriched xenon in our first module.
Future ton-scale multiple-module experiment will have 1 T of enrich xenon gas. 
\item[Live time $T$] determines both the number of signal and background events.
The nominal live time for NLDBD experiments is between 1 to 3 years.
\item[Energy resolution $\delta E$] is often defined as the full width half maximum (FWHM) of the detector at the NLDBD Q-value. 

$\delta E$ is also the region of interest (ROI) in which we search for possible NLDBD events.
For PandaX-III first module, the expected ROI width $\delta E$=74 keV (3\% at the Q-Value).
Future modules with \TM (see next section) will reach a 1\% energy resolution.
\item[Background index $r$,] in the unit of counts per keV$\cdot$kg$\cdot$y, indicates how radioactively clean the detector is and how well it is shielded from environmental background. 
Background rate in PandaX-III first module is estimated at $10^{-4}$ \ckky, which is based on the simulation results from Tab.~\ref{tab:results_trigger} and a modest background suppression factor of 30 to 35 from topological analysis.
Future modules, with the help of better selection of detector parts and better topological analysis efficiency (gain from better energy resolution and pixel instead of strip readout), will push the background index by another order of magnitude. 
Often times the background is also quoted as counts per year (CPY) in the ROI for the whole detector.
The baseline background for PandaX-III first module is 1.5 CPY.
\end{description}

We can deduce the above correlation by  constructing a single-bin counting experiment and counting the expected number of signal and noise events according to detector performance~\cite{Alessandria:2011rc}.
The number of possible NLDBD events $S$ is proportional to overall exposure $M\cdot T$, while the number of possible background events $B$ is proportional to $M\cdot T \cdot r \cdot \delta E$.
For a large number of background, fluctuation of background events is Gaussian, i.e. $\sqrt{B}$, and thus the Equation~\ref{eq:sens-exp} by calculating $S/\sqrt{B}$.
For small number of background, Poissonian is preferred. 
Deviation from Equation~\ref{eq:sens-exp}  is about 10\% for number of background events of about 10.
In our calculation, we follow the procedure outlined in Ref.~\cite{Alessandria:2011rc} and treat background fluctuation correctly for both cases.
 
 The final sensitivity projections as a function of live time $T$ for PandaX-III first module (blue curve) and PandaX-III 1~ton setup (red curve) are shown in Fig.~\ref{fig:sensitivity}.
 All parameters used in the calculation are as listed previously.
For the first module, the half-life sensitivity with 1 (3) year(s) of live time is $5\times10^{25}$($1\times10^{26}$) year at 90\% confidence level. 
 Considering the spread of phase space factor~\cite{Kotila:2012zza, Stoica:2013lka} and nuclear matrix elements~\cite{Barea:2015kwa, Simkovic:2013qiy, Hyvarinen:2015bda,
Neacsu:2014bia, Menendez:2008jp, Rath:2013fma, Rodriguez:2010mn} of NLDBD, the corresponding effective Majorana mass \mbb ranges from 90 to 230 (65 to 165)~meV. 
For the future 1~ton setup, the half-life sensitivity with 3 years of live time reaches $1\times10^{27}$ year and the corresponding effective Majorana mass range is 20 to 50 meV.

Fig.~\ref{fig:connectivity_plot} compares the sensitivity range of \mbb with PandaX-III with bands preferred by oscillation data. 
The left panel shows the Invented Hierarchy (IH) case where the smallest neutrino mass is $m_1$ and the right panel the Normal Hierarchy (NH) case.
With a 200-kg TPC and 3 years of live time, our sensitivity will approach to the IH band. 
Next stage ton-scale experiment will cover a large portion of the phase space in the IH band.

%% file: FutureRnD.tex

\section{R\&D prospects towards future upgrade}

\Piii will pursue a tonne-scale NLDBD experiment using an array of high pressure Xenon gas
chambers (modules) each containing $\SI{200}{kg}$ of active \xeots isotope in compressed
gas form at a pressure up to about \SI{10}{bar}.  It demands a charge readout in each module that
is capable of scaling up to cover an area of $\sim{1}{m}$ in diameter, while maintaining
excellent energy resolution at double-beta decay \Qbb value and sufficient spatial resolution for
tracking purposes.  Microbulk Micromegas, the baseline readout
option of \Piii, that electronically reads out the amplified charge through controlled
gas-electron avalanche, while in principle could scale up easily and has a spatial resolution as
good as the chosen pixel pitch $d/\sqrt{12}$,  the energy resolution at \Qbb of \xeots is currently demonstrated only at 3\% level.
Improvements to the charge readout option is the primary direction of the R\&D for the future
upgrade. 

\subsection{TopMetal: Integrated Pixel Plane for Gain-less Charge Readout}
As a future upgrade of \Piii, we plan to realize a charge readout plane with a tiled array of
CMOS charge sensors, without gas-electron multiplication.  Charge collection electrodes,
front-end amplifiers, as well as data processing circuits, are integrated in the CMOS sensors
placed directly at the site of the charge measurement.  The plane will simultaneously achieve the
necessary low electronic noise for the highly demanding energy resolution for NLDBD and the high
spatial resolution for ionization charge tracking, while satisfying the stringent radiopurity and
scalability requirements.

A pixelated charge readout plane without gas-electron avalanche will eliminate the avalanche fluctuations, 
yet the charge sensing electronics must have exceedingly low internal noise to achieve sufficient
signal-to-noise ratio.  The front-end must be placed very close to the charge collection site and
the signal must be digitized as early as possible and transmitted digitally to minimize noise
pickup.  A coarse estimate of important parameters for such a plane to achieve a \SI{1}{\percent}
FWHM energy resolution is shown in Tab.~\ref{tab:param}.  All technical requirements point
towards a charge-collecting analog-digital mixed integrated circuitry.  The core technology
enabling such instrumentation is the recently developed CMOS pixel direct charge sensor, the
\TM\cite{TopmetalII-2016}.

\begin{table}[tb]
  \centering
  \begin{tabular}{ c  L{0.2\linewidth}   L{0.13\linewidth}  L{0.55\linewidth}}
    \hline    \hline
    & {\bfseries Parameters} & {\bfseries Initial value} & {\bfseries Considerations}\\
    \hline
    1 & ENC of a single node & $<\SI{30}{e^-}$
      & Highly dependent on the node capacitance $C$, therefore the node size $D$.\\
    \hline
    2 & $D$ -- node size (diameter) & $\sim\SI{1}{mm}$
      & Impacts the charge collection efficiency and node capacitance $C$.\\
    \hline
    3 & $P$ -- pitch size & $\sim\SI{5}{mm}$
      & $P$ determines the spatial resolution, however $D/P$ is limited by the focusing electric
        field strength in order not to start charge multiplication.\\
    \hline
    4 & $N$ -- number of nodes that see charge & $60$
      & Smaller $N$ is preferred, since the total electronic noise is ENC$\times\sqrt{N}$.  Gas additives that
        reduces diffusion will help.\\
    \hline    \hline
  \end{tabular}
    \caption{Design parameters and their considerations}
    \label{tab:param}
\end{table}

\begin{figure}[!tb]
  \centering
    \includegraphics[height=0.35\textwidth]{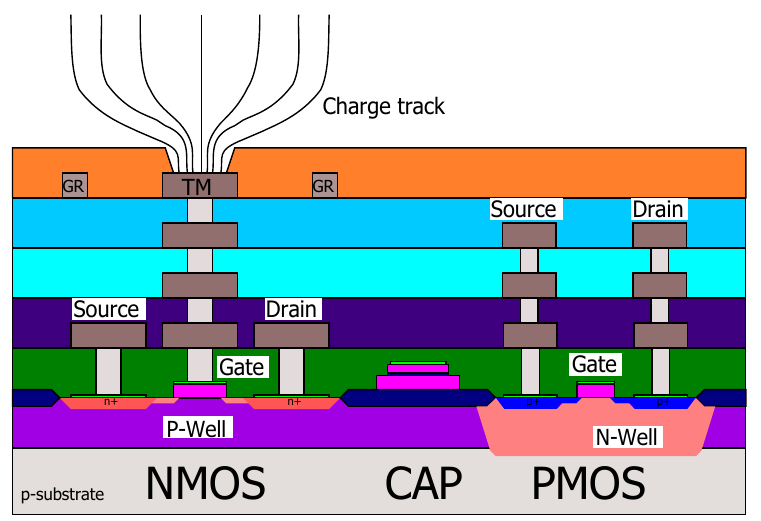}
    \hspace{1ex}
    \includegraphics[height=0.35\textwidth]{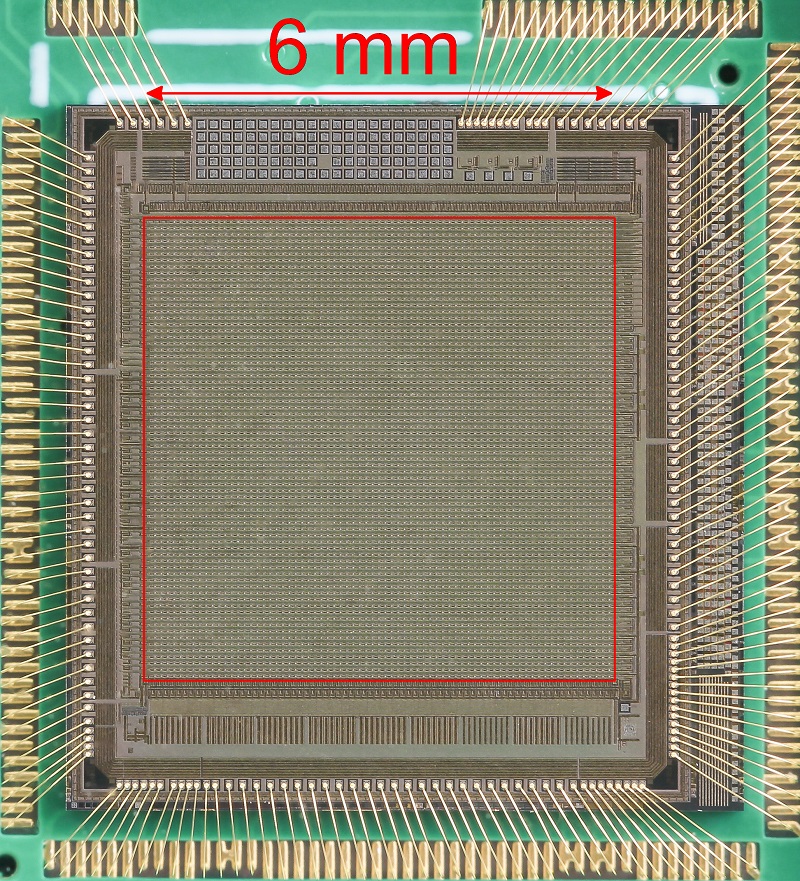}
   \caption{(Left) Cross-section view of a single \TM CMOS pixel for
    charge collection. (Right) Photograph of a \TMIIm sensor.}
  \label{fig:tm}
\end{figure}

\begin{figure}[!tb]
  \centering
      \includegraphics[height=0.33\textwidth]{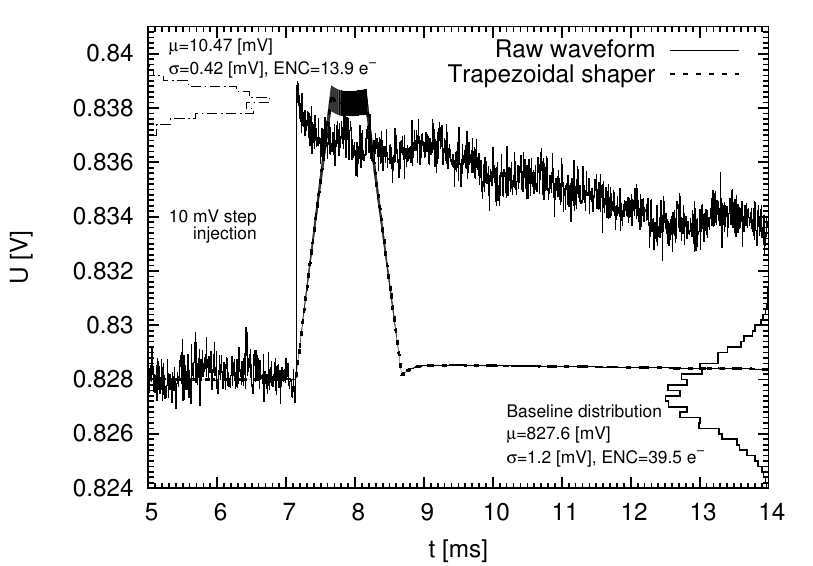}
   \includegraphics[height=0.33\textwidth]{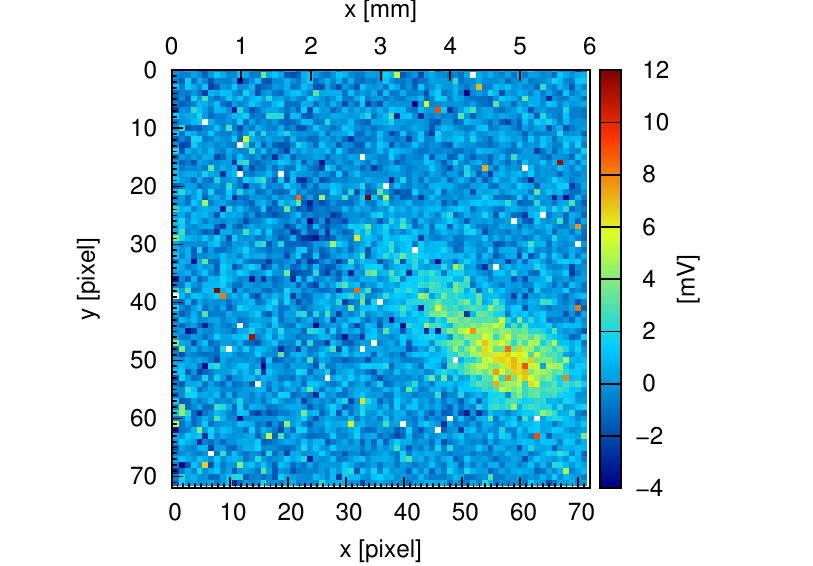}
  \caption{(Left) Noise characterization of a charge sensitive amplifier in a pixel. 
  (Right) A time slice of a single track of alpha producing ionization in air, imaged by the \TMIIm sensor. }
  \label{fig:tmNoiseSignal}
\end{figure}

\TM is a series of CMOS pixel direct charge sensors with a small metal patch placed on the top of
each pixel for direct charge collection (see Fig.~\ref{fig:tm}~(Left)).  \TM is manufactured using
industrial standard \SI{0.35}{\micro m} CMOS technology and requires no additional
post-processing.  To date (2016), two versions of \TM sensors (\emph{-I} and
\emph{-II\raise0.5ex\hbox{-}}), both with high density pixel array, have been produced and proven
to work successfully~\cite{TopmetalII-2016}.  Fig.~\ref{fig:tm}~(Right) shows a photograph of one fully
fabricated and wire-bonded \TMIIm sensor.  Performance tests show that \TMIIm achieved a
$<\SI{14}{e^-}$ noise per pixel, and its signal retention time is well into the many milliseconds
regime (Fig.~\ref{fig:tmNoiseSignal}~(Left)).  These characteristics allow the direct observation of ions
liberated by alpha particles ionizing air, then drifted slowly towards the sensor, without the gas
gain.  A time slice of such an alpha track is shown in Fig.~\ref{fig:tmNoiseSignal}~(Right).  It has also been
demonstrated elsewhere that much lower noise can be achieved with standard CMOS
technology~\cite{OConnor2002}.  Also, CMOS sensors have low mass and are produced with relatively
radiopure processes therefore are expected not to contribute significantly to the overall experimental backgrounds.


\begin{figure}[!htb]
  \centering
    \includegraphics[width=0.4\textwidth]{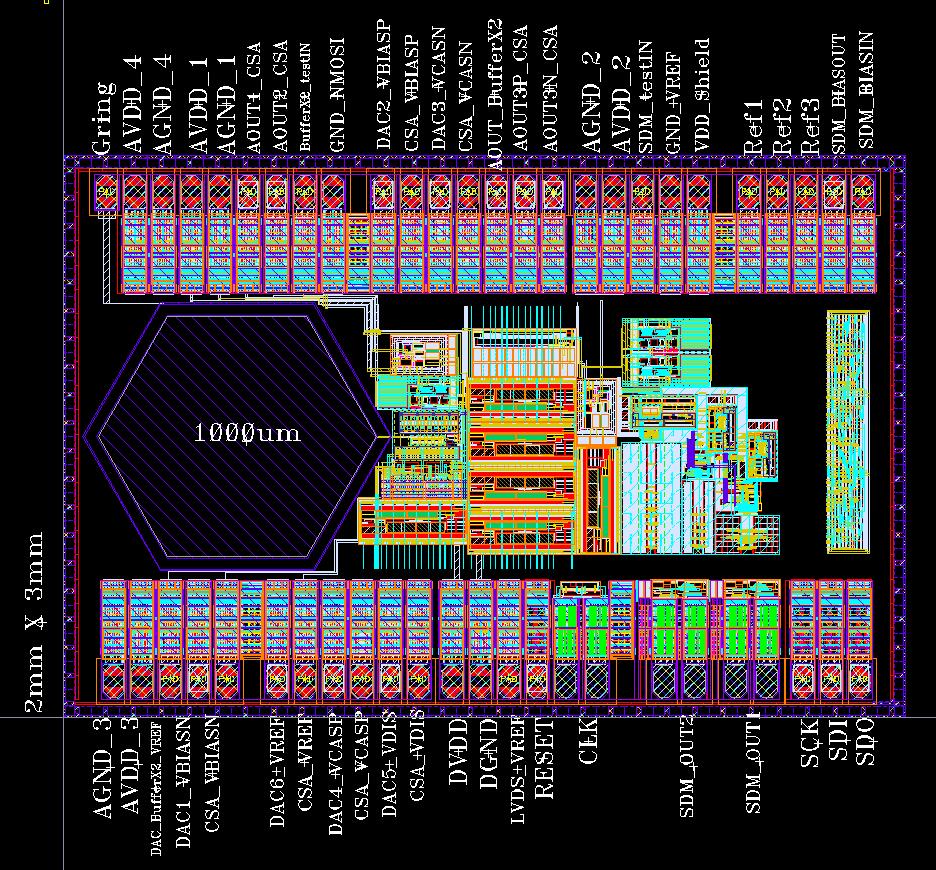}
\hspace{0.5em}%
    \includegraphics[width=0.5\textwidth]{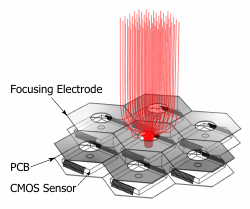}
  \caption{(Left) Design layout of the \TMS CMOS sensor.  
  (Right) \TMS CMOS sensors tiled in a hexagonal pattern to form a charge readout plane without gas gain.  
  Focusing electrode with hole-pattern matching the sensor array is placed above the array to improve the charge collection efficiency.  Red lines are simulated charge drifting paths.}
  \label{fig:tmTPC}
\end{figure}

\TM was conceived for imaging the charge cloud in a TPC with high
spatial resolution, therefore in its first two versions, a high density pixel array
($\sim\SI{80}{\micro m}$ pitch size between pixels) was implemented in each CMOS sensor.  The
sensor development has been lead collaboratively by Lawrence Berkeley National Lab (LBNL)  and  Central
China Normal University (CCNU).  For \Piii, since NLDBD tracks are as long as tens of \si{cm},
pixel density could be greatly reduced provided charge collection efficiency and noise do not
deteriorate.  A new sensor version, \TMS, with only one charge collection node on a single
sensor, with the intention to tile many such sensors on a large plane, is being developed and
optimized specifically to meet the needs of this experiment.  The layout of the very first
version is shown in Fig.~\ref{fig:tmTPC}~(Left).

A conceptual design with \TMS arranged in hexagonal pattern is shown in Fig.~\ref{fig:tmTPC}~(Right).
The electrode size is limited to $<\SI{1}{mm}$ diameter to reduce its capacitance to ground which in
turn reduces the noise.  The pitch is determined under the guidelines in Tab.~\ref{tab:param}.  A
focusing electrode with perforated round-hole pattern matching the array is placed above the
plane with openings aligned with each electrode concentrically.  When biased correctly, virtually
every charge drifting in the gas will eventually land on one of the \TMS electrodes, thus
minimizing the charge loss.  Digitized signals coming out of each sensor are routed through the
PCB substrate.

A $\sim\SI{10}{cm}$ diameter prototype of \TM charge plane is being produced and tested in a high
pressure chamber at LBNL as of 2016.  Once the technology is proven, a full-scale system will be
built and deployed in the \Piii as an upgrade to the default Micromegas readout system,
targeting a \SI{1}{\percent} FWHM energy resolution with full tracking information.

\subsection{Alternative readout technologies}
Several developments are taking place in Zaragoza and IRFU groups on the different electron read-out detection technologies.
The aim is to improve the energy resolution of the Micromegas detector, with an objective to reach a resolution of 1\% at 2~MeV.
A study on direct read-out of primary electrons is also conducted using pads on kapton boards read by low-noise read-out chips.
In case of success full scale detector planes will be developed to be installed in future PandaX-III modules.

\subsubsection{Evolution of Microbulk detectors}

\begin{figure}[tb]
\centering
\includegraphics[width=\textwidth]{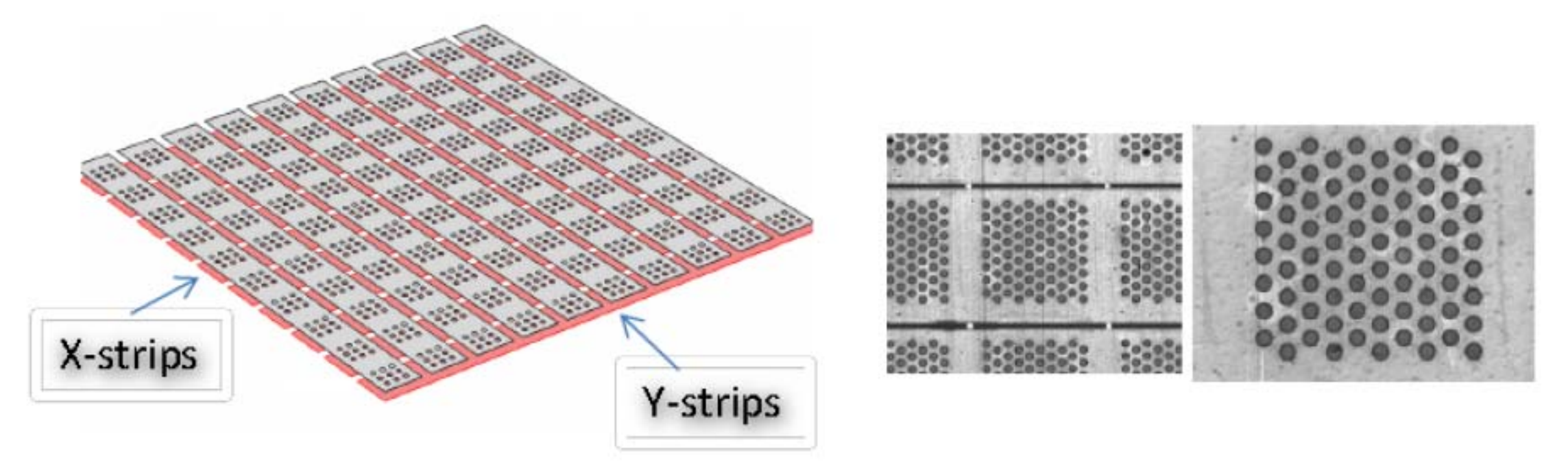}
\caption{(Left) Scheme of segmented mesh Microbulk. 
    (Right) The mesh strips give particle position in Y direction while the anode strips give position in X. Zoomed picture of Microbulk holes on mesh strips.}
\label{fig:Seg_Mesh}
\end{figure}

The baseline Microbulk Micromegas detectors are based on 50~$\mu$m-thick Kapton foils, which are easily available and well optimized for gaseous amplification at atmospheric pressure.
For high-pressure detection, better performance could be expected with thinner gaps~\cite{Giomataris:1998rc,Attie:2014pra}. 
More systematic studies on performance of amplification processes in xenon-based gas mixture with different gap thicknesses are foreseen at IRFU.
Copper-clad Kapton foils with thickness of 25 or 12.5~$\mu$m can only be obtained if a large quantity of such materials is ordered at once.
Alternatively, more standard, commercially available thin polyimide material can be used, but the etching would have to be done with laser or plasma etching, instead of chemical etching. 

Segmented mesh Microbulk Micromegas is also considered independently.  
The mesh is segmented on 1 or 2 mm-wide strips in the X direction while the anode is segmented on strips of the same size in Y direction (Fig.~\ref{fig:Seg_Mesh}). 
The energy measurement can then be done on both electrodes independently, which would enhance its resolution. 
Preliminary tests~\cite{Geralis:2014sxa} were done with 1~mm-wide segmented structure which gave energy resolution of 13.4\% at 5.9~keV. 
The difficulty of such a structure lies to the fact that high voltage has be to be applied on the mesh strips. 
Specific connections have to be foreseen to apply this high voltage value and to decouple it from the readout electronics.
Both thinner gap and segmented mesh evolutions will also be combined together in a second phase of the R\&D.

\subsubsection{Thin mesh bulk Micromegas}
\begin{figure}[tb]
\centering
\includegraphics[width=.7\textwidth]{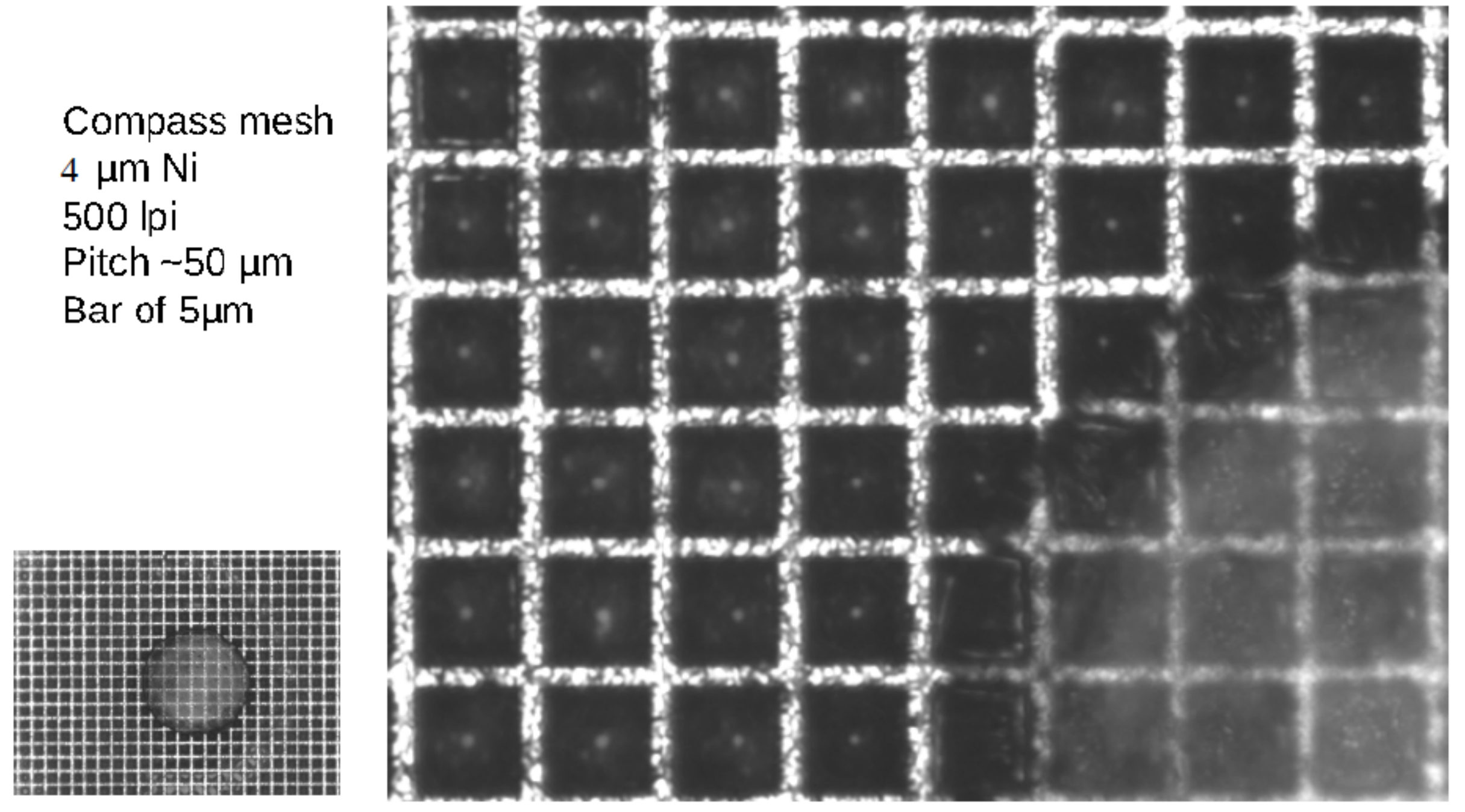}
\caption{Picture of bulk Micromegas detector with a 4~$\mu$m thick electro-formed nickel mesh with
500 LPI pitch. 10~cm-large prototypes were already produced.}
\label{fig:Bulk_Mesh}
\end{figure}

IRFU is also developing the bulk technology to produce Micromegas detectors with thin non-woven meshes, as an alternative to Microbulk Micromegas.
Compared to woven stainless-steel meshes presently used in bulk Micromegas, a thin flat mesh would define a more homogeneous amplification field and thus an well defined detector gain.
The bulk technology can use several types of photosensitive layers of different thicknesses like 25 or 12.5~$\mu$m,
and avoids the availability issue of thin ($\textless 50~\mu$m) copper-clad Kapton foils.

10~cm-large prototypes were produced with 4~$\mu$m-thick electro-formed nickel meshes placed 50 and 100~$\mu$m
above the anode plane (Fig.~\ref{fig:Bulk_Mesh}). Very preliminary performance measurements show energy resolutions at the level
of 13\% at 5.6~keV with 50~$\mu$m amplification gap and 100~$\mu$m-diameter support pillars.
R\&D is ongoing to develop larger anode planes up to a size of 20 to 30~cm. 
Evaluation of other mesh materials such as copper are also foreseen. 

\subsubsection{Read-out boards without amplification}
Direct read-out of primary ionization electrons without prior amplification is also considered at IRFU.
The set-up would be based on an anode plane formed of small pads etched on a Kapton board, and connected to a low-noise read-out
electronics directly mounted on the other side of the board. 
Such low-noise multi-channel read-out chips already exist, such as the 32-channels IDeF-X chips~\cite{Gevin:2007gug}
which have a intrinsic noise of 30~$e^-$, and an overall noise lower than 100~$e^-$ when they are connected to a 10~pF input capacitance. 
The read-out plane design will feature small read-out pads (1 to 2~mm large) in order to keep a low input capacitance,
placed at a pitch of $\sim$5~mm, each pad being surrounded by HV field strips to focus electrons on the pads.
The set-up would be completed by a Frisch mesh a few millimeter above the board.
Compared to a collection of ionization electrons estimated at the level of 10000~$e^-$ per pad for a deposited energy of 2.5~MeV,
an electronic noise lower than 100~$e^-$ would lead to an energy resolution at the level of 1\%, which would be competitive
compared to Micromegas read-out. 
With this solution the difficulty to produce large-size Microbulk or bulk boards is avoided, and a design with larger read-out boards
can be considered. 
As read-out electronics has to be placed closer to the TPC compared to the Micromegas solution, a particular attention should be paid
on the radio-purity of the read-out chips.

This solution is still conjectural, and should be first evaluated with Garfield simulations and with tests on small prototypes. 
Low-noise chips could be developed for this specific usage or in the framework of the development of a chip dedicated to TPC read-out,
with an important attention payed on the radiopurity of the chip. 
Optimization of the anode design would have also an important impact on the detector performance.

\subsection{Absolute timing with IR measurement}
As described in Section~\ref{sec:TPC}, the admixture of TMA to the xenon converts all the direct VUV scintillation light into free drifting electrons. 
The absolute timing, or the $t_0$ of an event, is not measured in \Piii and will not be available for the reconstruction of the event and the background discrimination. 
However it is possible that for some cases the knowledge of the $t_0$ could make a decisive difference. 
Xenon is known to also emit photons in the 850 nm to 1200 nm range additional to the 178 nm VUV light~\cite{Neumeier:2015ori}. 
Additional IR sensitive photo detector could add the missing determination of the $t_0$.  
Given the relatively high energy of about 2.5~MeV, the 
IR detection does not have to be very efficient and a full coverage of the surface is therefore not required. 
Acrylic transmits IR quite well up to 2800 nm with a maximum in the 800 to 1200 nm range
The resistive coating on the inside of the panels can be a mesh structure with transparent gaps.
IR sensors can be envisioned to view the active volume through the acrylic field cage. 
This option will not be implemented in the base design, but might be added in future when deemed necessary.

%% file: Conclusions.tex

\section{Conclusions}
\Piii, the new neutrinoless double beta decay projection in China, aims to search for NLDBD of \xeots with high pressure gas TPC at CJPL.
The collaboration will take a phased approach. 
In the first phase of the experiment, we will construct a 200-kg TPC module. 
The half-life sensitivity to \xeots NLDBD will reach $10^{26}$ year after 3 years of run time. 
In the second phase, the collaboration will build more modules, with potential improvements in terms of energy resolution and background control, to have a total of one ton active mass.
With three years of live time, the full experiment will reach a half-life limit of $10^{27}$ year.

The collaboration has built a prototype, 20-kg scale TPC with Micromegas readout.
Commissioning of this demonstrator is currently under way at SJTU.
Meanwhile, efforts on high pressure vessels, water shielding, electronics, calibration system, simulation, and CJPL infrastructure are being carried out at collaborating institutions. 
The present plan is to start assembling the first TPC module in late 2017 and early 2018 and to start commissioning the experiment in late 2018 and early 2019.
Data taking of the first phase will take place in 2019.

%% file: Acknowledgment.tex
\section*{Acknowledgments}
The authors thank Shanghai Jiao Tong University (SJTU) for their financial and technical support. We also appreciate technical and administrative assistance from China Jinping Underground Laboratory (CJPL). The authors especially thank Xiang Xiao, a former graduate student at the SJTU group, who made significant contribution to the design and construction of the prototype TPC. Much of the R\&D work on low background Microbulk Micromegas readout on which this proposal is based has been performed by the Zaragoza group under the T-REX project, funded by the European Research Council under grant ERC-2009-StG-240054. This work was supported by National Key Programme for Research and Development (NKPRD) Grant \#2016YFA0400300 from Ministry of Science and Technology (MOST), China and by Ministry of Education (MOE), China.